\documentclass[fleqn,usenatbib]{mnras}
\usepackage{amsmath}
\usepackage{graphicx}
\usepackage{amssymb,txfonts}
\usepackage{breakurl}
\newcommand{\g}{$\gamma$}
\newbox\grsign \setbox\grsign=\hbox{$>$} \newdimen\grdimen \grdimen=\ht\grsign
\newbox\simpropbox
\setbox\simpropbox=\hbox{\raise.5ex\hbox{$\propto$}\llap
 {\lower.5ex\hbox{$\sim$}}}\ht2=\grdimen\dp2=0pt
\def\simprop{\mathrel{\copy\simpropbox}}

\title[Spectral models for relativistic reflection]{Improved spectral models for relativistic reflection}
\author[A. Nied{\'z}wiecki et al.]
{Andrzej Nied{\'z}wiecki,$^1$\thanks{E-mail: niedzwiecki@uni.lodz.pl (AN), mitsza@camk.edu.pl (MS), aaz@camk.edu.pl (AAZ)} Micha{\l} Szanecki$^2$\footnotemark[1] and Andrzej A. Zdziarski$^2$\footnotemark[1]\\ 
$^1$Department of Astrophysics, {\L}{\'o}d{\'z} University, Pomorska 149/153, 90-236 {\L}{\'o}d{\'z}, Poland\\
$^2$Nicolaus Copernicus Astronomical Center, Polish Academy of Sciences, Bartycka 18, PL-00-716 Warszawa, Poland\\
}

\pagerange{\pageref{firstpage}--\pageref{lastpage}}
\pubyear{2019}

\begin{document}
\maketitle

\label{firstpage}

\begin{abstract}
We have developed improved spectral models of relativistic reflection in the lamppost and disc-corona geometries. The models calculate photon transfer in the Kerr metric and give the observed photon-energy spectra produced by either thermal Comptonization or an e-folded power law incident on a cold ionized disc. Radiative processes in the primary X-ray source and in the disc are described with the currently most precise available models. Our implementation of the lamppost geometry takes into account the presence of primary sources on both sides of the disc, which is important when the disc is truncated. We thoroughly discuss the differences between our models and the previous ones. 
\end{abstract}

\section{Introduction}
\label{intro}

Accreting sources around compact objects often contain both hot, mildly relativistic, plasma and a cold medium, usually an optically-thick accretion disc. Then the emission of the hot plasma not only reaches the observer but also irradiates the cold medium. This process is usually called Compton reflection. The pioneering paper studying this process was that of \citet*{basko74}, who considered reflection of X-rays emitted by the accretion flow from the atmosphere of the companion star in a close binary. Since then, a lot of work on this subject has been done. In particular, \citet*{white88} derived angle-averaged Green's function for reflection from a fully ionized medium, which were extended to include bound-free absorption in \citet{lightman88}. \citet{mz95} then derived the corresponding Green's functions dependent on the viewing angle.

However, those works treated the absorption probability as constant throughout the reflecting medium, while the ionization structure and temperature of that medium do depend on the depth from the surface. Calculations taking into account the effect of that were performed, e.g., in \citet{ross93,ross05,ross07}. The current most comprehensive publicly available treatment of those effects (though for a constant density medium) appears to be that of \citet{garcia10} and \citet{garcia14,garcia16}. In the present work, we use their numerical model.

In the vicinity of black holes (BHs) and neutron stars, these spectra are modified by relativistic effects. \citet{fabian89} gave an approximate treatment of those, considering the main X-ray feature in reflection spectra, the fluorescent Fe K$\alpha$ line. Since then, there have been many studies of the relativistic effects in reflection, e.g., those of \citet*{laor91,dovciak04,nm10,dovciak11,wf12} and \citet{dauser10,dauser13,dauser16}. A popular family of \texttt{xspec} \citep{arnaud96} models, \texttt{relxill}, is based on the latter work. 

The relativistic effects were treated in two main geometries. In one, the primary source of X-rays was assumed to cover an accretion disc, and emit with a prescribed radial profile. In the other, a point-like X-ray source was assumed to be located on the BH rotation axis, and irradiate a surrounding flat disc \citep{mm96,mf04}. Both geometries are included in the \texttt{relxill} model family. The latter has become a popular model for accreting systems in both binaries containing either a BH or a neutron star and for active galactic nuclei (e.g., \citealt{parker14,parker15,degenaar15,keck15,furst15,beuchert17,basak17,xu18, tomsick18}). 

\begin{figure*}
\centering
\includegraphics[height=7cm]{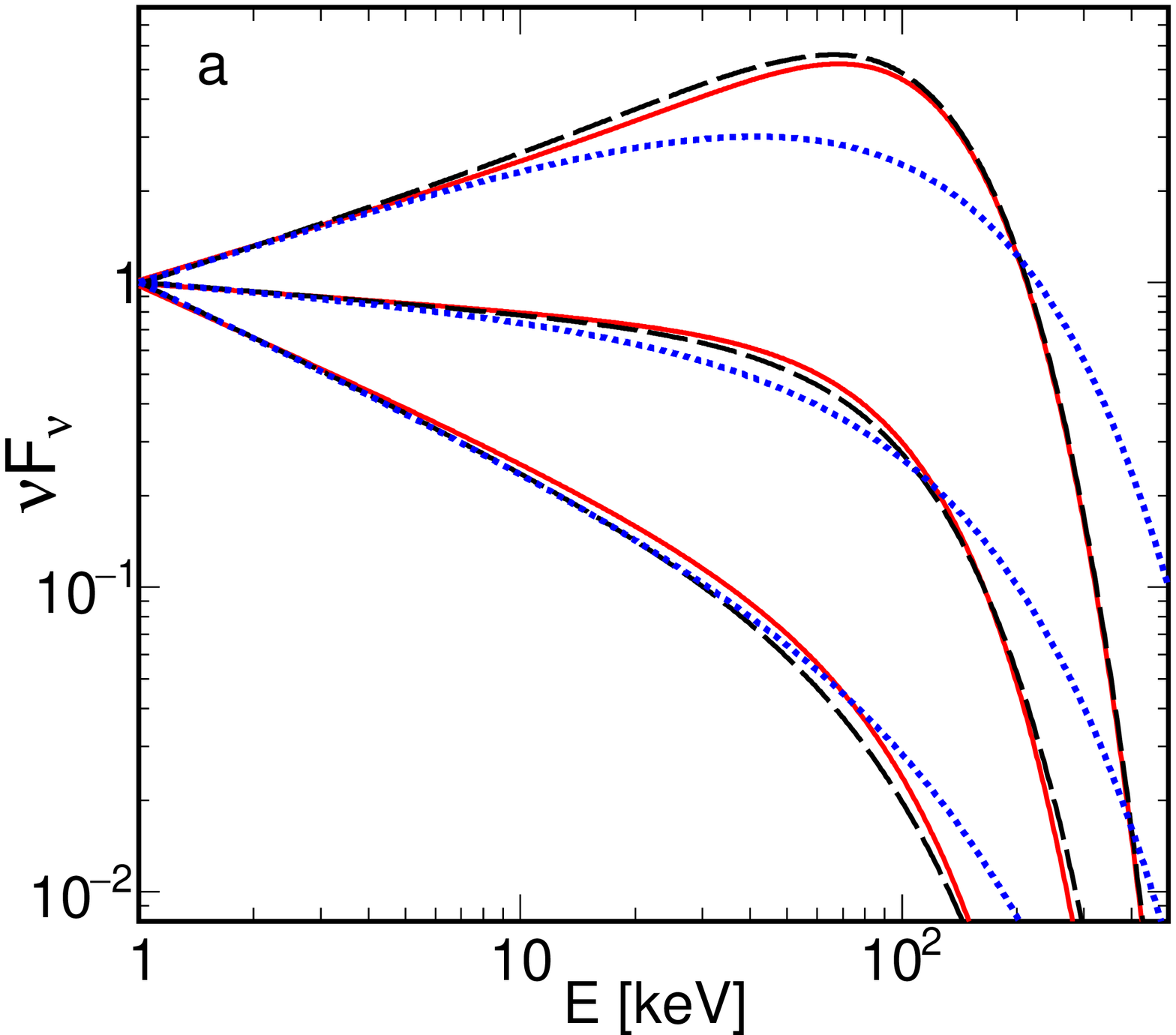}\includegraphics[height=7cm]{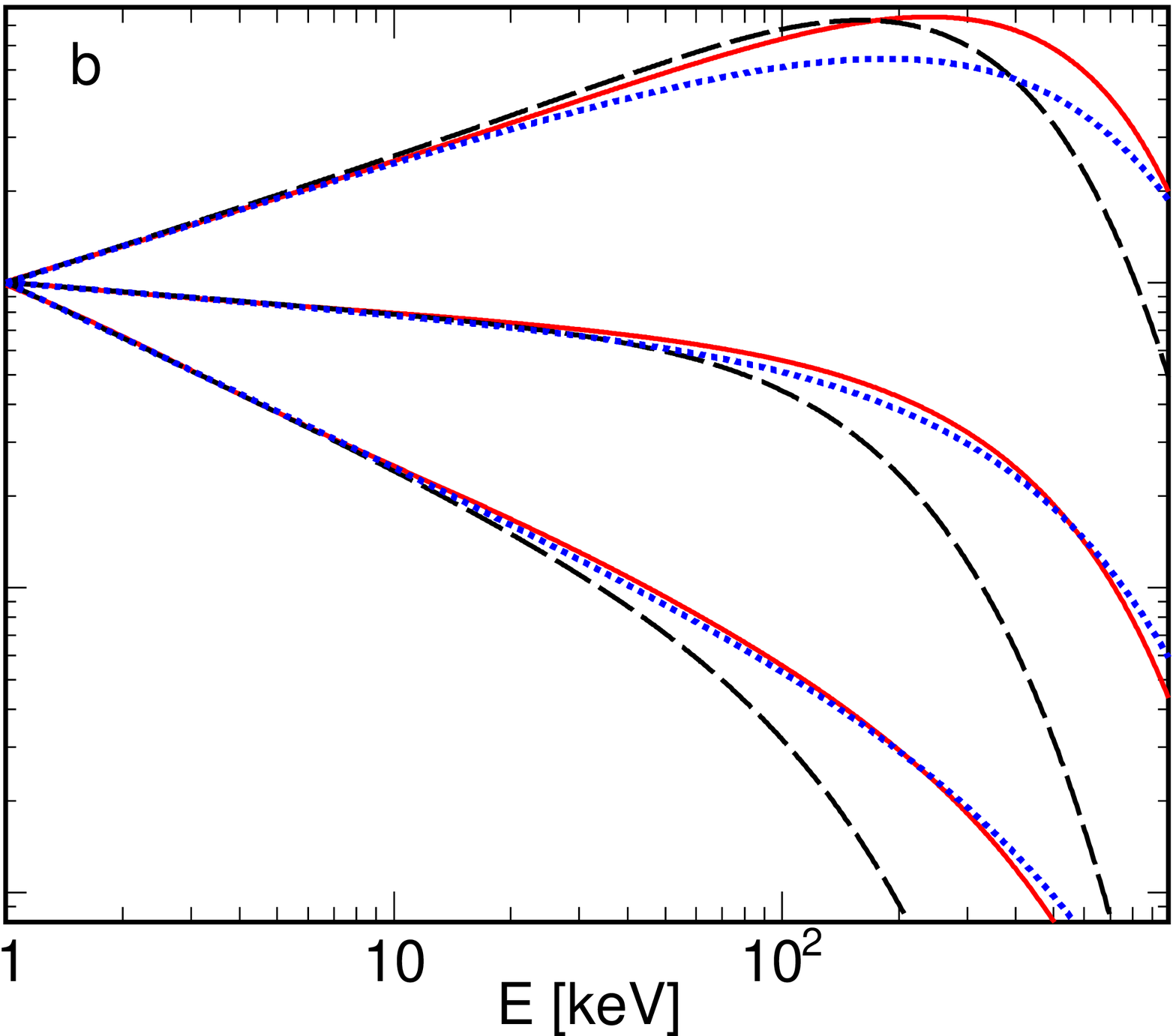}
\caption{Rest-frame thermal Comptonization spectra for $kT_{\rm bb}=1$ eV and $\Gamma=2.6$, 2.1 and 1.6 (from bottom to top) computed with \texttt{compps} in spherical geometry (red solid curves)  and with \texttt{nthcomp}  (black dashed curves). (a) \texttt{compps} uses $kT_{\rm e}=30$ keV and \texttt{nthcomp} uses temperatures representing the mapping shown in Fig.\ \ref{map}, i.e.\ $kT_{\rm e}=53$, 42 and 34 keV for $\Gamma=2.6$, 2.1 and 1.6, respectively. (b) \texttt{compps} uses $kT_{\rm e}=200$ keV and \texttt{nthcomp} uses $kT_{\rm e}=450$ keV. We see that \texttt{nthcomp} significantly underestimates the positions of the high-energy cutoff; correction of the $kT_{\rm e}$ parameter of \texttt{nthcomp}  makes this model consistent with the actual Comptonization spectra only for subrelativistic electron temperatures ($\la 100$ keV). The blue dotted curves show the e-folded power law spectrum for $\Gamma=2.6$, 2.1 and 1.6 and (a) $E_{\rm cut}=123$, 113 and 104 keV and (b) $E_{\rm cut}=551$, 461 and 463 keV, respectively, from bottom to top, which give the same Compton temperatures as the corresponding \texttt{compps} spectra (see Section \ref{rest_frame}).
}
\label{compps_nthcomp}
\end{figure*}

\begin{figure}
\centerline{\includegraphics[height=5.5cm]{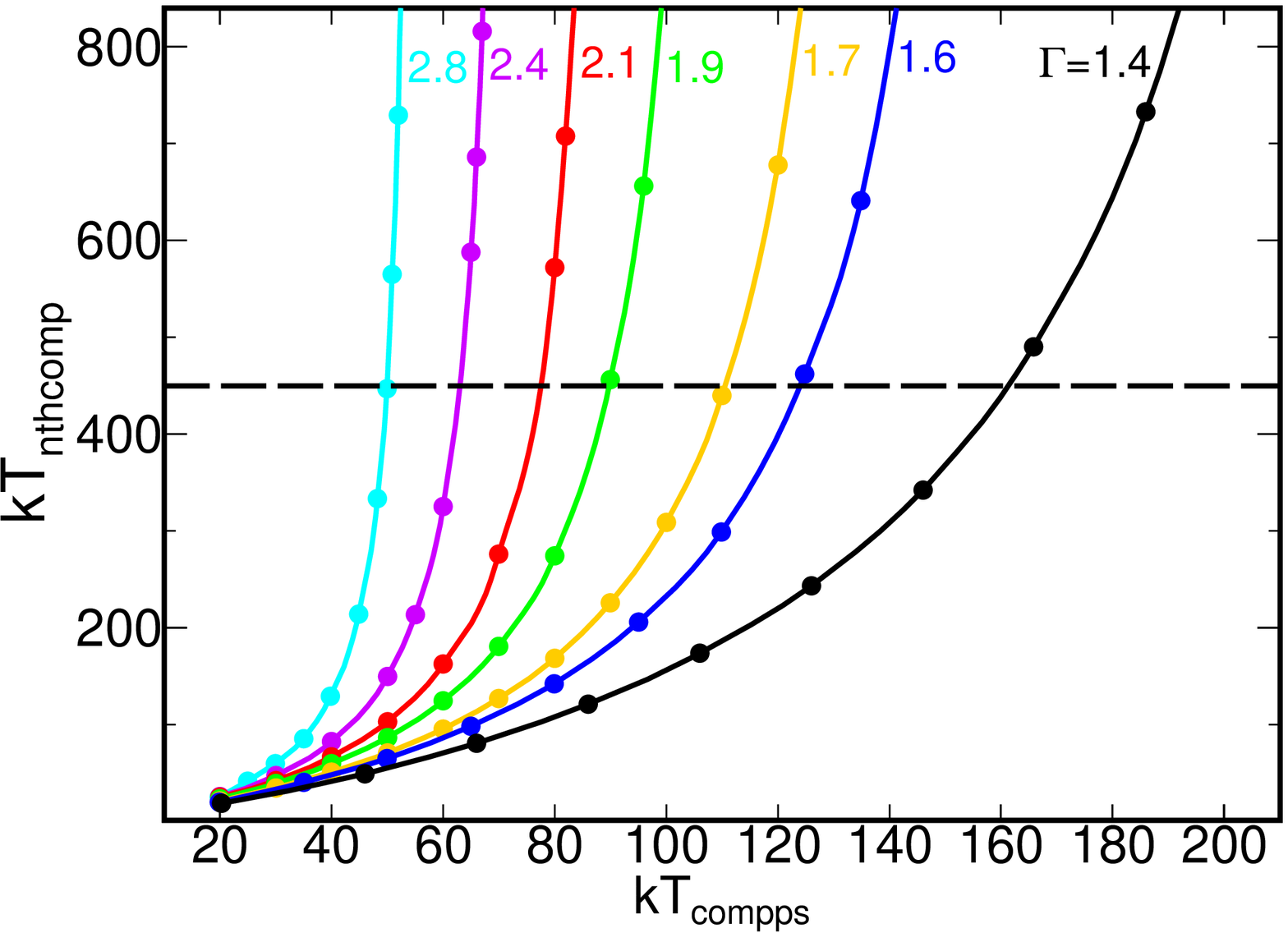}}
\caption{Mapping between the $kT_{\rm e}$ parameters of \texttt{compps} in spherical geometry and \texttt{nthcomp} obtained by fitting the latter model to the former at a fixed $\Gamma$. The curves from left to right present the mapping for $\Gamma=2.8$, 2.4, 2.1, 1.9, 1.7, 1.6 and 1.4. The dashed horizontal line indicates the maximum temperature of 450 keV used for tabulating the reflection spectra in \texttt{xillverCp}.
} 
\label{map}
\end{figure}

However, certain inaccuracies of the \texttt{relxill} models were found by \citet*{niedzwiecki16}. Given those, we have developed new \texttt{xspec} models for relativistic reflection\footnote{The models can be downloaded at \url{users.camk.edu.pl/mitsza/reflkerr}.}. One is \texttt{reflkerr}, which computes the observed primary and reflection spectra for a broken power-law radial emissivity profile, approximating the case of a disc corona, as in the \texttt{relxill} model. Another is \texttt{reflkerr\_lp}, which computes the observed primary and reflection spectra in the lamppost geometry, as in \texttt{relxilllp}. One modification with respect to the previous models is taking into account the primary sources on both sides of the accretion disc, which is important in the presence of disc truncation \citep{nz18}. In this work, we present our formalism and discuss the main improvements with respect to the previous treatments. We then compare our spectra with those of \texttt{relxill} and \texttt{relxilllp}. 

\section{The incident spectrum}
\label{incident}

We choose Comptonization of soft blackbody photons by mildly relativistic thermal electrons as the physical incident spectrum in the rest frame. This process appears to be the dominant one in coronae in the vicinity of inner accretion discs in X-ray binaries (e.g., \citealt*{zg04,dgk07,burke17} and references therein). The accuracy of hard X-ray/soft \g-ray spectra measured from Seyferts and radio galaxies is lower than that for X-ray binaries. Still, high-energy cutoffs compatible with thermal Comptonization are commonly observed, e.g., \citet{madejski95}, \citet*{zjm96}, \citet{gondek96}, \citet*{zpj00}, \citet{zg01}, \citet{malizia08}, \citet{lubinski10,lubinski16}, \citet{marinucci14}, \citet{brenneman14}, \citet{ballantyne14}, \citet{balokovic15}, \citet{fabian15}.

In order to reproduce this spectrum, we use the \texttt{compps} model \citep{ps96}, which gives an excellent agreement with our Monte Carlo simulations for any set of thermal plasma parameters with the Thomson optical depth of $\tau\la 3$, e.g., \citet{zpj00,z03}. That model allows a choice of the plasma geometry and the distribution of seed photons, see Appendix \ref{compps}. In the case of the lamppost source, our relativistic model, \texttt{reflkerr\_lp}, assumes a spherical plasma geometry. In the case of the coronal geometry, with a corona above a disc, our model of relativistic effects, \texttt{reflkerr}, assumes either a slab or a local spherical plasma geometry. The latter corresponds to a large number of spherical active regions above the disc surface.

\begin{figure}
\centering
\includegraphics[width=8.4cm]{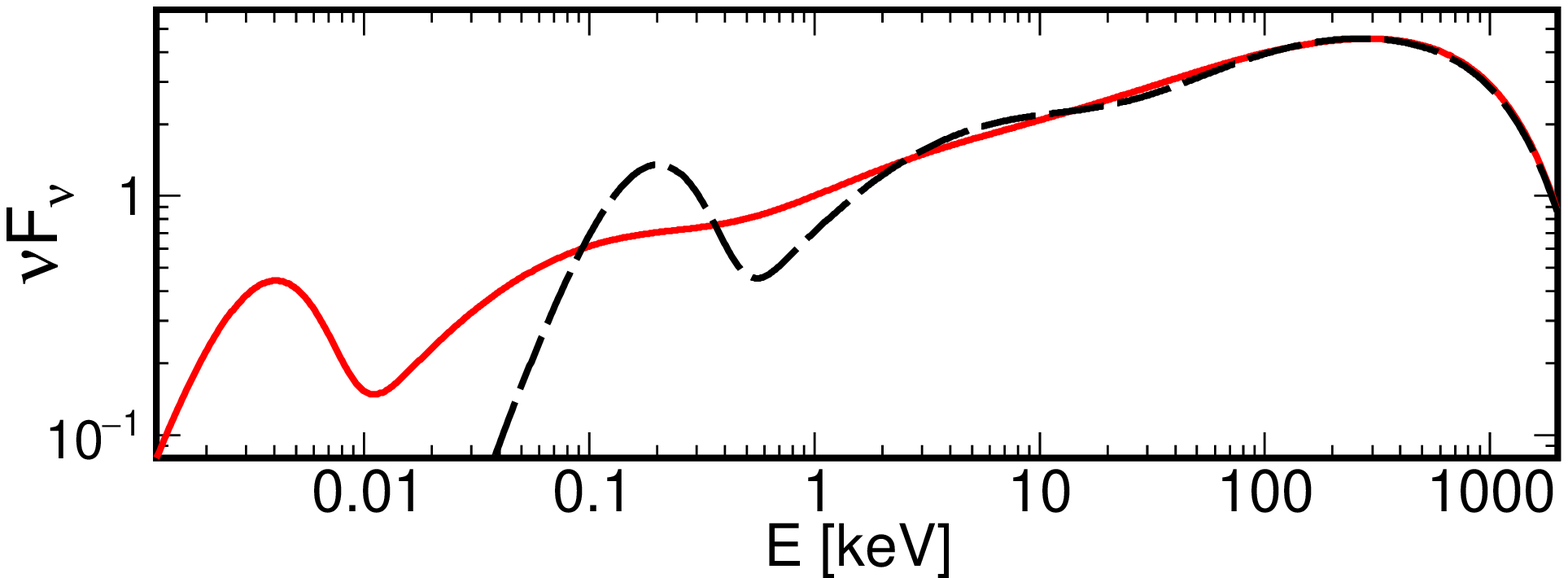}
\caption{Rest-frame thermal Comptonization spectra in spherical geometry calculated with \texttt{compps} for $kT_{\rm bb}=1$ eV (red solid curve) and 50 eV (black dashed curve), at $kT_{\rm e}=400$ keV and $\tau = 0.32$, yielding $\Gamma \simeq 1.7$. We see the seed-photon blackbody spectra (slightly attenuated due to up-scattering) on the left-hand side. The first two scattering orders are clearly visible, which effect is much less visible at $E\ga 1$ keV in the case of $kT_{\rm bb}=1$ eV.
}
\label{orders}
\end{figure}

\begin{figure}
\centering
\includegraphics[width=8cm]{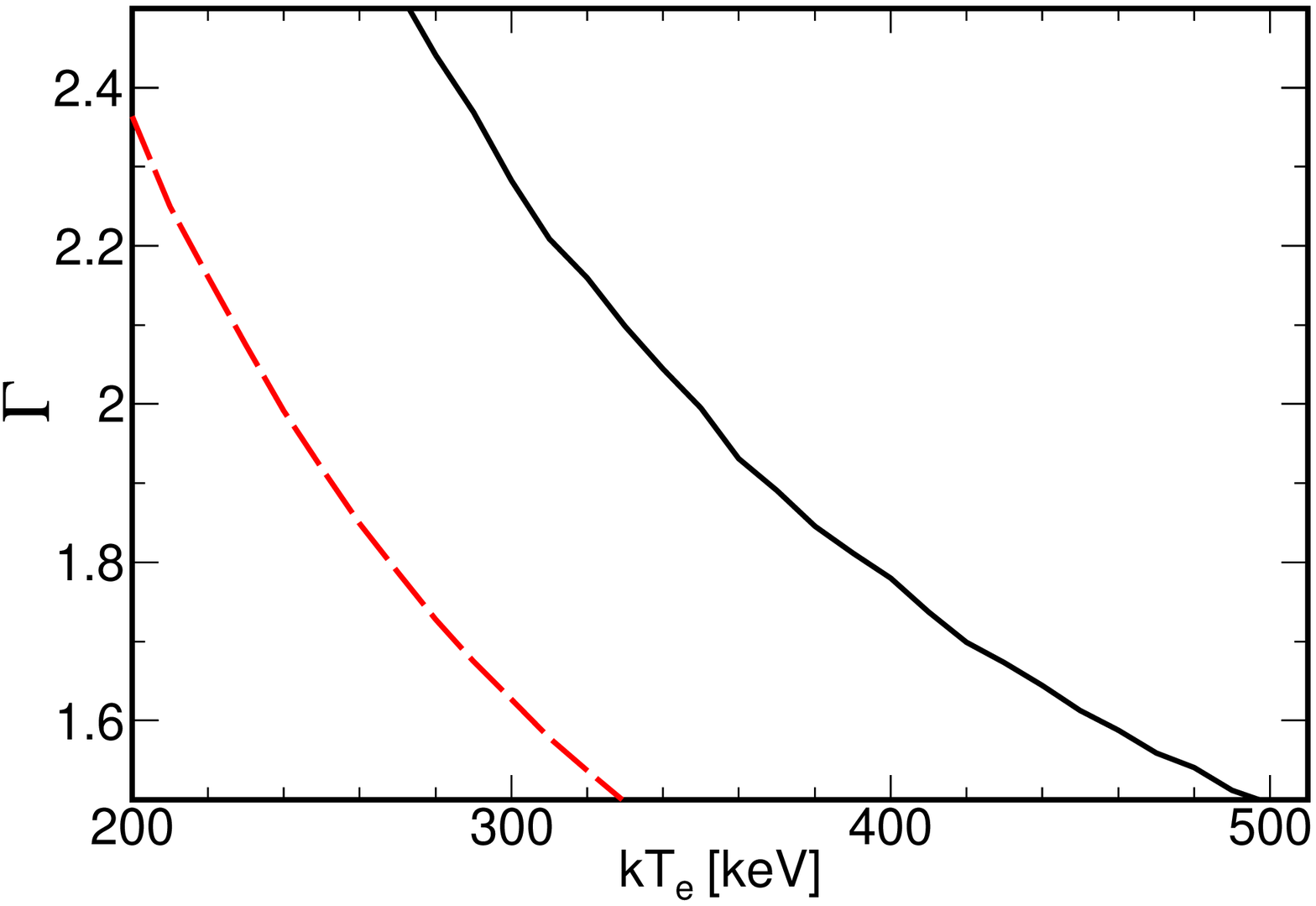}
\caption{Approximate limits for a power-law shape of thermal Comptonization spectra in the spherical geometry for $kT_{\rm bb}=1$ eV (black solid) and $kT_{\rm bb}=50$ eV (red dashed). At the electron temperatures above those given by the curves, the difference of the spectral indices fitted in the 2--5 keV and 8--15 keV ranges exceeds 0.05.
}
\label{boundary}
\end{figure}

We note that \texttt{compps} is the most accurate thermal Comptonization model currently implemented in \texttt{xspec}. The model \texttt{nthcomp} \citep{zjm96}, which is used in \texttt{relxilllpCp}, correctly approximates the Comptonization spectrum only for $\tau \ga 1$ and it strongly underestimates the high-energy extent of the spectrum for electron temperatures $kT_{\rm e} \ga 100$ keV, see Fig.\ \ref{compps_nthcomp}. This problem can be partially reduced by applying a correction for the $T_{\rm e}$ parameter in \texttt{nthcomp}; however, such a corrected model still agrees with the actual Comptonization spectra only below some maximum, $\Gamma$-dependent temperature, see Fig.\ \ref{map}, where $\Gamma$ is the X-ray photon spectral index ($N(E)\propto E^{-\Gamma}$), approximately determined by the product of $T_{\rm e}$ and $\tau$.   This is a shortcoming for lamppost models with a low height above the horizon, where we would not be able to properly model the observed spectra extending up to $\sim 100$ keV in the rest-frame with \texttt{nthcomp} when the gravitational redshift of the direct lamppost emission is taken into account.

Furthermore, \texttt{compps} properly describes departures from a power-law at relativistic electron temperatures, with individual scattering orders seen in the Comptonization spectra. In such cases, the spectra in the X-ray range depend on $T_{\rm bb}$, as illustrated in Fig.\ \ref{orders}. For $kT_{\rm bb} \ga 10^2$ eV, which is the case of seed photons produced by thermal emission from an accretion disc in a binary, such departures are seen at $kT_{\rm e} \ga 300$ keV. For $kT_{\rm bb} \la 1$ eV, which case approximates Comptonization of seed photons produced by cyclo/synchrotron emission, the X-ray spectrum is formed by mixing of the 2nd and higher scattering orders, which gives a shape closer to a power-law. In Fig.\ \ref{boundary}, we show an approximate limit on $kT_{\rm e}$ for a given $\Gamma$ for the spherical geometry and two values of $kT_{\rm bb}$, above which the Comptonization spectrum deviates significantly from a power-law. The assumed criterion of the difference of the spectral indices fitted in the 2--5 keV and 8--15 keV ranges exceeding 0.05 appears relevant for spectral fitting with data quality available from {\it Suzaku} and {\it NuSTAR} observations of bright Seyfert galaxies.

In order to facilitate comparison with other models parametrized by the photon spectral index, $\Gamma$, we have also tabulated the $\Gamma (\tau, T_{\rm e})$ relationship at $kT_{\rm bb}=1$ eV for a sphere and slab (\texttt{geometry}=0, 1, respectively), where $\Gamma$ is the photon index fitted in the 1--20 keV range. Below the limit shown in Fig.\ \ref{boundary}, the difference between the \texttt{compps} spectrum and the fitted power-law is $\la 5$ per cent above 1 keV. Inverting the fitted relationship to $\tau(\Gamma, T_{\rm e})$ allows us to use $T_{\rm e}$ and $\Gamma$ as free parameters in a version of our model. The number intensity of the scattered photons in the local rest-frame is denoted below by $N_{\rm PS}(E)$, and the explicit dependence on the model parameters is indicated as either $N_{\rm PS}(E; T_{\rm e}, T_{\rm bb}, \tau)$ or $N_{\rm PS}(E; T_{\rm e}, \Gamma)$.

\begin{figure}
\begin{center}
\includegraphics[width=7.2cm]{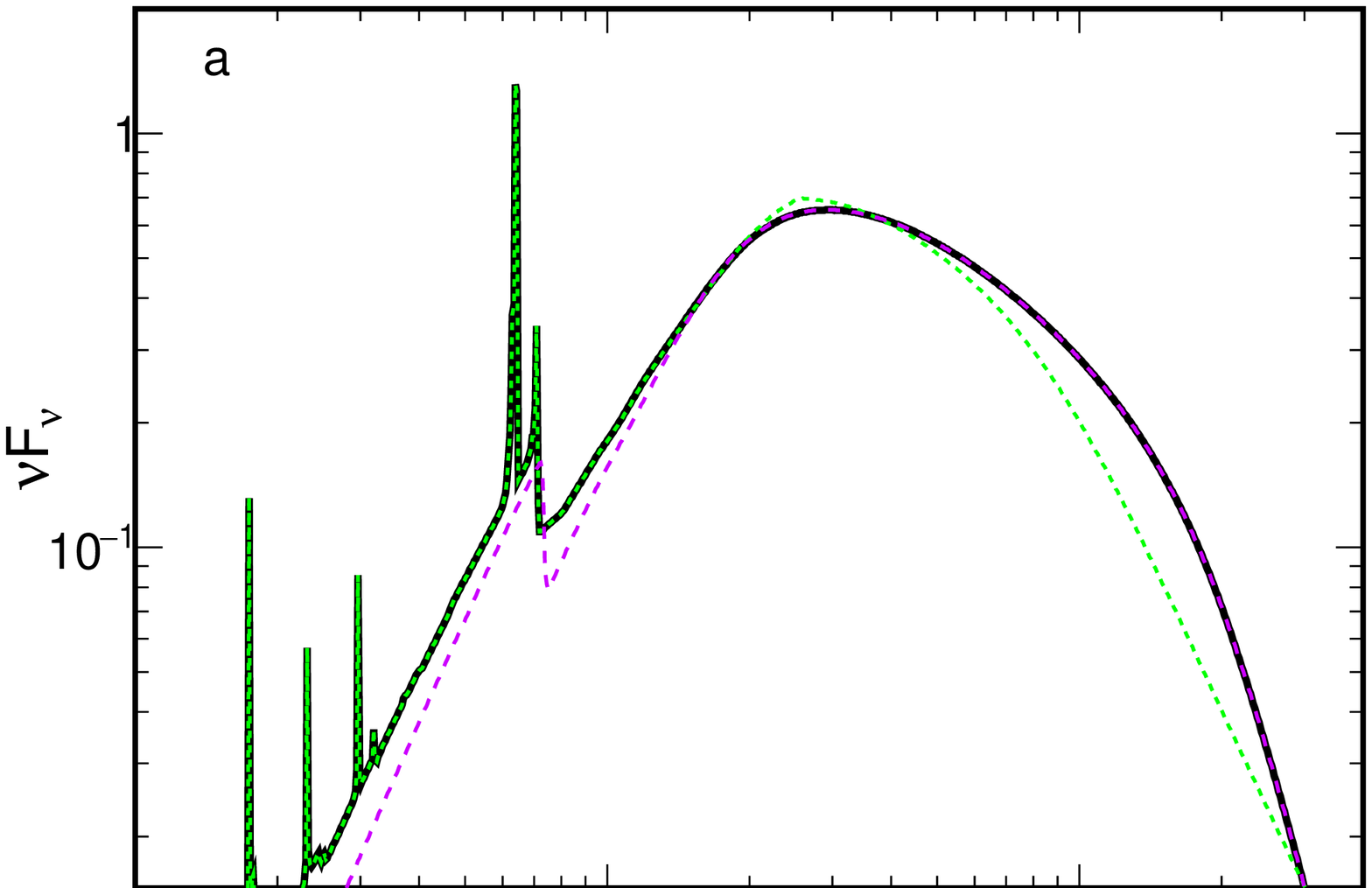}
\includegraphics[width=7.2cm]{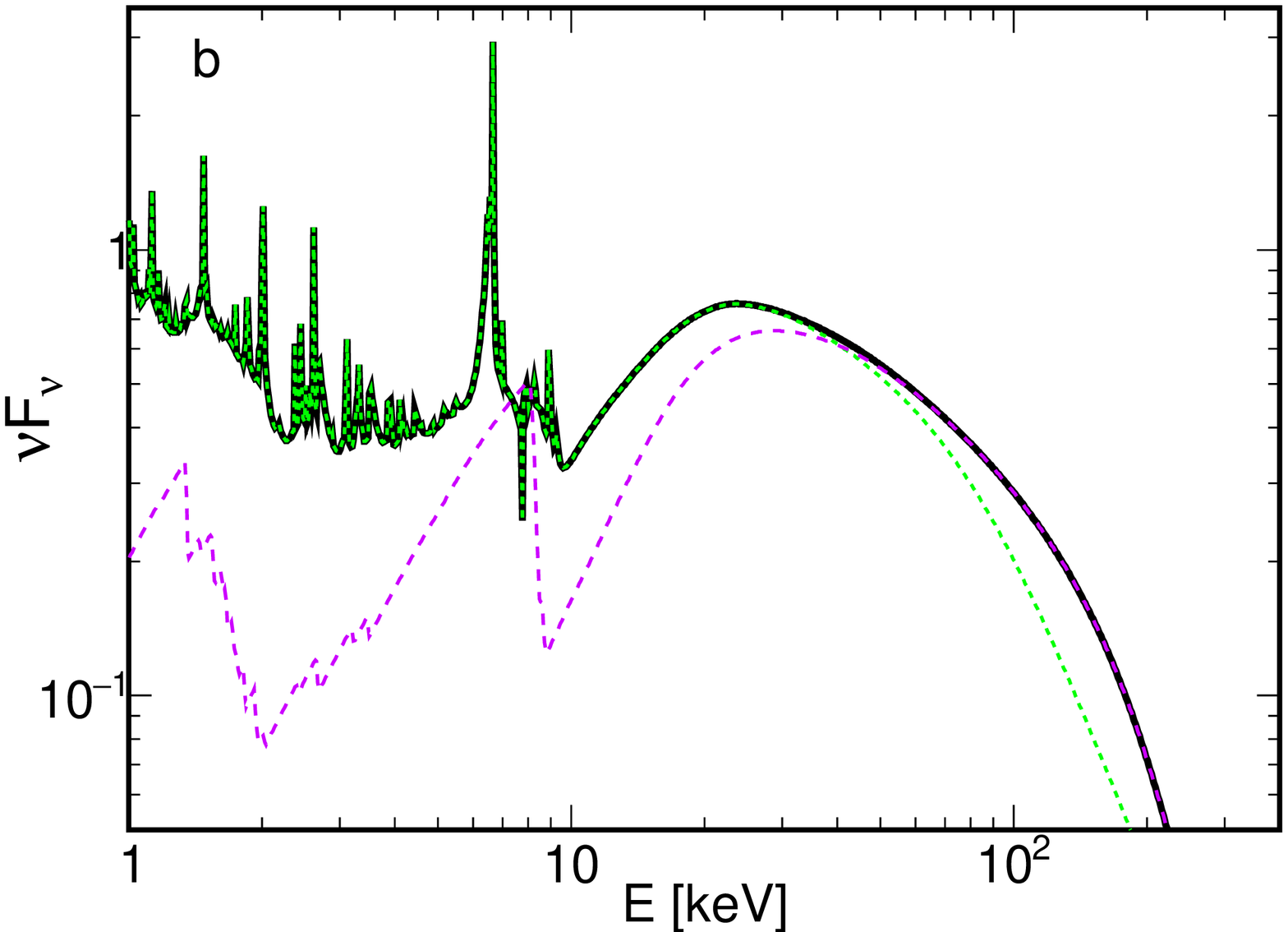}
\end{center}
\caption{Example rest-frame reflection spectra for $\Gamma=2.1$, $kT_{\rm bb}=1$ eV, $\theta_{\rm o} = 30\degr$ and (a) $\xi=1$ and (b) $\xi=1000$. The dashed magenta and solid black curves show the reflection spectra of \texttt{ireflect} and our model \texttt{hreflect} (using \texttt{xillverCp}), respectively, for $kT_{\rm e}=200$ keV. The green dotted curves show the reflection spectra of \texttt{xillverCp} for $kT_{\rm e}=450$ keV. In all models, the irradiating radiation spectra are normalized at 1 keV.}
\label{merging} 
\end{figure}

\begin{figure}
\centering
\includegraphics[width=8cm]{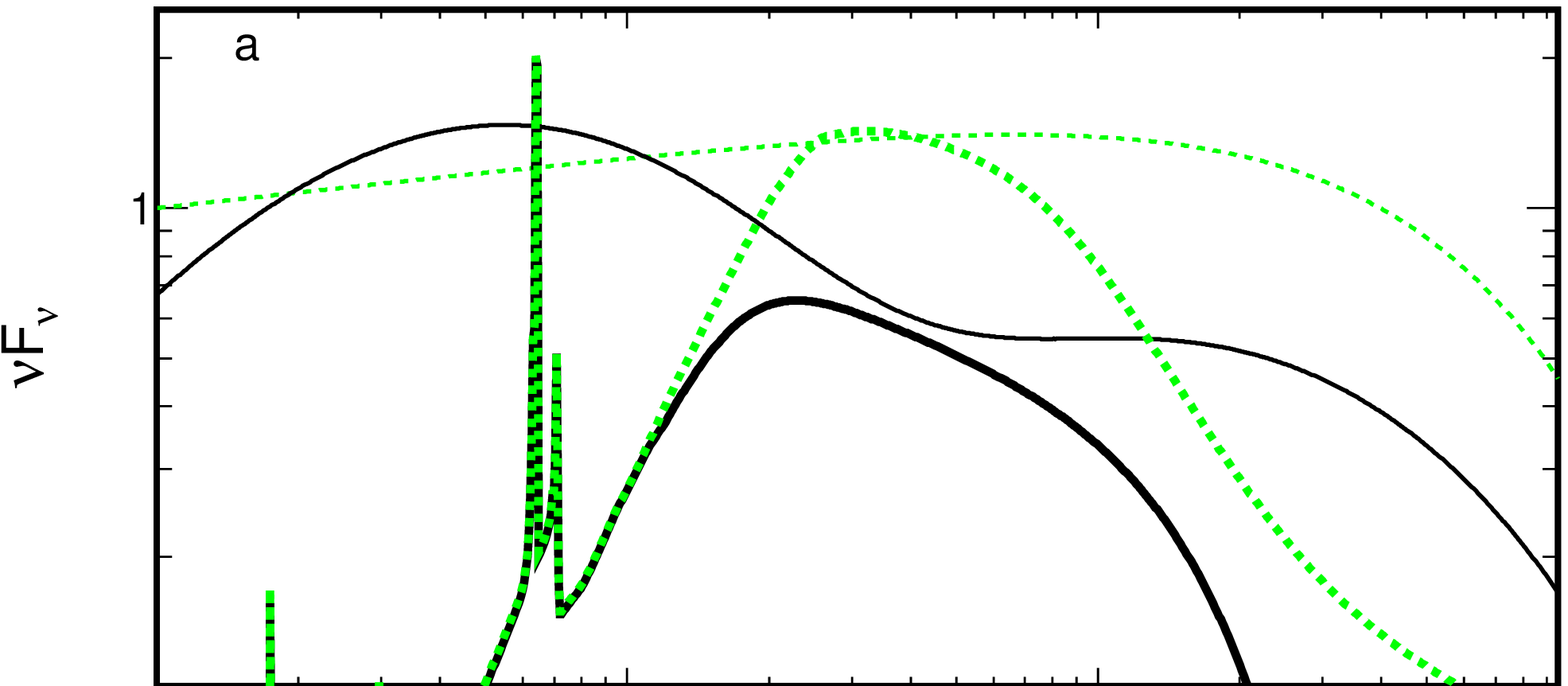}
\includegraphics[width=8cm]{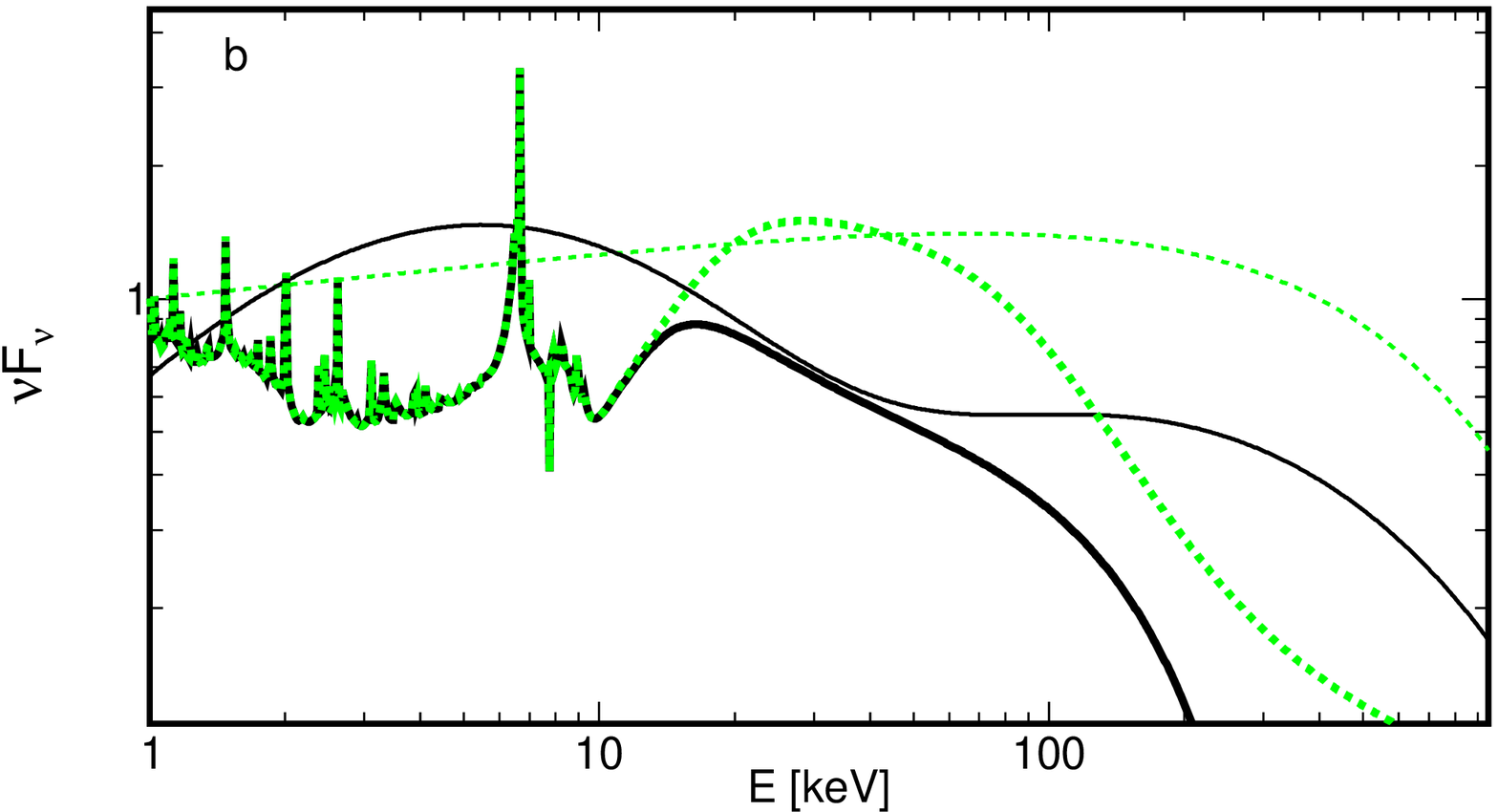}
\caption{Example rest-frame reflection spectra at a high electron temperature, $kT_{\rm e}=450$ keV. The thick solid black curves show the reflection spectrum of our model \texttt{hreflect} (using \texttt{xillver} which is more suitable than \texttt{xillverCp} at high $T_{\rm e}$, see Appendix \ref{hreflect}) for (a) $\xi=1$ and (b) $\xi=1000$ and the incident spectrum with $\Gamma\simeq 1.9$ (fitted in the 1--20 keV range and corresponding to $\tau \simeq 0.05$), $kT_{\rm bb}=40$ eV and $\theta_{\rm o} = 30\degr$. The thin solid black and dotted green curves show the incident spectra of \texttt{compps} (used in \texttt{hreflect}), and the e-folded power-law with $\Gamma=1.9$ and $E_{\rm cut}=630$ keV (used in \texttt{xillver}; the $E_{\rm cut}$ value follows from the equal-$T_{\rm C}$ condition, see text), respectively. The thick dotted green curve shows the reflection spectrum of \texttt{xillver}. We see that at high electron temperatures, the incident spectrum of \texttt{xillver}  strongly deviates from the actual Comptonization spectrum, and consequently its reflected spectrum strongly deviates from that of \texttt{hreflect}.
}
\label{high_T_e}
\end{figure}

\section{Rest-frame reflection}
\label{rest_frame}

The spectrum of the reflected radiation in the disc rest frame used by us at low X-ray energies, namely below $E_{\rm merge}\simeq 10$--30 keV (with the exact value dependent on the ionization parameter), is given by that of the constant-density model \texttt{xillver} (\citealt{garcia10} and following work). The ionization parameter is defined in a usual way (e.g., \citealt{garcia10}),
\begin{equation}
\xi\equiv\frac{4\upi F_{\rm irr}}{n_{\rm e}},
\label{xi}
\end{equation}
where $F_{\rm irr}$ is the irradiating flux in the 13.6 eV--13.6 keV photon energy range, and $n_{\rm e}$ is the electron density of the reflecting medium. In the current version, $n_{\rm e}=10^{15}$ cm$^{-3}$ (characteristic of inner discs in active galactic nuclei) is assumed, following the value used in \texttt{xillver}. For a given $\xi$, $F_{\rm irr}\propto n_{\rm e}$, and thus the effective temperature of the reprocessed emission (neglecting any intrinsic dissipation), is $T_{\rm eff}\simprop n_{\rm e}^{1/4}$. While the total reflected/reprocessed emission is far from a blackbody, the reprocessed part still forms a hump at low energies with $\langle E\rangle\simprop n_{\rm e}^{1/4}$, as seen, e.g., in Fig.\ 4 of \citet{garcia16}. At low densities, $n_{\rm e}\ll 10^{15}$ cm$^{-3}$, the reflector ionization state is independent of $n_{\rm e}$, and the reflected spectra are also only weakly dependent on it, with the main effect being the average energy of the soft, quasi-thermal, hump. However, processes with rates proportional to $n_{\rm e}^2$ (e.g., bremsstrahlung heating and cooling, collisional de-excitation, three-body recombination) become important at high densities, and the ionization state is then significantly dependent on $n_{\rm e}$, see \citet{garcia16}.

For $n_{\rm e}= 10^{15}$ cm$^{-3}$, the reflected spectrum of \texttt{xillver} is the most accurate one available in \texttt{xspec}. At $E\geq E_{\rm merge}$, we use the model \texttt{ireflect} convolved with \texttt{compps}. The convolution model \texttt{ireflect} \citep{mz95} gives an exact reflection for any shape of primary radiation for given reflector opacities. However, its built-in treatment of ionization is rather approximate for any value of $\xi$ (of \citealt{done92}), and it fails at high values of $\xi$. However, at $E\geq E_{\rm merge}$, there are virtually no lines or edges in the reflected spectrum, and that shortcoming is of no importance. Our hybrid model of rest-frame reflection is called \texttt{hreflect}.

For thermal Comptonization models we considered two versions of \texttt{xillver} tables to check how the details of their incident spectra, which only roughly approximate the actual Comptonization spectrum of \texttt{compps} in some range of parameters, see Fig.\ \ref{compps_nthcomp}, affect the soft X-ray part of \texttt{hreflect}. First, we use \texttt{xillverCp} with the incident spectrum of \texttt{nthcomp}, with $T_{\rm e}$ fitted to match the \texttt{compps} spectrum. These fitted incident spectra of \texttt{xillverCp} approximately agree with \texttt{compps} for the (actual) electron temperatures $\la (60$--160) keV, depending on $\Gamma$; however, they underestimate the high-energy extent of the \texttt{compps} spectrum at larger temperatures, see Fig.\ \ref{compps_nthcomp} and \ref{map}. Secondly, we use \texttt{xillver} version \texttt{a-Ec5}, which assumes the e-folded power-law incident spectrum. Here we use the $E_{\rm cut}$ parameter for which  the e-folded power-law gives the same Compton temperature as the \texttt{compps} spectrum\footnote{We use here the non-relativistic definition of the Compton temperature, $4 kT_{\rm C}\equiv \int E F(E) {\rm d}E/\int F(E) {\rm d}E$.}. As we see in Fig.\ \ref{compps_nthcomp}, at high $T_{\rm e}$ such e-folded power-law spectra give a much better approximation of the \texttt{compps} spectra than \texttt{nthcomp}.

We found that after slight rescaling of \texttt{xillver} reflection spectra, they can be robustly combined with \texttt{ireflect} for a large range of parameters. See Appendix \ref{hreflect} for details of our procedure and verification of its accuracy. In most cases the \texttt{hreflect} model using \texttt{xillverCp} gives very similar spectra to that using \texttt{xillver}. For low ionization parameters, where photoionization is the dominant process heating the reflecting material (e.g., \citealt{garcia13}), the soft X-ray part of reflection depends only on the shape of the incident spectrum below $\sim 10$ keV and then using either of the \texttt{xillver} models we get exactly the same \texttt{hreflect} spectrum. For high ionization parameters, where Compton heating dominates, the soft X-ray part of the reflection component may depend on details of the high energy part of the incident spectrum. However, we found that the difference between the \texttt{hreflect} spectra obtained using the two versions of \texttt{xillver} is $\la 10$ per cent even at high $T_{\rm e}$, where the incident spectra of \texttt{xillverCp} and \texttt{xillver} differ  substantially at high energies. The two cases where we could not formally validate our procedure include: (i) $\Gamma \ga 2.4$ and $\xi \ga 1000$, for which  \texttt{xillver} (both versions) predict much flatter spectra than \texttt{ireflect}, and (ii) $\Gamma \la 1.7$ and $\xi \ga 3000$, for which  thermal Comptonization effects are likely overestimated in \texttt{xillver}.

Fig.\ \ref{merging} illustrates the merging of the reflection spectra of \texttt{ireflect} and \texttt{xillverCp} for a $kT_{\rm e}$ below the limit shown in Fig.\ \ref{boundary}, where the low energy part of the \texttt{compps} spectrum has a power-law shape and unambiguously determines the $\Gamma$ parameter of \texttt{xillverCp}, but above the limit where \texttt{nthcomp}, which is the incident spectrum of \texttt{xillverCp}, can reproduce the actual high-energy extent of the \texttt{compps} spectrum. On the other hand, for these parameters the incident spectrum of \texttt{xillver} closely approximates the \texttt{compps} spectrum and although it strongly deviates from \texttt{nthcomp} (see the spectra for $\Gamma=2.1$ in Fig.\  \ref{compps_nthcomp}(b)), the corresponding reflected spectra differ negligibly for any $\xi$ (the spectra for $\xi = 1000$ are compared in Fig.\ \ref{exp_cp}).

At temperatures above the limit shown in Fig.\ \ref{boundary}, the high-energy part of our rest-frame reflection (i.e., that computed with \texttt{ireflect}) is still exact. In order to find the low energy part of reflection, we fit $N_{\rm PS}(E)$ with a power law in the 1--20 keV range and use the fitted $\Gamma$ as the \texttt{xillver} parameter, as illustrated in Fig.\ \ref{high_T_e}. We see that now the actual incident spectrum and its reflection calculated with \texttt{hreflect} are very different from those calculated with \texttt{xillver}. This is a shortcoming of \texttt{xillver}, especially important for cases of the primary source located close to the BH horizon, in which case the primary radiation is blueshifted by a large factor with respect to the observed emission. We also note that even in this regime, the \texttt{hreflect} spectrum formed using \texttt{xillverCp} with $kT_{\rm e}=450$ keV (i.e.\ the largest value available in that model) only weakly differs from that calculated using \texttt{xillver}. We regard the latter as a better approximation in the high-temperature case, with our Compton-temperature criterion better approximating the reflector thermal and ionization structure for large $\xi$.

We note that in computing the rest-frame reflection, we lose the information about the actual angular distribution of the irradiating photons in the disc frame. This is because \texttt{xillverCp} assumes a fixed incidence angle of $45 \degr$, while \texttt{ireflect} convolves the primary spectrum with Green functions averaged over the incidence angle for an isotropic source above a slab.

\section{Relativistic modifications to the incident spectrum and reflection}
\label{relativity}

We have then developed models taking into account the relativistic effects on emission from a given point in the Kerr metric. We take into account those effects acting on the primary spectrum and on the reflection, and we compute them strictly following the method of \citet{niedzwiecki08}. Our convolution of the general-relativistic (GR) effects with the rest-frame radiation spectra makes use of the transfer functions, which, following \citet*{laor90} and \citet{laor91}, are constructed by tabulating a large number of photon trajectories.

We allow the disc to extend down to an arbitrary inner radius above the horizon. We assume the disc to have no intrinsic dissipation, similarly to most of previous works on the subject (but with the exception of, e.g., \citealt{ross07}).

\subsection{The lamppost}
\label{lamppost}

\begin{figure}
\centering
\includegraphics[width=8cm]{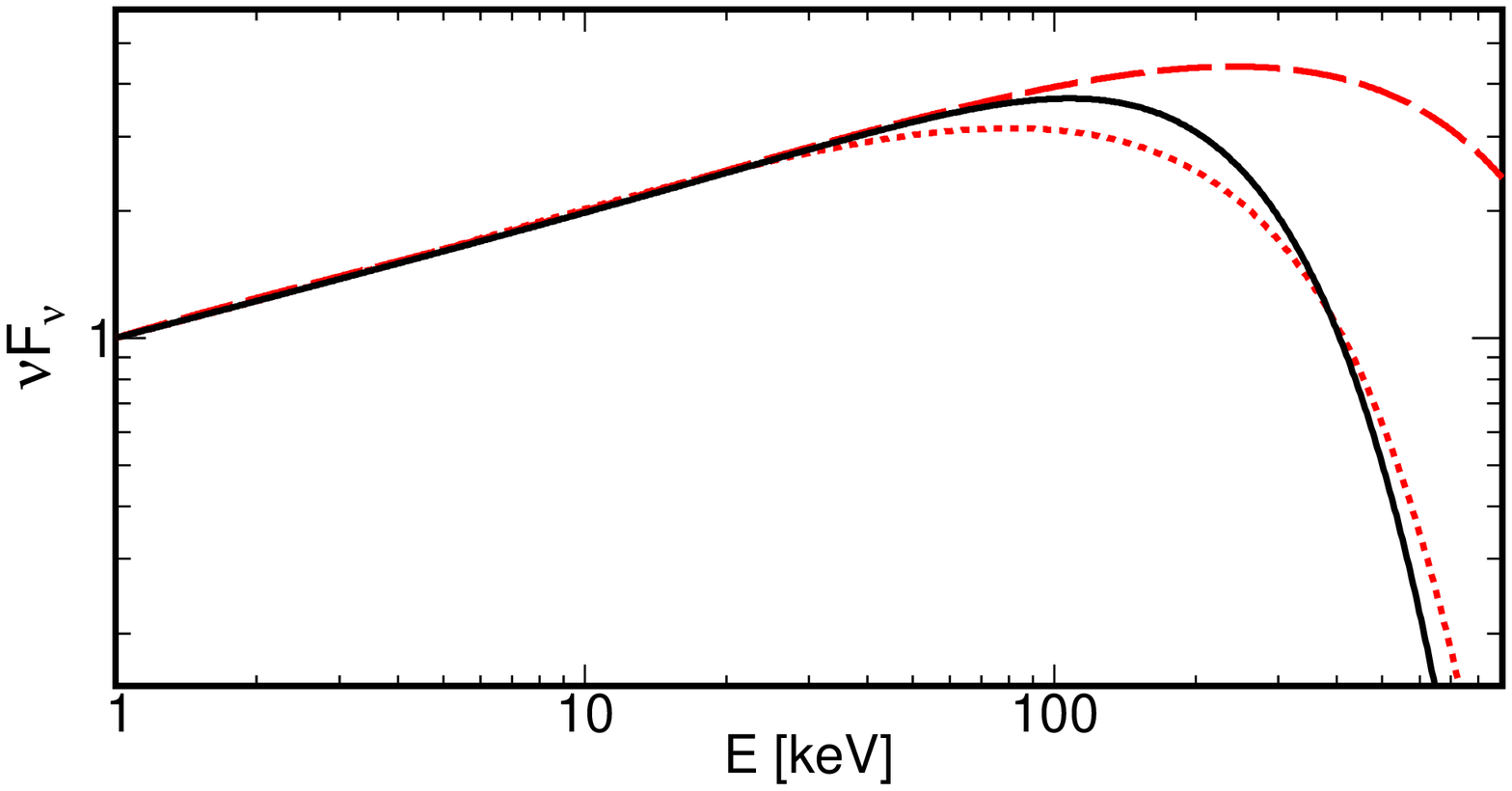}
\caption{The effect of the gravitational redshift on thermal Comptonization spectra. The solid black and dashed red curves show the rest-frame spectra of our model, $E^2 N_{\rm PS}$, for $k T_{\rm e}=100$ and 300 keV, respectively, $\Gamma=1.7$ and $kT_{\rm bb}=1$ eV. The solid red curve shows the spectrum for $k T_{\rm e}= 300$ keV redshifted by $g_{\rm so}=1/3$. We see it significantly differs from the spectrum calculated for $g_{\rm so}k T_{\rm e}$.
}
\label{redshift}
\end{figure}

\begin{figure}
\centering
\includegraphics[width=8cm]{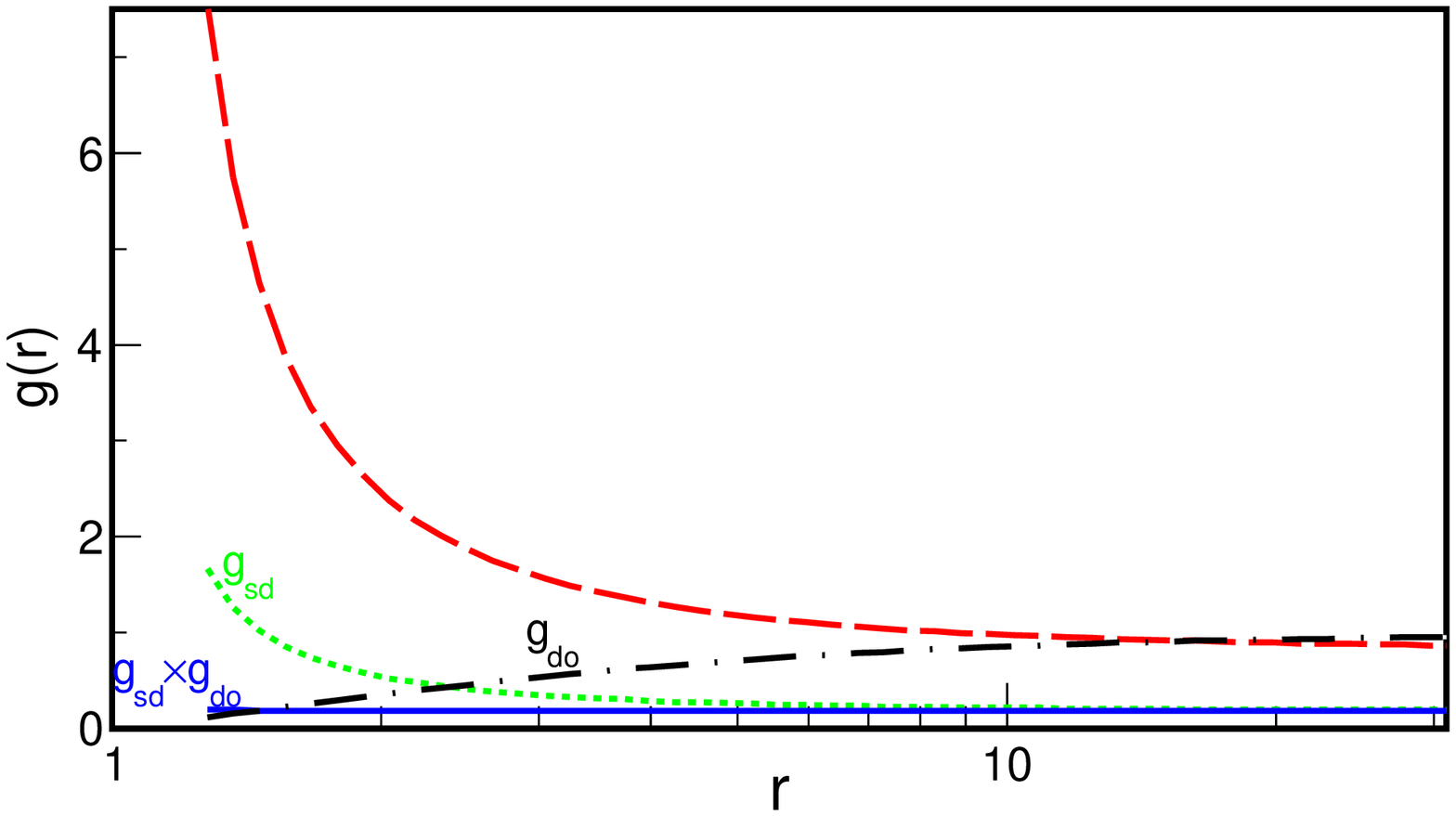}
\caption{Example energy-shift factors. The red dashed, black dot-dashed and blue solid curves show the lamp-to-disc, $g_{\rm sd}$, and disc-to-observer, $g_{\rm do}$, factors and their product, respectively, for $h=1.3$ and $a=0.998$. The $g_{\rm do}$ factor is at $\theta_{\rm o} = 9\degr$ and weighted by the number of observed photons. The dashed red curve shows the lamp-to-disc energy-shift factor, $g_{\rm sd}$, for $h=6$. We see the energy-shift factors can be both $<1$ and $>1$.
}
\label{g_factor}
\end{figure}

In the lamppost geometry, two identical, point-like X-ray sources are symmetrically located on the BH rotation axis perpendicular to a surrounding flat disc. A further assumption adopted in all previous works is that the lamppost is not rotating, and we follow it as well. Our \texttt{xspec} model for the observed spectra of thermal Comptonization and its reflection is called \texttt{reflkerr\_lp}. Usually, we consider the disc to be truncated at an inner radius $r_{\rm in} \geq r_{\rm ISCO}(a)$, where ISCO is the innermost stable circular orbit and $a$ is the BH dimensionless angular momentum. However, the free-falling medium below the ISCO can be still optically thick at high accretion rates \citep{reynolds97}, and therefore we allow $r_{\rm in} > r_{\rm hor}(a)$, where $r_{\rm hor}$ is the horizon radius. The height of the source, $h$, and the radial distance in the disc plane, $r$, are in units of the gravitational radius, $R_{\rm g} = GM/c^2$, where $M$ is the BH mass. Our model uses the rest-frame spectra of the primary and reflected radiation calculated with \texttt{hreflect}, and convolves them with the Kerr metric transfer functions to compute the spectra of radiation received by a distant observer as well as that irradiating the disc surface. 

In \citet{nz18}, we investigated effects related to optical thinness of the region below $r_{\rm in}$ in models with a truncated disc. We found that the bottom lamp may contribute significantly both to the directly observed and to the reflected radiation when $r_{\rm in} \ga 3$. In the general
version of our model, we take into account these effects, as well as a secondary effect related with radiation of the top lamp crossing twice the equatorial plane at $r < r_{\rm in}$ (see fig.\ 4b in \citealt{nz18}). However, we consider also an attenuation, parametrized by $\delta$, of these effects by a tenuous matter which may be located at $r < r_{\rm in}$. The model with $\delta=0$  takes into account only radiation of the top lamp which does not cross the equatorial plane and $\delta=1$ corresponds to the region within $r < r_{\rm in}$ being fully transparent. The user can  set $\delta = 0$ to compare our model with other computations, e.g.\ of \texttt{relxilllpCp}, or to estimate the contribution of the additional effects included in our model.

The flux of the primary radiation seen by a distant observer is computed by means of a photon transfer function, ${\cal T}_{\rm \! so}$, which describes travel of photons from the lamps to a distant observer. It takes into account the reduction of the observed flux due to light bending, photon trapping and time dilation. We consider separately the transfer of radiation from the top lamp which does not cross the equatorial plane, described by ${\cal T}_{\rm \! so,1}(a, h)$, and radiation from the bottom lamp plus radiation from the top lamp deflected by the BH \citep{nz18}, described by ${\cal T}_{\rm \! so,2}(a, h, r_{\rm in}, \theta_{\rm o})$. In general, we consider both components, in which case the observed photon flux is
\begin{equation}
N_{\rm direct}(E_{\rm o}, \theta_{\rm o}) = {1 \over D^2} \left({\cal T}_{\rm \! so,1}+ \delta \,{\cal T}_{\rm \! so,2} \right) \, N_{\rm PS}(E_{\rm o} / g_{\rm so}),
\label{eq:direct1}
\end{equation}
where $D$ is the distance to the source, $g_{\rm so} = E_{\rm o} / 
E_{\rm s}$ [$=1/(1+z)$, where $z$ is the redshift] is the photon energy-shift factor, and the subscripts 's' and 'o' denote quantities measured in the local source frame and those observed by a distant observer, respectively.

\begin{figure}
\centerline{\includegraphics[trim=0 1.71cm 0 0,clip=true,height=4.4cm]{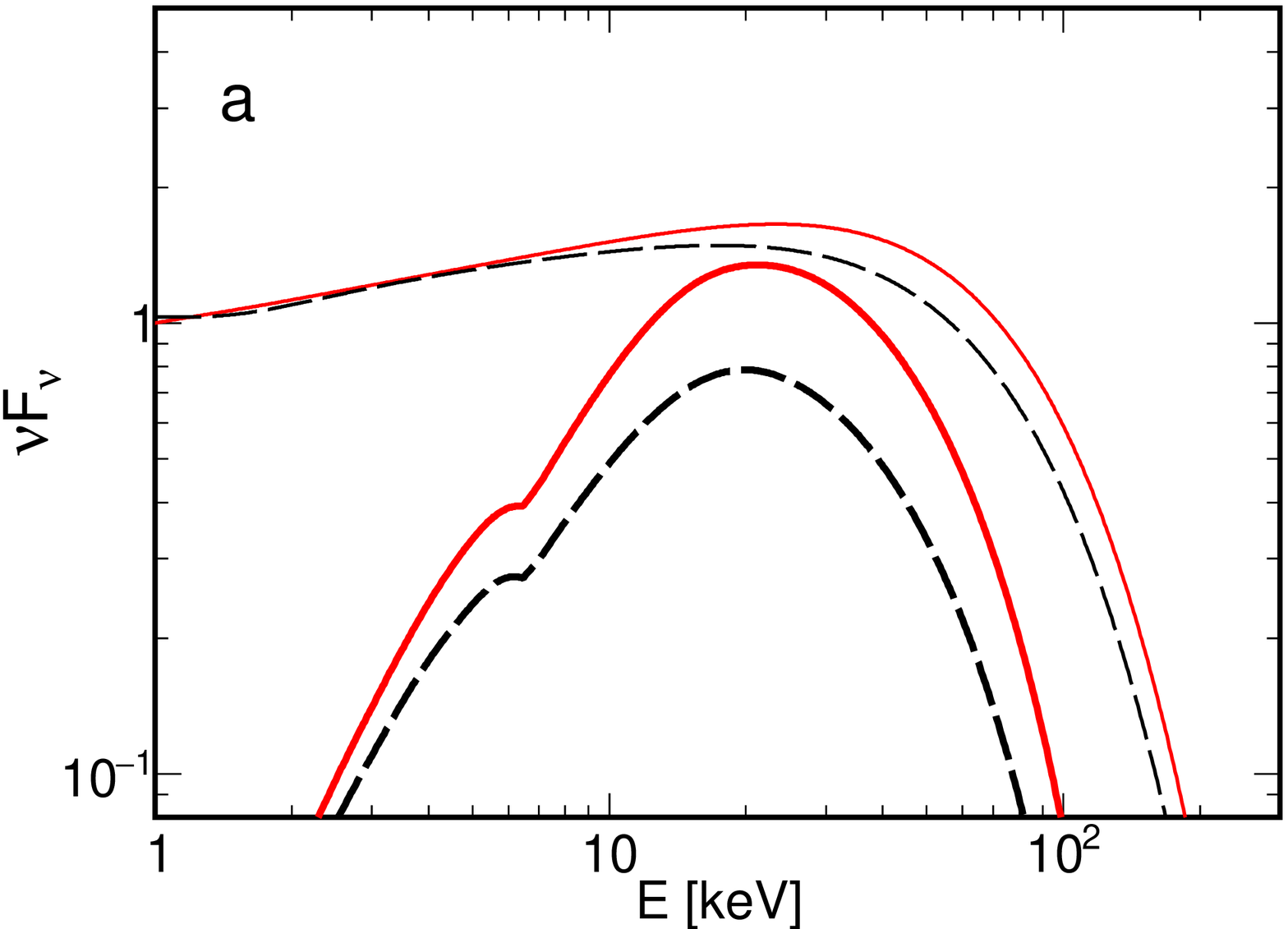}} 
\centerline{\includegraphics[height=5.cm]{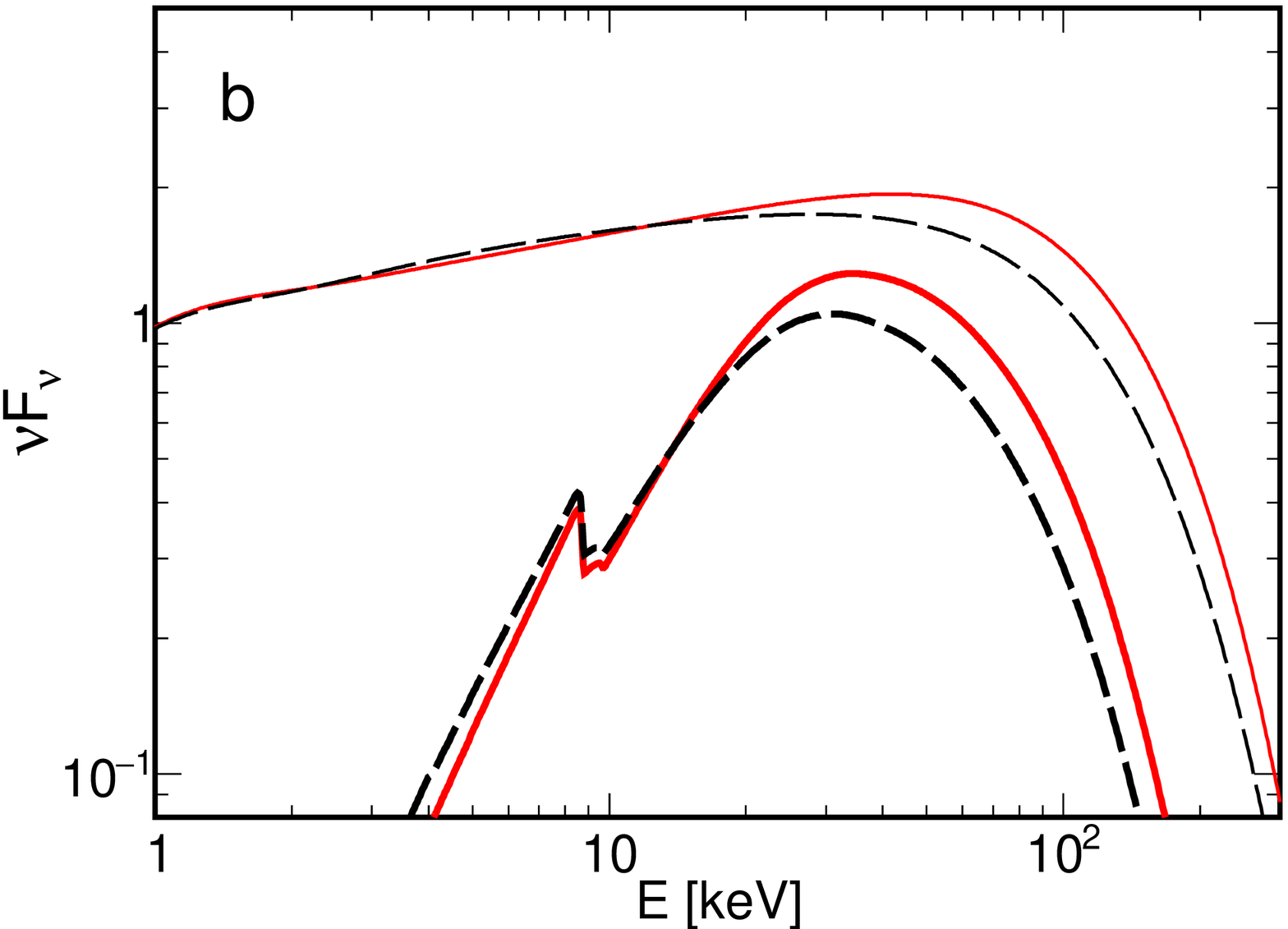}}
\caption{The observed primary (thinner curves) and reflection (thicker curves) spectra for coronal reflection computed  using \texttt{reflkerr} for
$kT_{\rm e} = 30$ keV and $\Gamma=1.8$  for (a) $\theta_{\rm o} = 9\degr$ and (b) $\theta_{\rm o} =75\degr$ and $a=0.998$, $r_{\rm in} = r_{\rm ISCO}\simeq 1.24$, $r_{\rm out} = 1000$, $\xi=1$, $q=4$. The red solid curves are for the spherical geometry, \texttt{geometry=0}, and the black dashed curves are for the slab, \texttt{geometry=1}. The difference of the reflection components results from the difference between the incident spectra shown in Fig.\ \ref{local}. 
} 
\label{reflkerr_geom}
\end{figure}

\begin{figure*}
\begin{center}
\includegraphics[height=4.4cm,scale=0.1]{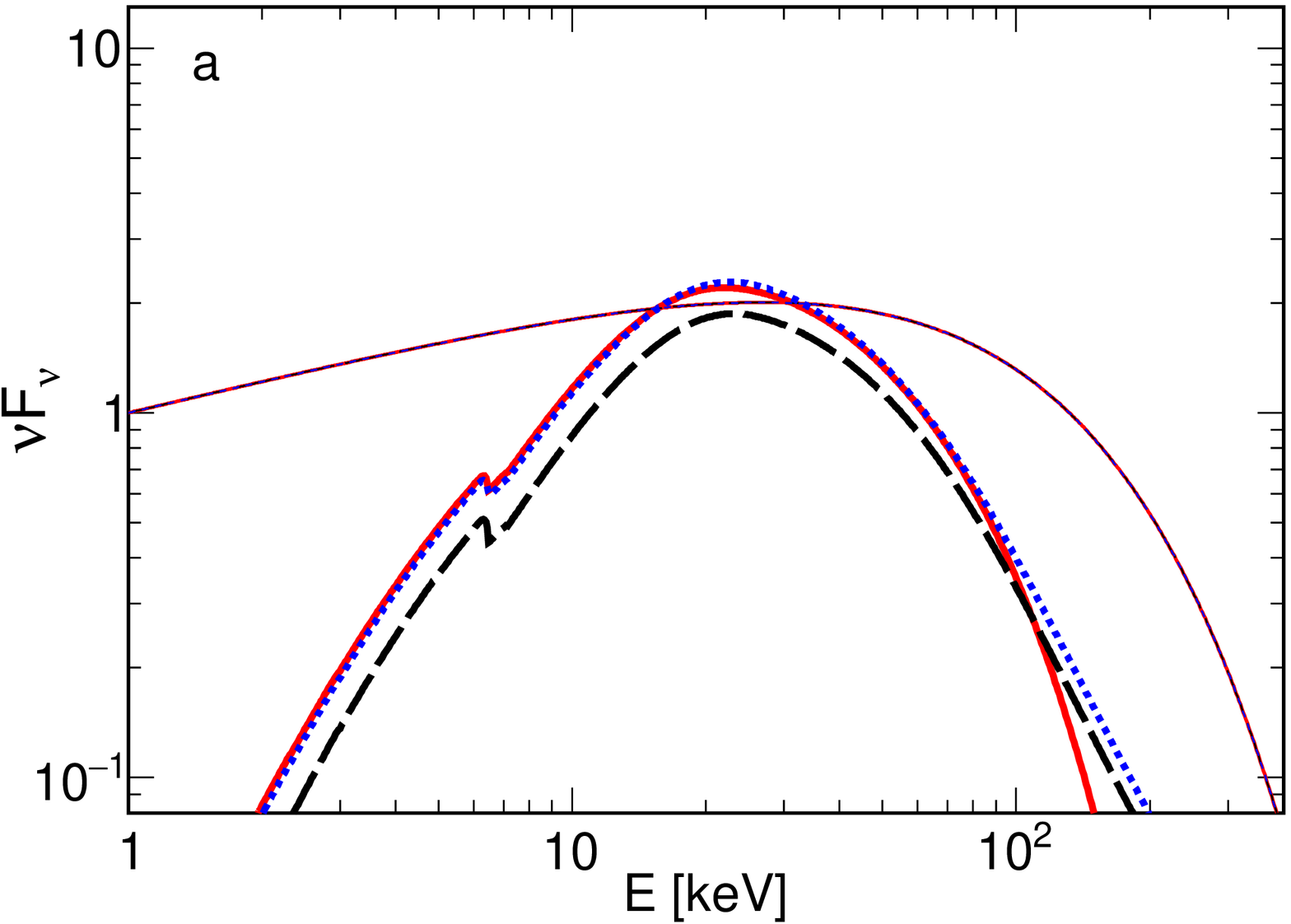}
\includegraphics[height=4.4cm,scale=0.1]{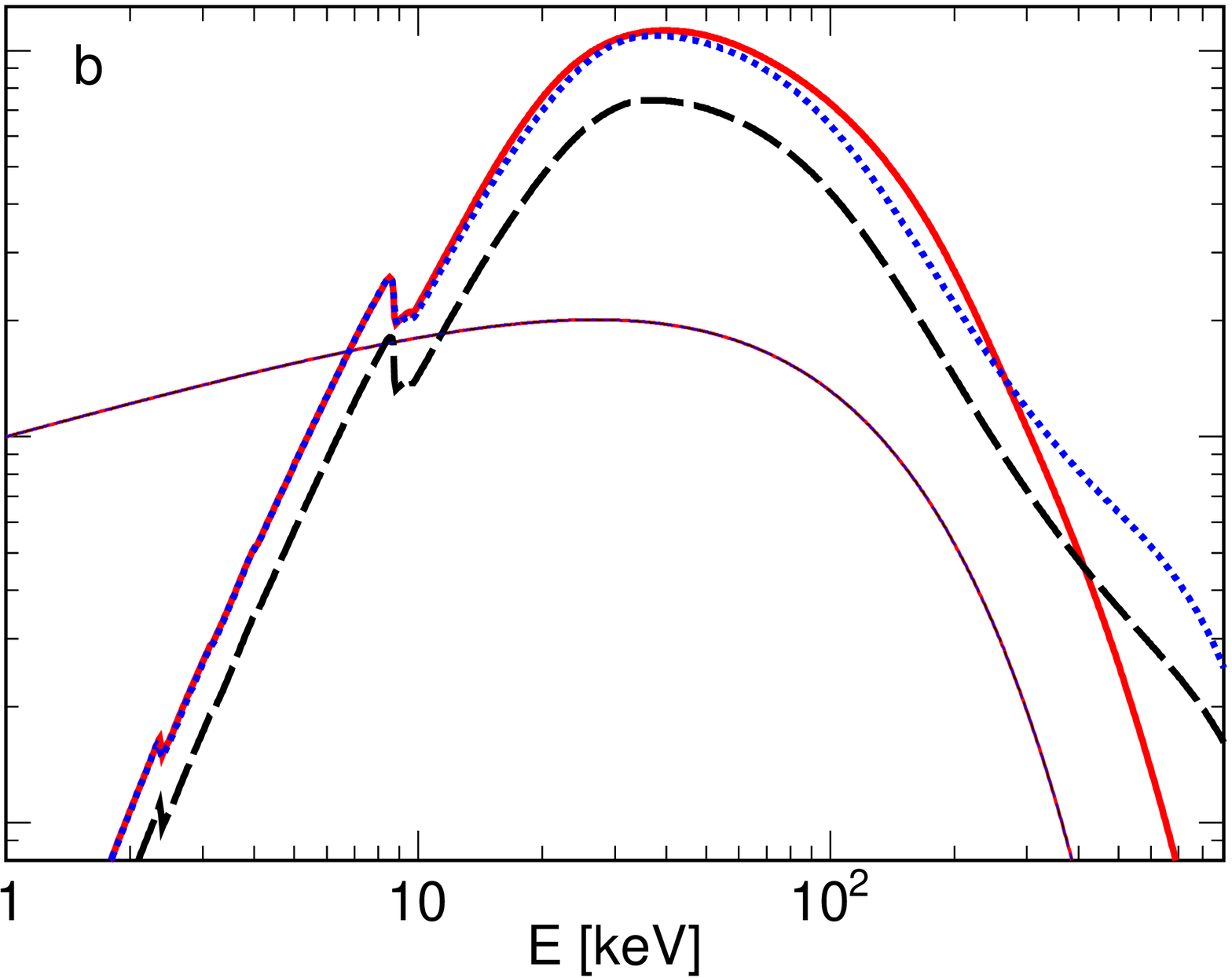}
\includegraphics[height=4.4cm,scale=0.1]{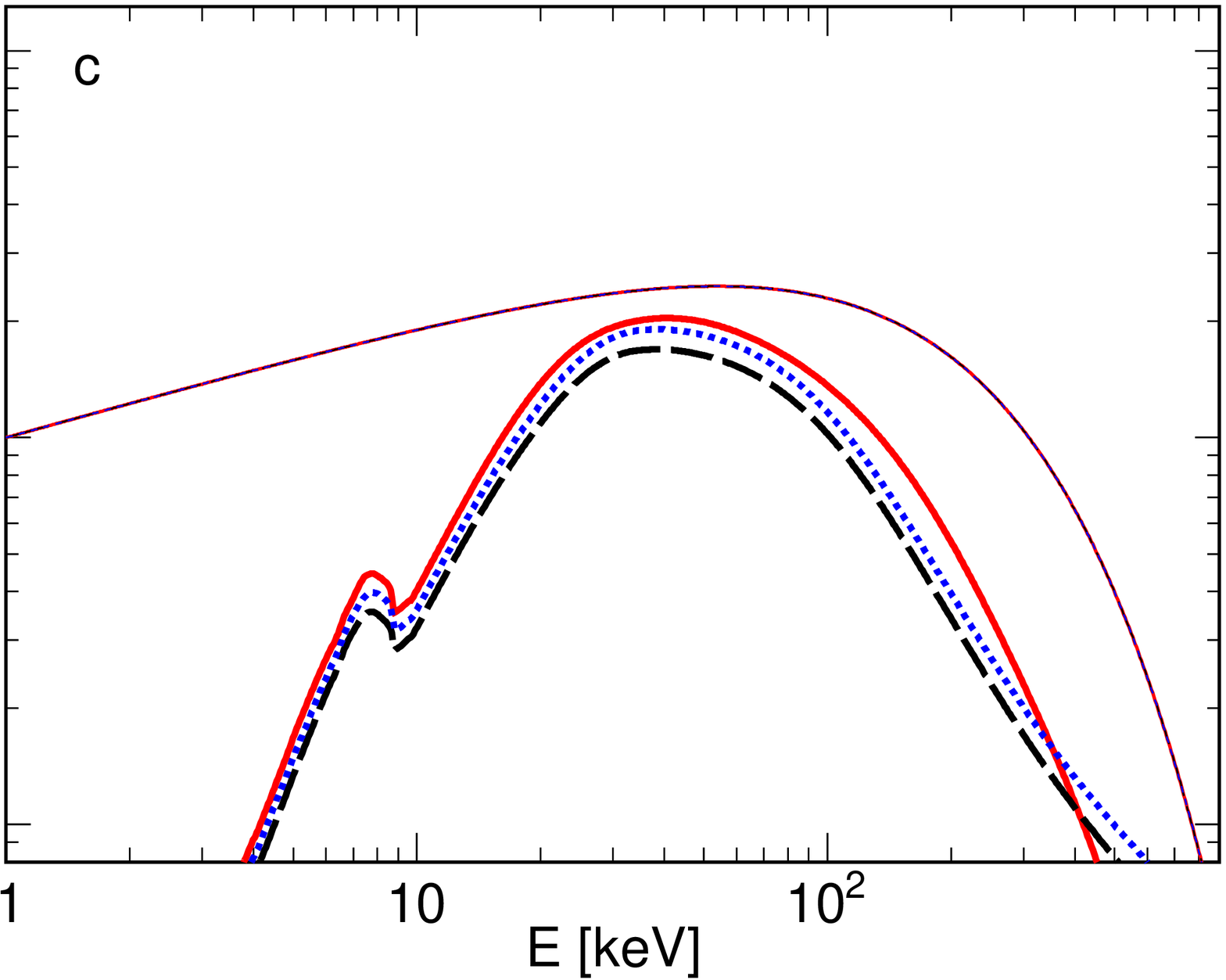}
\caption{A comparison of observed lamppost spectra predicted by the \texttt{reflkerrExp\_lp} (solid red), \texttt{KYNxillver} v.\ 1.4.3 (dotted blue) and \texttt{relxilllp} v.\ 1.2 (black dashed) models with an e-folded power-law at (a) $h=2$ and $\theta_{\rm o} = 9\degr$, (b) $h=2$ and $\theta_{\rm o} = 75 \degr$, (c) $h=10$ and $\theta_{\rm o} = 75\degr$. Both \texttt{reflkerrExp\_lp} and \texttt{KYNxillver} use the rest-frame $E_{\rm cut}=200$ keV; \texttt{relxilllp} uses $E_{\rm cut}= g_{\rm so} 200$ keV, which equals $90$ keV at $h=2$ and 180 keV at $h=10$. The other parameters are: $a=0.998$, $r_{\rm in} = r_{\rm ISCO}\simeq 1.24$, $r_{\rm out} = 1000$, $\Gamma=1.7$, $\xi=1$. All models use the actual lamppost reflection fraction, i.e., $\texttt{fixReflFrac}=1$ is set in \texttt{relxilllpCp} and  $\texttt{Np:Nr}=1$ is set in \texttt{KYNxillver}. The thick curves show the observed reflection and the thin solid curve shows the observed primary component, the latter is the same in all models. The strong reflection component in (b) is due to light bending of the photons emitted by the lamp at low $h$, which causes many more emitted photons to hit the disc than go to the observer. Since the reflected photons are preferentially emitted at large angles with respect to the axis, due to both the Kerr metric effects and Doppler boosting, the observed reflection strength is much higher in (b) than in (a). 
}
\label{compare1}
\end{center}
\end{figure*}

\begin{figure}
\begin{center}
\centerline{\includegraphics[height=4.3cm]{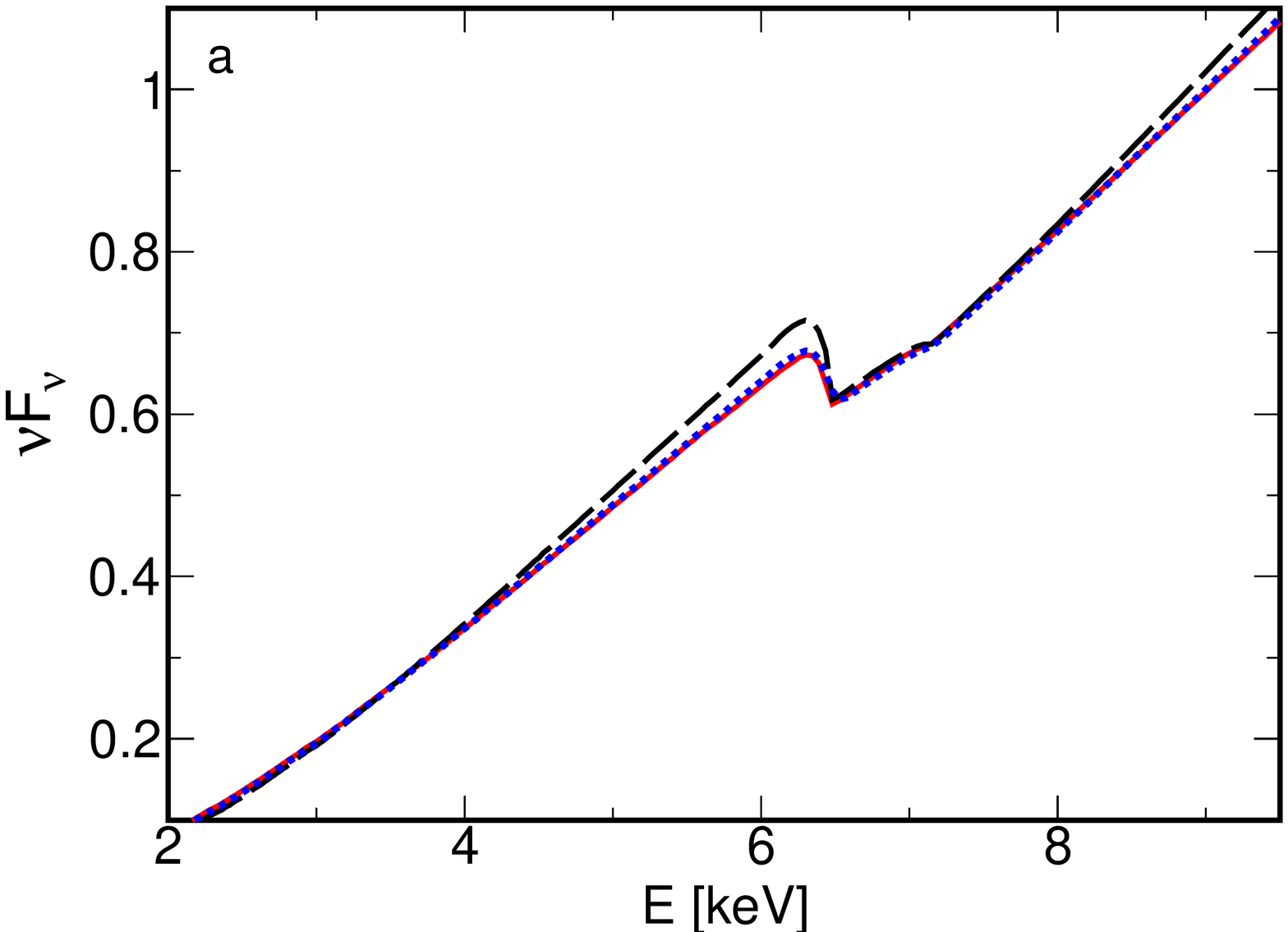}}
\centerline{\includegraphics[height=4.3cm]{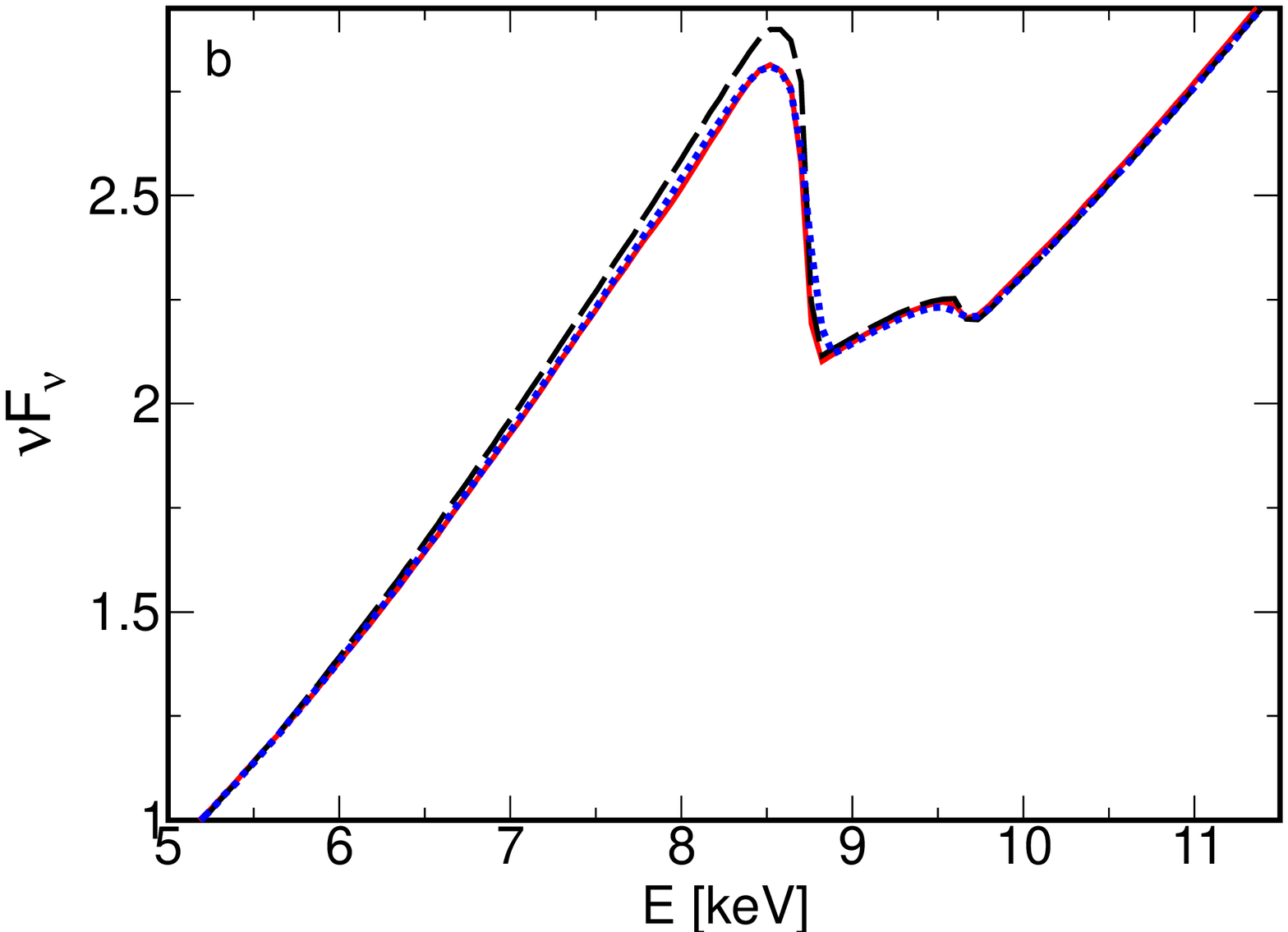}}
\caption{The reflection spectra for $h=2$ and (a) $\theta_{\rm o} = 9\degr$ and  (b) $\theta_{\rm o} = 75 \degr$,  the same as in Figs \ref{compare1}(a,b), but focused on the Fe K$\alpha$ line and edge. The reflection spectra of \texttt{relxilllp} are rescaled by a factor of $\sim 1.5$.
}
\label{compare2}
\end{center}
\end{figure}

\begin{figure*}
\begin{center}
\centerline{\includegraphics[trim=0 1.71cm 0 0,clip=true,height=4.4cm]{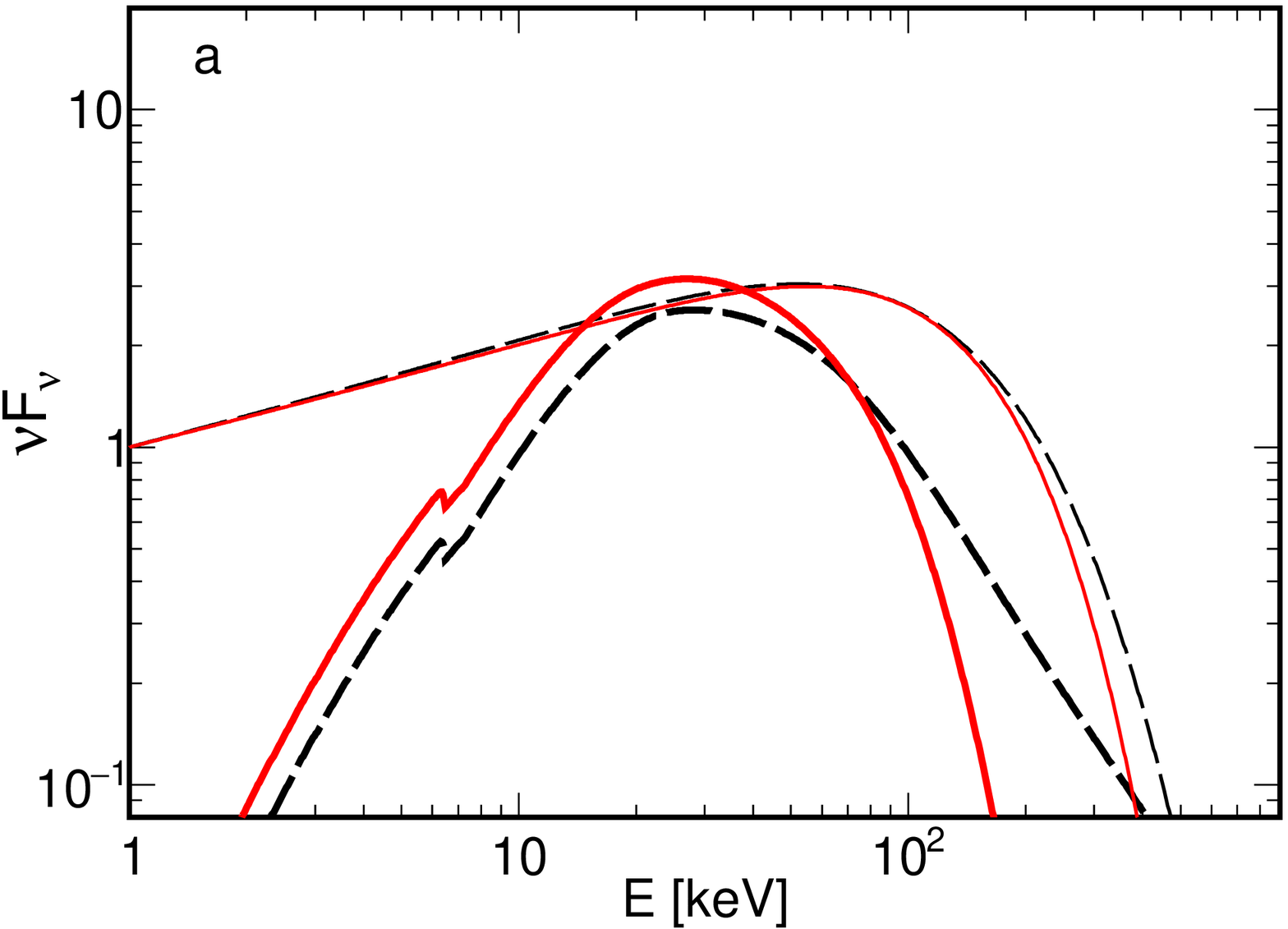}
\includegraphics[trim=1.92cm 1.71cm 0 0,clip=true,height=4.4cm]{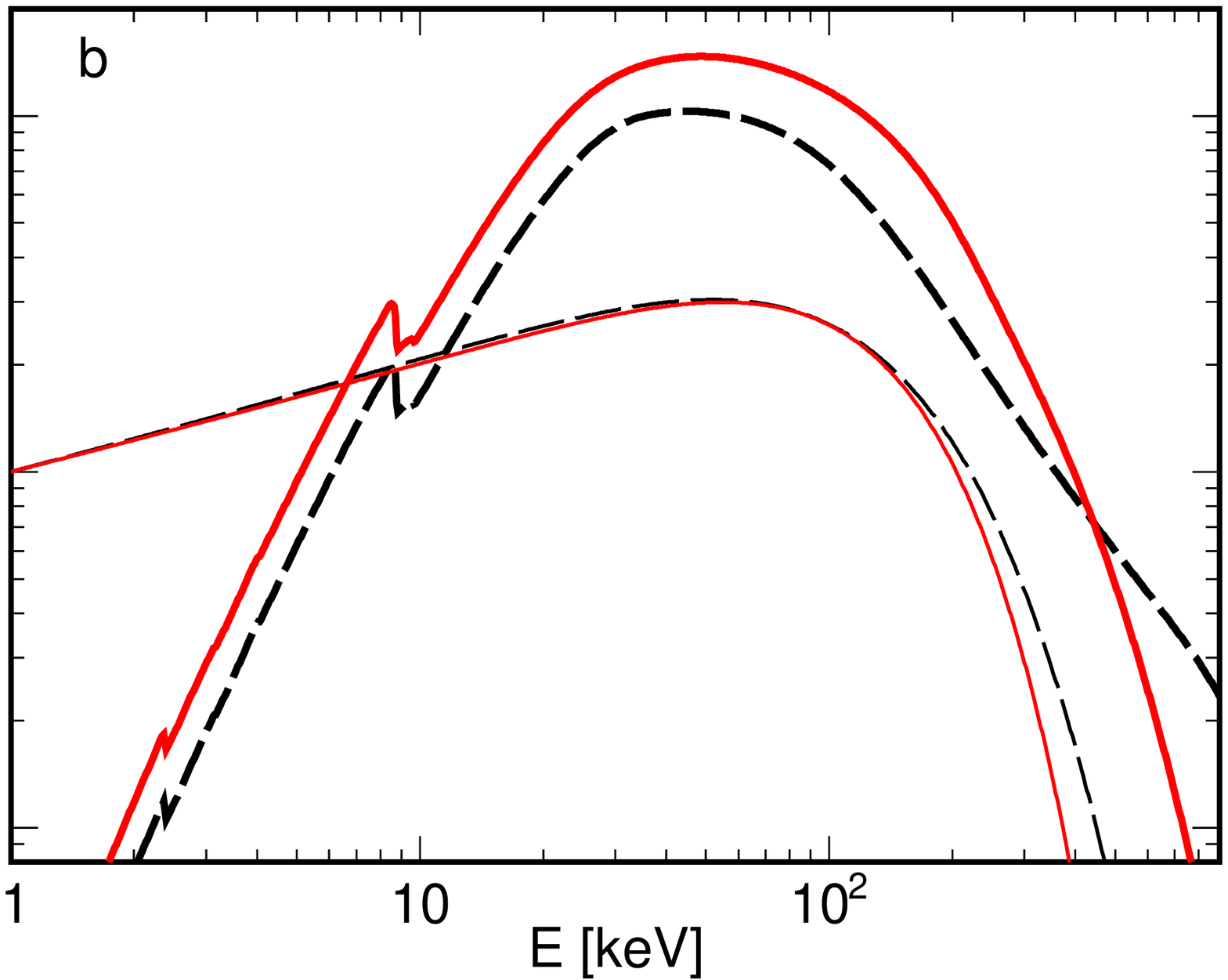}}
\centerline{\includegraphics[height=5.cm,clip=true]{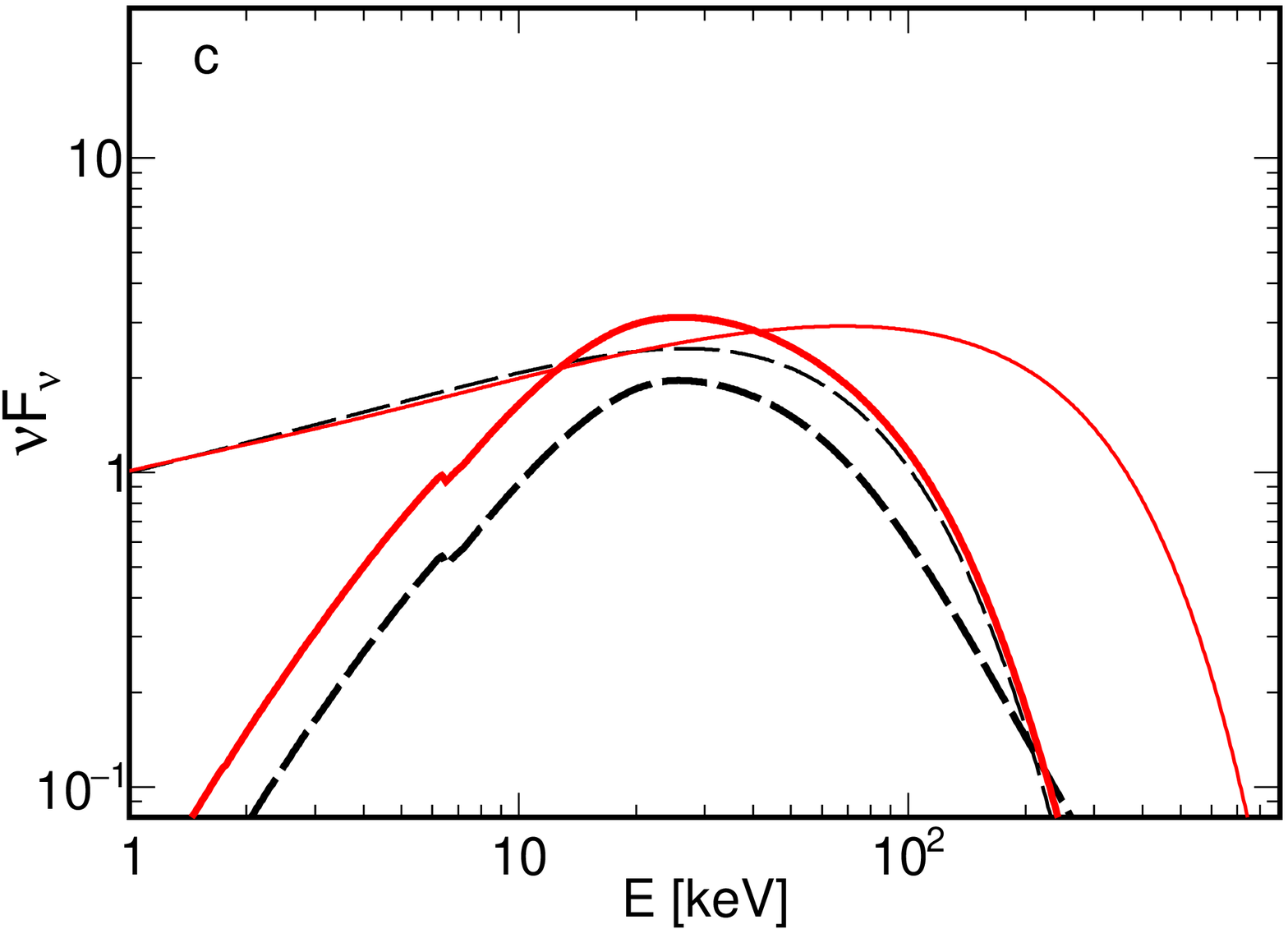}
\includegraphics[trim=1.92cm 0 0 0,clip=true,height=5.cm]{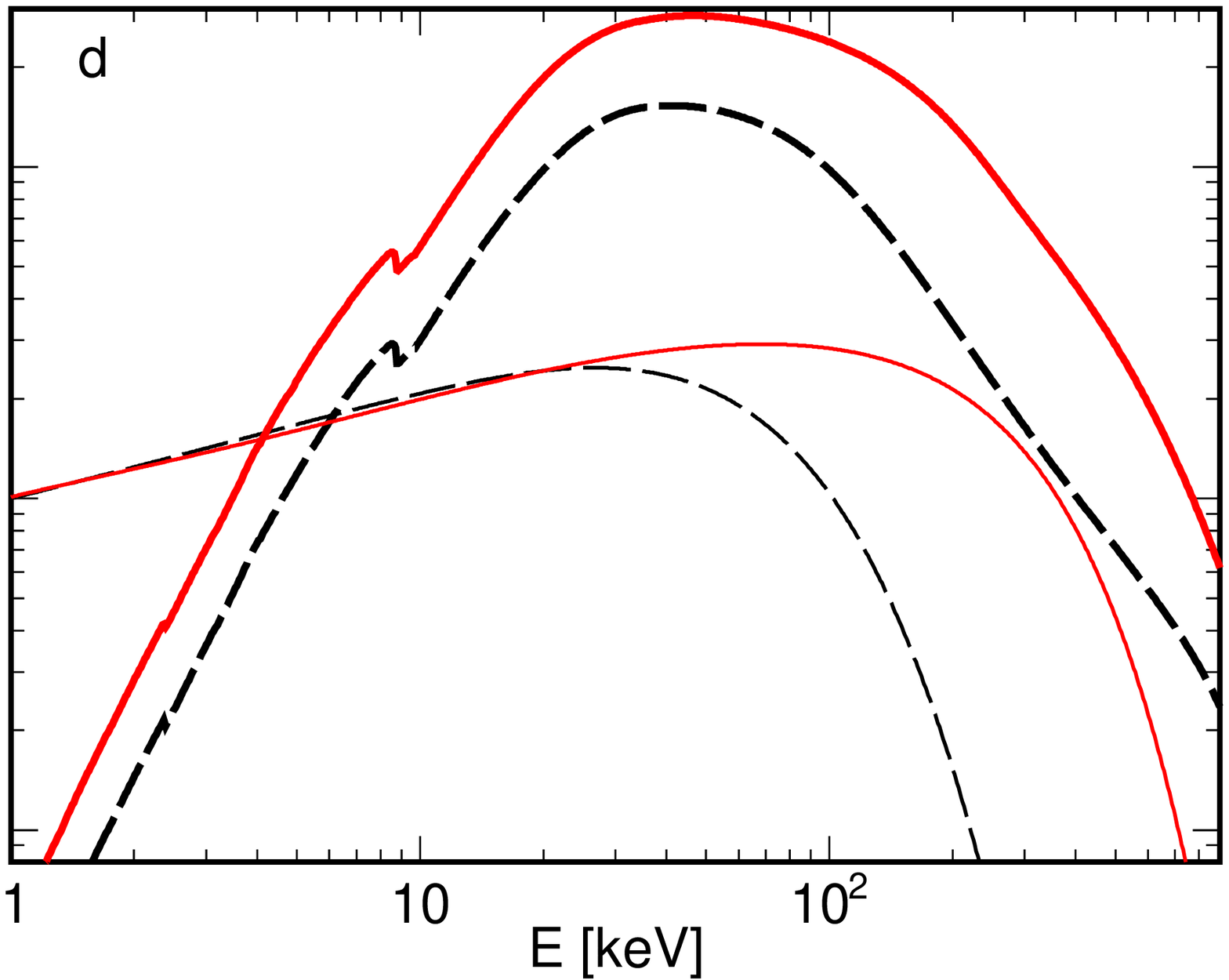}}
\caption{A comparison of observed lamppost spectra predicted by the \texttt{reflkerr\_lp} (solid red) and \texttt{relxilllpCp} v.\ 1.2 (black dashed) thermal Comptonization models. Top: $h=2$, $kT_{\rm e}=100$ keV in \texttt{reflkerr\_lp} and 450 keV in \texttt{relxilllpCp}, bottom: $h=1.4$, $kT_{\rm e}=400$ keV in \texttt{reflkerr\_lp} and 450 keV in \texttt{relxilllpCp}. The other parameters are: (a,c) $\theta_{\rm o} = 9\degr$ and (b,d) $75\degr$, and $a=0.998$, $r_{\rm in} = r_{\rm ISCO}\simeq 1.24$, $r_{\rm out} = 1000$, $\Gamma=1.7$, $\xi=1$. Both models use the actual lamppost reflection fraction, i.e., $\texttt{fixReflFrac}=1$ is set in \texttt{relxilllpCp}. The thick curves show the observed reflection and the thin  curves shows the observed primary components.
}
\label{compare3} 
\end{center}
\end{figure*}

\begin{figure*}
\centerline{\includegraphics[trim=0 1.71cm 0 0,clip=true,height=4.4cm]{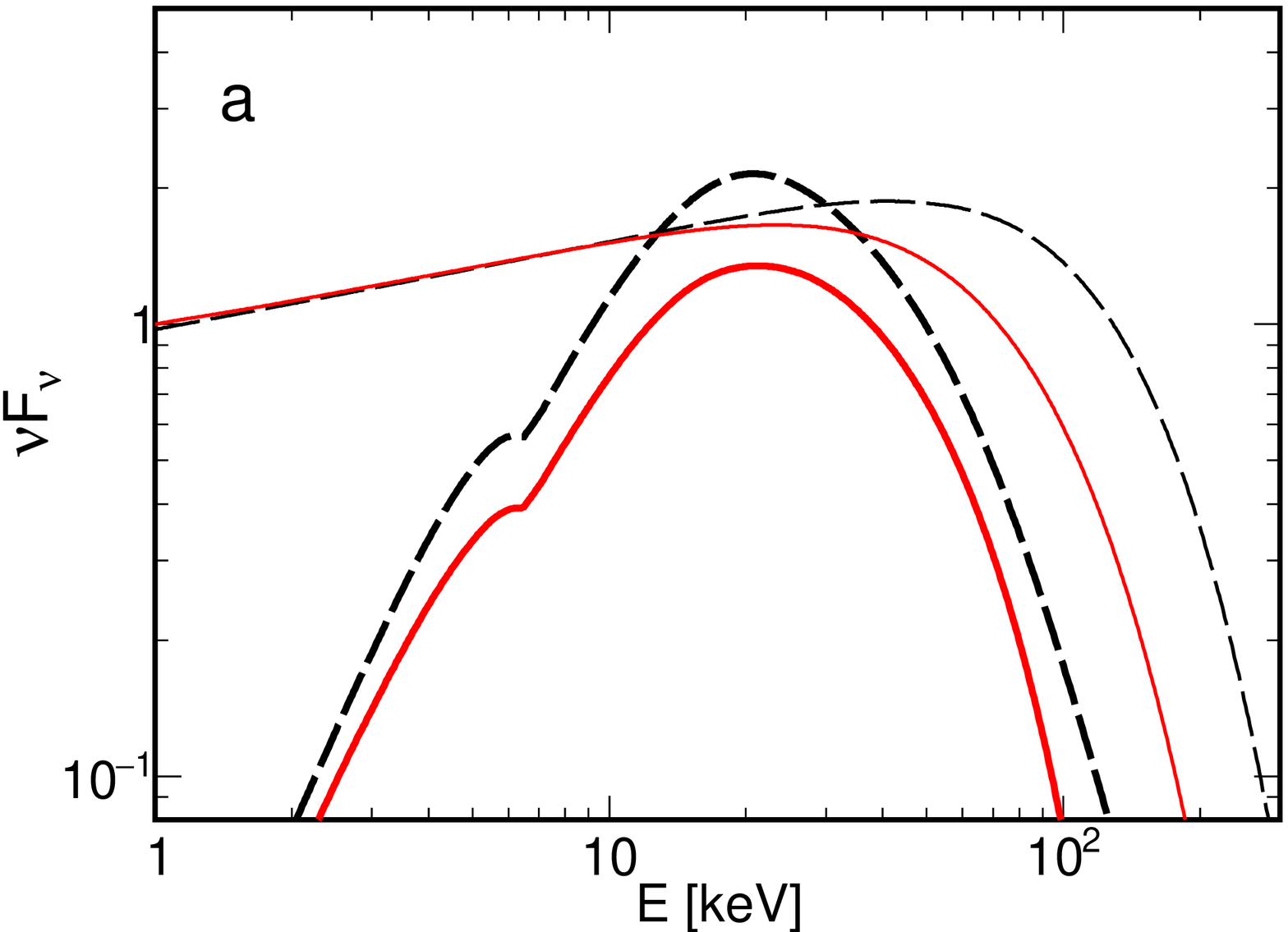} 
\includegraphics[trim=2.3cm 1.71cm 0 0,clip=true,height=4.4cm]{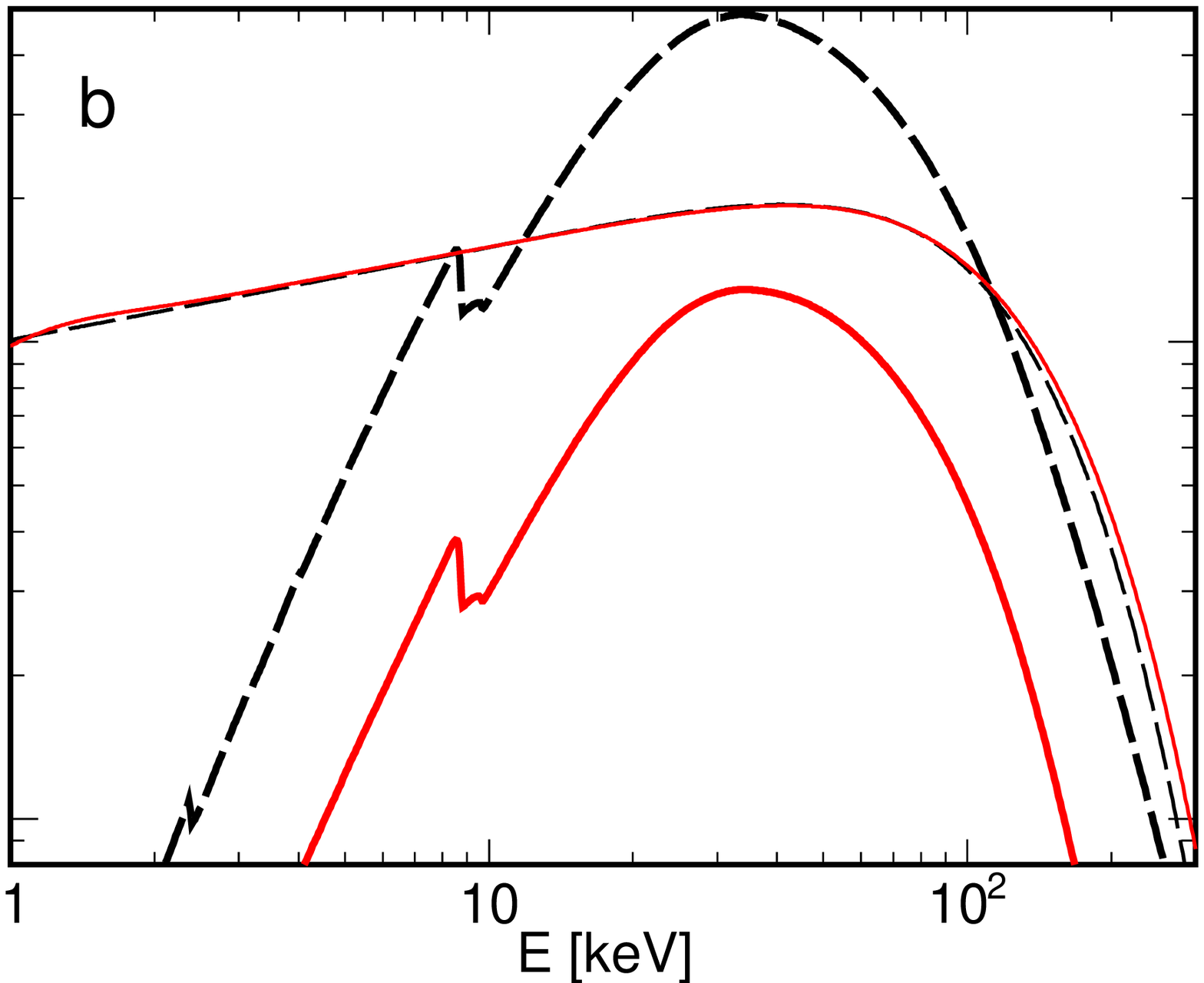}}
\centerline{\includegraphics[height=5.cm]{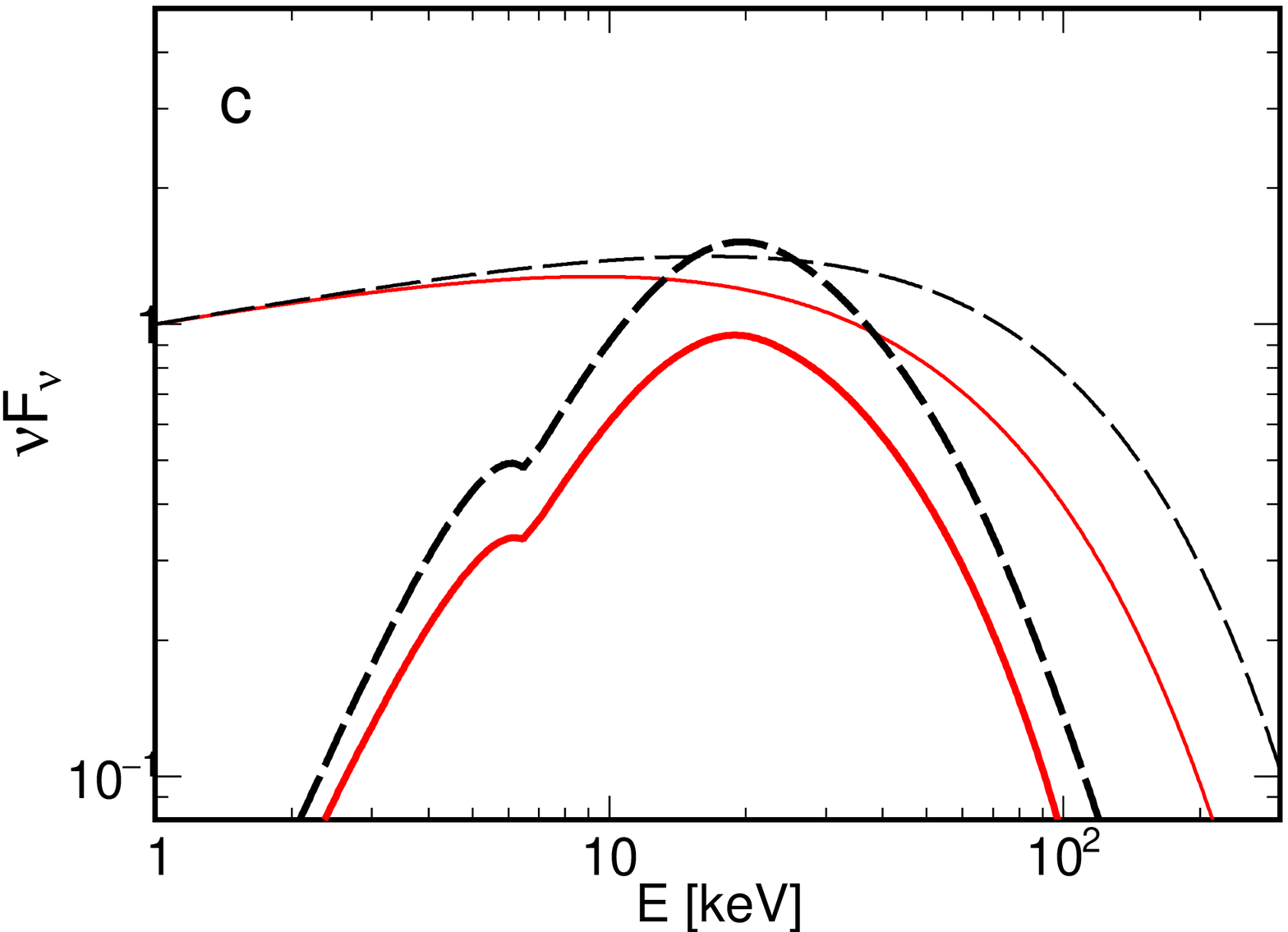} 
\includegraphics[trim=2.3cm 0 0 0,clip=true,height=5.cm]{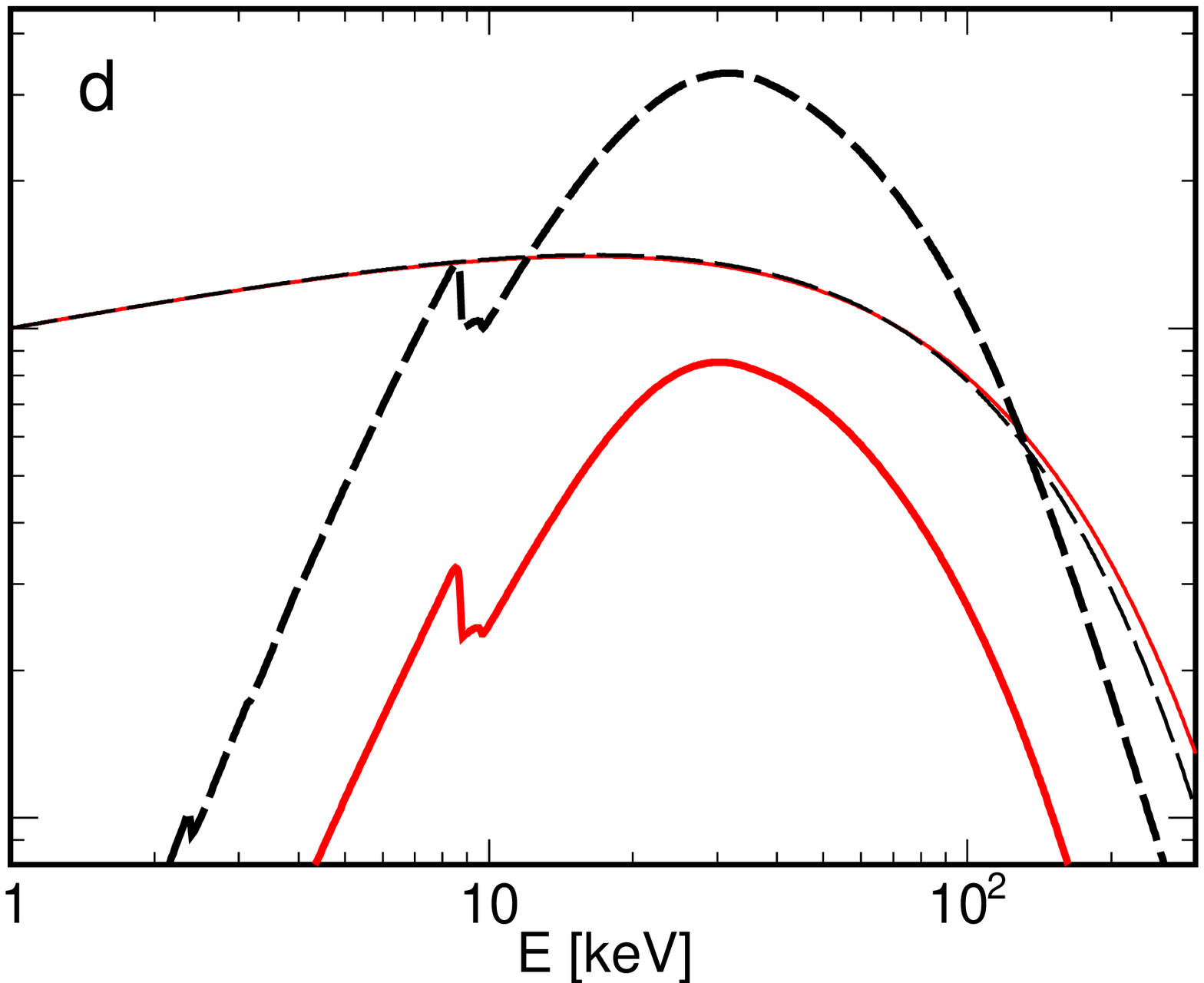}}
\caption{Comparison of the observed primary and reflection spectra for coronal reflection in thermal Comptonization (top) and e-folded power-law (bottom) models for $a=0.998$, $r_{\rm in} = r_{\rm ISCO}\simeq 1.24$, $r_{\rm out} = 1000$, $\xi=1$, $q=4$ and $\theta_{\rm o} = 9\degr$ (left) and $75\degr$  (right). Top: the red solid curves show the spectra computed  using \texttt{reflkerr} for \texttt{geometry=0} with $kT_{\rm e} = 30$ keV and $\Gamma=1.8$. The dashed black curves show the spectra computed with \texttt{relxillCp} v.\ 1.2 for $\Gamma=1.8$ and $kT_{\rm e} = 34$ keV (which gives the same position of the high energy cutoff in \texttt{nthcomp} as $kT_{\rm e} = 30$ keV in \texttt{compps} with \texttt{geometry=0}).  Bottom: the black dashed and red solid curves show the spectra  computed with \texttt{relxill} v.\ 1.2 and \texttt{reflkerrExp}, respectively,  for $E_{\rm cut}=90$ keV and $\Gamma=1.8$.  The thick curves show the observed reflection and the thin  curves shows the observed primary components.
} 
\label{reflkerr_relxill}
\end{figure*}

Our treatment of the directly observed radiation is exact, i.e., we use the actual rest-frame spectrum redshifted by $g_{\rm so}$. We note that the shape of the high energy cutoff in the redshifted spectrum differs from the shape of the spectrum computed for temperature scaled by the same $g_{\rm so}$ factor, i.e., $N_{\rm PS}(E_{\rm o};g_{\rm so}T_{\rm e},\Gamma)$ differs from $N_{\rm PS}(E_{\rm o}/ g_{\rm so};T_{\rm e},\Gamma)$, as illustrated in Fig.\ \ref{redshift}. 

The spectrum irradiating the disc is given by the \texttt{compps} spectrum shifted by $g_{\rm sd}$, $N_{\rm PS}(E_{\rm d}/g_{\rm sd})$, where $g_{\rm sd} = E_{\rm d} / E_{\rm s}$ is the radius-dependent energy-shift factor and the subscript 'd' denotes quantities measured in the local frame co-moving with the disc. For the transfer of photons from the lamps to the disc, we again consider separately direct illumination by the top lamp, described by ${\cal T}_{\rm \! sd,1}(a, r, h)$, and illumination by the bottom lamp, described by ${\cal T}_{\rm \! sd,2}(a, r, h, r_{\rm in})$. Illumination by the top lamp radiation deflected by the BH and crossing twice the equatorial plane is always negligible. The photon flux irradiating the disc at distance $r$ is
\begin{equation}
N_{\rm d}(E_{\rm d},r) = {\cal T}_{\rm \! sd}(a, r, h) \, N_{\rm PS}(E_{\rm d}/g_{\rm sd}),
\label{eq:illum}
\end{equation}
where ${\cal T}_{\rm \! sd} = {\cal T}_{\rm \! sd,1} + \delta \,{\cal T}_{\rm \! sd,2}$. The spectrum of reflected radiation from a unit area of the disc at distance $r$ seen by a distant observer is
\begin{align}
&N_{\rm o}(E_{\rm o},\theta_{\rm o},r) =\cr
&\quad {1 \over D^2 } \!\int\! {\cal T}_{\rm do}(a, r, \theta_{\rm d}, \theta_{\rm o}, g_{\rm do}) \, N_{\rm ref} ({E_{\rm o} / g_{\rm do}}, r, \theta_d) {\rm d}\mu_{\rm d} {\rm d}g_{\rm do}, 
\label{eq:ref}
\end{align}
where $\mu_{\rm d} = \cos \theta_{\rm d}$, $\theta_{\rm d}$ is the emission angle in the disc frame, $g_{\rm do} = E_{\rm o} / E_{\rm d}$ and $N_{\rm ref}$ is the photon number intensity of the reflected radiation in the disc frame, corresponding to the irradiating flux given by equation (\ref{eq:illum}). The transfer functions ${\cal T}_{\rm \! sd}$, ${\cal T}_{\rm \! do}$ and the energy-shift factors $g_{\rm sd}$, $g_{\rm do}$ treat the special and GR effects affecting both the irradiating and observed fluxes. The total observed photon flux of the reflected radiation is 
\begin{equation}
N_{\rm reflected}(E_{\rm o},\theta_{\rm o}) = \!\int_{r_{\rm in}}^{r_{\rm out}}\! N_{\rm o}(E_{\rm o},\theta_{\rm o},r) 2 \upi r \, {\rm d}r.
\label{eq:totrefl}
\end{equation}
Our tabulated transfer functions allow us to compute the observed disc reflection up to $r_{\rm out} = 1000$.

Fig.\ \ref{g_factor} shows example values of energy-shift factors affecting the irradiating and observed disc radiation. We note that for low $\theta_{\rm o}$, the effective redshift, i.e.\ $g_{\rm sd} g_{\rm do}$, approximately equals that affecting the direct radiation, i.e.\ $g_{\rm so}$, e.g., $\simeq 0.18$ for $h=1.3$. In models with large values of $h$, the radiation illuminating the disc close to ISCO is strongly blueshifted, e.g.\ $g_{\rm sd} \ga 4$ at $r<1.5$ for $h>6$, as shown by the dashed curve. This effect is typically not important for the observed reflected radiation, as for large $h$ the contribution of radiation reflected from this inner edge of the disc is small.

Except for the approximations noted above in computing the rest-frame reflection, \texttt{reflkerr\_lp} involves the exact implementation of the model of \citet{niedzwiecki08}. However, it does not include the second-order reflection, as the spectrum of radiation returning to the disc strongly deviates from a power-law and therefore \texttt{xillverCp} cannot be applied to compute the reflection in the disc frame. This is an important shortcoming of this model for some cases, see fig.\ 4 of \citet{niedzwiecki16}.

\subsection{Relativistic reflection from a corona above a disc}
\label{corona}

Our second model, \texttt{reflkerr}, approximates the geometry of a corona covering the accretion disc at a low scale-height, and co-rotating with it. Following previous works, we approximate it by a broken-power law emissivity profile of the corona. As in the lamppost model, no intrinsic dissipation in the disc is taken into account. The locally produced radiation is calculated using \texttt{compps}, 
assuming it to have the rest-frame spectrum constant with radius. 

The observed spectrum from a unit area is given by
\begin{align}
&N_{\rm o}(E_{\rm o},\theta_{\rm o},r) =\cr
&\quad {1 \over D^2 } \!\int\!
 {\cal T}_{\rm do}(a, r, \theta_{\rm d}, \theta_{\rm o}, g_{\rm do})
\, N_{\rm em} (E_{\rm o} / g_{\rm do}, \theta_{\rm d})\epsilon(r) {\rm d}\mu_{\rm d} {\rm d}g_{\rm do},\label{local_obs}\\
&N_{\rm em} (E_{\rm d}, \theta_{\rm d}) = N_{\rm PS} (E_{\rm d}, \theta_{\rm d}) + R N_{\rm ref} (E_{\rm d}, \theta_{\rm d}),
\label{local_corona}\\
&\epsilon(r)=\begin{cases}(r/r_{\rm br})^{-q_1},&r \leq r_{\rm br};\cr
(r/r_{\rm br})^{-q_2},&r \geq r_{\rm br},\end{cases}
\label{epsilon}
\end{align}
where the normalization of $N_{\rm PS}$ is at $r_{\rm br}$, $q_1$ and $q_2$ are the parameters of the broken power-law emissivity, and $R$ is the reflection fraction, defined in the same way as that in the codes \texttt{compps} and \texttt{ireflect}, i.e., $R=1$ corresponds to a local reflection of an isotropic point source from a semi-infinite slab, neglecting at this point any attenuation of the reflected component due to scattering in the corona. Equation (\ref{local_obs}) is similar to equation (\ref{eq:ref}), but now the locally emitted spectrum includes both the primary and reflected components. The total observed photon flux is 
\begin{equation}
N_{\rm obs}(E_{\rm o},\theta_{\rm o}) = \!\int_{r_{\rm in}}^{r_{\rm out}}\!\!
N_{\rm o}(E_{\rm o},\theta_{\rm o},r) 2 \upi r \, {\rm d}r.
\label{eq:totobs}
\end{equation}
The corona may either be formed by a large number of isotropically emitting active regions, in which case the local $N_{\rm em}$ is calculated with the \texttt{compps} spherical geometry, or it may be a uniform, plane-parallel plasma described by the \texttt{compps} slab geometry. Note that in the slab geometry, the Comptonization photon intensity, $N_{\rm PS}$ depends on the local emission angle, $\theta_{\rm d}$, and it differs from the incident spectrum, as shown in Fig.\ \ref{local}.  Fig.\ \ref{reflkerr_geom} illustrates the effect of the anisotropy of the Comptonizing plasma by comparing the spectra calculated with the spherical and the slab geometries.

\section{Comparison with other models}
\label{comparison}

Figs \ref{compare1}--\ref{reflkerr_relxill} compare the spectra predicted by our model and two other publicly available models for relativistic reflection. In Fig.\ \ref{compare1} we show spectra for the lamppost geometry 
and for an untruncated disc computed with \texttt{reflkerrExp\_lp} (Appendix \ref{sect:exp}),  
the recent version, 1.2, of \texttt{relxilllp}\footnote{\url{www.sternwarte.uni-erlangen.de/~dauser/research/relxill/}} and version 1.4.3 of \texttt{KYNxillver}\footnote{\url{projects.asu.cas.cz/stronggravity/kyn/tree/v1.4.3}} of \citet{dovciak04}. These models use the e-folded power-law primary spectra and all three models use the same \texttt{xillver-a-Ec5.fits} tables for the low energy part of reflection. However, the cutoff energy of the directly observed primary spectrum is given in the local frame only in \texttt{reflkerrExp\_lp} and \texttt{KYNxillver}. Therefore, in \texttt{relxilllp} we use $g_{\rm so} E_{\rm cut}$, where $E_{\rm cut}$ is the rest-frame e-folding energy of the other two models. At low values of $h$, many more directly emitted photons hit the disc than go to the observer due to light bending. Also, the reflected photons are preferentially emitted at large angles with respect to the axis, due to both the Kerr metric effects and Doppler boosting. Therefore, the observed reflection strength is much higher for $\theta_{\rm o}=75\degr$ than for $\theta_{\rm o}=9\degr$. In addition, the redshift of the direct radiation at $\theta_{\rm o}=75\degr$ is stronger, by a factor of $\sim 2$ at $h=2$, than the effective redshift of the reflected emission. This causes a relative shift of the two spectra, and, e.g., at $h=2$ the observed reflected flux above 300 keV is a few tens times higher than the observed direct flux.

We note an excellent agreement between both the fluxes and the spectral shapes of the low energy parts (at $E\la 40$ keV) of the reflected components computed with \texttt{reflkerrExp\_lp} and \texttt{KYNxillver}; at $E\ga 40$ keV there is a difference in the spectral shape related with a more accurate treatment of the reflected component in \texttt{reflkerr\_lp} (see Section \ref{rest_frame}). Our 
rest-frame reflection model uses the angular dependence of \texttt{ireflect} which for large $\theta_{\rm em}$ predicts a stronger reflection than \texttt{xillver} (see Appendix \ref{hreflect}). Therefore, the \texttt{reflkerrExp\_lp} reflection including a strong contribution of radiation reflected at large  $\theta_{\rm em}$ may be stronger than that of \texttt{KYNxillver}. However, due to light bending and Doppler beaming, the reflection components observed from small $r$ are always dominated by radiation reflected at small $\theta_{\rm em}$ in the disc frame. Therefore, the difference between the reflection strength of \texttt{reflkerrExp\_lp} and \texttt{KYNxillver} is negligible for small $h$, and we see only a minor difference for $h=10$ in Fig.\ \ref{compare1}(c). Using the original angular dependence of \texttt{xillver} in \texttt{hreflect} (i.e.\ neglecting the modification described in Appendix \ref{hreflect}) we obtain the same reflection strength in \texttt{reflkerrExp\_lp} and \texttt{KYNxillver}  for any $h$.

We also note a good agreement with the spectra computed with \texttt{relxilllp}, with differences of the spectral shapes in the Fe K$\alpha$ energy range not exceeding a factor of $\sim 1.1$, see Fig.\ \ref{compare2}. The difference by a factor of $\sim$1.5 between the observed reflection strengths predicted by \texttt{relxilllp} (for $\texttt{fixReflFrac}=1$, giving the actual lamppost reflection fraction) and both \texttt{reflkerrExp\_lp} and \texttt{KYNxillver} occurs only at $h \la 10$ and it is now much smaller than differences with the previous versions of \texttt{relxilllp}.

Fig.\ \ref{compare3} shows a similar comparison of the lamppost models using the thermal Comptonization spectrum. The \texttt{KYN} models do not include such a variant, so only \texttt{reflkerr\_lp} and \texttt{relxilllpCp} are compared. The impact of GR effects in these models, in particular the difference of reflection strengths, is similar to that discussed above for the models with an e-folded power-law (but $T_{\rm e}$ in \texttt{relxilllpCp} v.\ 1.2 is measured in the local frame, like in \texttt{reflkerr\_lp}). At $h=2$, panels (a) and (b), we use a moderate electron temperature in \texttt{reflkerr\_lp}, for which the Comptonization spectrum can be approximated by \texttt{nthcomp} with the correction shown in Fig.\ \ref{map}. Then, the main difference between \texttt{reflkerr\_lp} and \texttt{relxilllpCp} is seen in the reflected components at $E \ga 30$ keV. At $h=1.4$, panels (c) and (d), we use a much larger $kT_{\rm e} = 400$ keV in \texttt{reflkerr\_lp} to compensate for the large gravitational redshift so that the observed high energy cutoff still occurs at $\sim 100$ keV. Such a compensation cannot be obtained in  \texttt{relxilllpCp}, which now predicts the primary spectra extending to a much lower energy than \texttt{reflkerr\_lp}.  
For even larger electron temperatures, above the limit shown in Fig.\ \ref{boundary}, also the spectral shapes of the reflected components of \texttt{reflkerr\_lp} and \texttt{relxilllpCp} will differ at all energies.

Fig.\ \ref{reflkerr_relxill} shows example coronal spectra calculated with \texttt{reflkerr} and \texttt{relxill} (coronal version of \texttt{KYN} using \texttt{xillver} is not available). We note that our model takes into account the gravitational redshift of the X-ray radiation produced in the corona, which effect has been neglected in \texttt{relxill} (including the current v.\ 1.2). In our model the corona co-rotates with the disc, so its radiation is also affected by the same kinematic effects as the reflected radiation.  We see substantial differences, including that in the amplitude of the reflected component. At $\theta_{\rm o} = 75\degr$, the gravitational redshift is approximately balanced by the Doppler blueshift, therefore the cutoff position is similar in both models in Figs \ref{reflkerr_relxill}(b) and (d). An extensive comparison of the \texttt{reflkerr} and \texttt{relxill} models as fitted to the BH binary GX 339--4 is given in \citet{dzielak18}.

\section{Summary}

We have presented the computational framework of our new relativistic reflection models, {\tt reflkerr\_lp} and {\tt reflkerr}, for two geometries, the lamppost and the disc corona, respectively. In both cases, we assume that the primary X-ray radiation is produced by thermal Comptonization and we compute it using the iterative code of \citet{ps96}, which is one of the most accurate publicly available models of this process. It allows us to correctly describe spectra produced at relativistic electron temperatures, which is particularly important for the lamppost geometry, where such temperatures are required for a low height above the horizon of the irradiating source. This Comptonization model also allows us to study the dependence on the seed photons energy as well as on the source geometry. We have chosen a sphere as an appropriate case for the lamppost. For the disc corona, the model can use either a slab or a sphere geometry, the former gives the strongest local anisotropy effects of all coronal geometries. In addition, we have developed versions of the models for the phenomenological incident spectrum given by an e-folded power law.

We use a hybrid model of the rest-frame reflection from a photoionized medium, which combines the model of \citet{garcia10} in the soft X-ray range with the exact model for Compton reflection of \citet{mz95} in the hard X-ray and soft $\gamma$-ray range. This probably gives the most accurate treatment of ionized reflection using currently available models.

We apply the photon transfer functions in the Kerr metric to compute the directly observed and reflected photon flux and, for the lamppost geometry, the flux irradiating the disc. The transfer functions for the lamppost geometry include photons crossing the equatorial plane within the truncation radius, which allows us to take into account the full contributions of primary sources located on both sides of the accretion disc to the directly observed and to the reflected radiation.

We have compared our {\tt reflkerr\_lp} model with the previous lamppost reflection models of \texttt{relxilllp} and \texttt{KYNxillver} and found an excellent agreement with the latter in the treatment of GR transfer, which is reflected in the full agreement of their predicted reflection strengths as well as the shapes of their soft X-ray spectra. We also note  a rather good agreement with the most recent version 1.2 of {\tt reflkerr\_lp}, with some modest differences occurring  at small $h$. For the coronal geometry we have compared {\tt reflkerr} for the sphere case with \texttt{relxill}. The latter neglects the GR redshift of the direct coronal radiation, which may result in significant differences between these two models, especially for a  steep inner emissivity profile (i.e.\ for a large $q$). The slab geometry, for which the local spectra of both the direct and reflected radiation may strongly deviate from those of the sphere, is not implemented in \texttt{relxill}.  Also, above $\sim$30 keV, the spectral shapes of {\tt reflkerr} or {\tt reflkerr\_lp} differ from those of \texttt{relxill}, \texttt{relxilllp} or \texttt{KYNxillver} due to a more accurate treatment of both the reflected and primary component in the {\tt reflkerr}-family models. 

\section*{ACKNOWLEDGEMENTS}

We thank Thomas Dauser, Javier Garc{\'{\i}}a and Elias Kammoun for discussions. Our model uses a set of routines originally included in the \texttt{relxill} package, which are necessary to compute the \texttt{xillver} model. This research has been supported in part by the Polish National Science Centre grants 2013/10/M/ST9/00729, 2015/18/A/ST9/00746 and 2016/21/B/ST9/02388.

\appendix

\section{Code implementation}
\label{code}

Our relativistic transfer functions (Section \ref{relativity}) are tabulated in a grid including 12 values of $a$, spaced uniformly at $a \le 0.9$ (0.9, 0.8, ...) and more densely at $a > 0.9$ (0.95, 0.98, 0.998), and 50 logarithmically spaced values of $h$ up to $h=100$. The calculated trajectories were tabulated in 10 uniform bins of $\mu_{\rm d}$ (which corresponds to the angular accuracy of both \texttt{ireflect} and \texttt{xillverCp}), 100 uniform bins of $\cos \theta_{\rm o}$, 100 logarithmic bins of $r$ and 2000 uniform bins in $g$. The total number of calculated trajectories was $10^8$ for each pair of $a$ and $h$ in computing ${\cal T}_{\rm \! so}$ and ${\cal T}_{\rm \! sd}$, and $10^8$ for each pair of $a$ and $r$ in computing ${\cal T}_{\rm \! do}$. 

\begin{figure}
\includegraphics[height=6.cm]{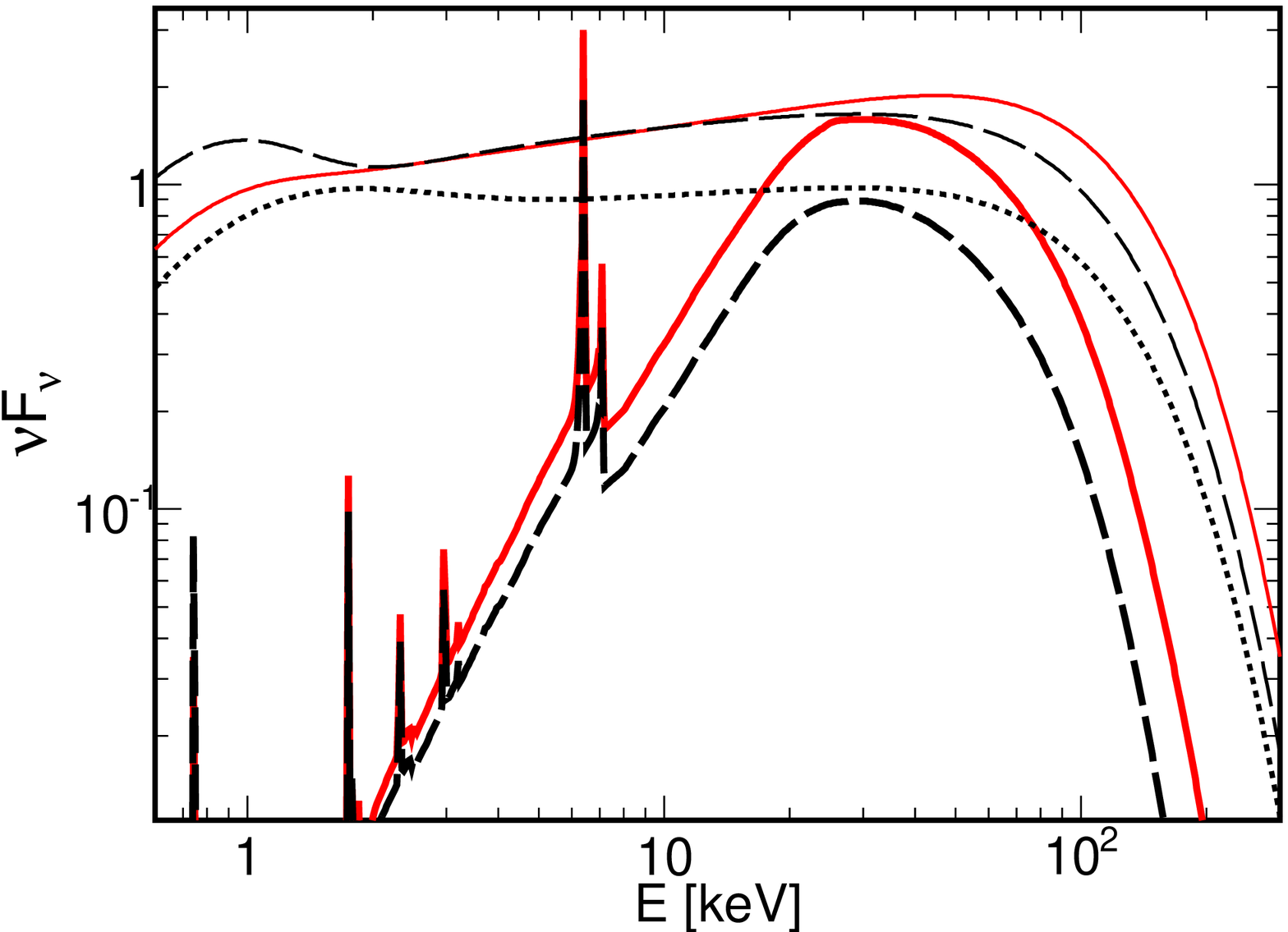}
\caption{A comparison of rest-frame reflection computed for two geometries of the Comptonizing plasma for $kT_{\rm bb}=0.2$ keV, $kT_{\rm e}=30$ keV, $\Gamma=1.8$ (determined in the 1--20 keV range) and $\theta_{\rm o}=30\degr$. The solid red curves are for the spherical geometry, \texttt{geometry = 0}, and the black dashed  curves are for the slab, i.e.\ \texttt{geometry = 1}. 
The thinner curves show the observed primary spectra, computed with \texttt{compps}, and the thicker curves show reflection, computed with \texttt{hreflect}. The incident spectrum for \texttt{geometry = 1} is shown by the black dotted curve and it differs from the observed primary component due to a significant anisotropy of the slab geometry. For \texttt{geometry = 0} the incident spectrum is the same as the observed primary.
} 
\label{local}
\end{figure}

\subsection{Thermal Comptonization}

\label{compps}

In our \texttt{reflkerr} models the user can choose between various versions of either spherical or slab geometry available in \texttt{compps} by setting the \texttt{geometry} parameter, which determines the shape of the scattering plasma and the spatial distribution of seed photons. This parameter has the same meaning as in \texttt{compps}. In the case of the lamppost geometry, in the \texttt{reflkerr\_lp} model, the choices are 4 for a sphere with seed photons emitted from the centre and $-4$ for homogeneous seed photon distribution, and $-5$ for isotropic seed photons distributed radially along $\tau'$ proportional to $\sin(\upi\tau'/\tau)/ (\upi\tau'/\tau)$, and 0 using a fast approximate method based on escape probability from a sphere (default). In the coronal  case (\texttt{reflkerr}), the user can choose 1 for slab geometry with the seed photons emitted from the bottom of the slab with a constant specific intensity (default) or either of the above spherical geometries.
The model parameters of thermal Comptonization are the optical depth, $\tau$ (measured from either the bottom of a slab or the centre of a sphere), the electron temperature, $T_{\rm e}$, and the blackbody seed photons temperature, $T_{\rm bb}$.

We note also that in the slab geometry the incident spectrum\footnote{That incident spectrum is not available as the standard output of \texttt{compps}; it is stored internally in the array \texttt{spref}.} is different from the directly observed one. In the spherical geometry the two spectra (incident and observed) are identical. The related difference in the reflected spectra is shown in Fig.\ \ref{local}. The difference between the incident and observed spectrum in the slab geometry increases with increasing $T_{\rm bb}$; we note also that here the observed spectrum always includes contribution of unscattered seed photons, whereas the incident spectrum does not include such a contribution. For the spherical geometry the unscattered seed photons may be present in the incident spectrum if they are internally produced in the X-ray source, e.g.\ by cyclo/synchrotron emission. Here the user can choose between including or neglecting this contribution.

\subsection{E-folded power-law models}
\label{sect:exp}

In addition to thermal Comptonization, we have also developed versions of our models for the (phenomenological) e-folded power-law shape of the incident spectrum, which specific intensity is given by
\begin{equation}
N_{\rm EPL} \propto E^{-\Gamma} \exp \left( {-E \over E_{\rm cut}} \right),
\label{eq:exp}
\end{equation}
where $E_{\rm cut}$ is the e-folding, or cutoff, energy. Here, the rest-frame reflection at high energies (given by \texttt{ireflect}) is the same as that of the \texttt{pexriv} model \citep{mz95}. We then merge it with the \texttt{xillver} model (Appendix \ref{hreflectexp}), which also uses the e-folded power-law incident spectrum. This version of our hybrid reflection model is referred to as \texttt{hreflectExp}.

In the lamp-post model for this form of the incident spectrum, $N_{\rm PS}$ in equations (\ref{eq:direct1}--\ref{eq:illum}) is to be replaced by $N_{\rm EPL}$. The directly observed cut-off energy equals $g_{\rm so} E_{\rm cut}$ and the radius-dependent cut-off of the incident spectrum is $g_{\rm sd} E_{\rm cut}$. We refer to this version the lamp-post model as \texttt{reflkerrExp\_lp}.
In the coronal case, \texttt{reflkerrExp}, $N_{\rm PS}$ in equation (\ref{local_corona}) is replaced by $N_{\rm EPL}$.

\begin{figure}
\begin{center}
\includegraphics[height=6.cm]{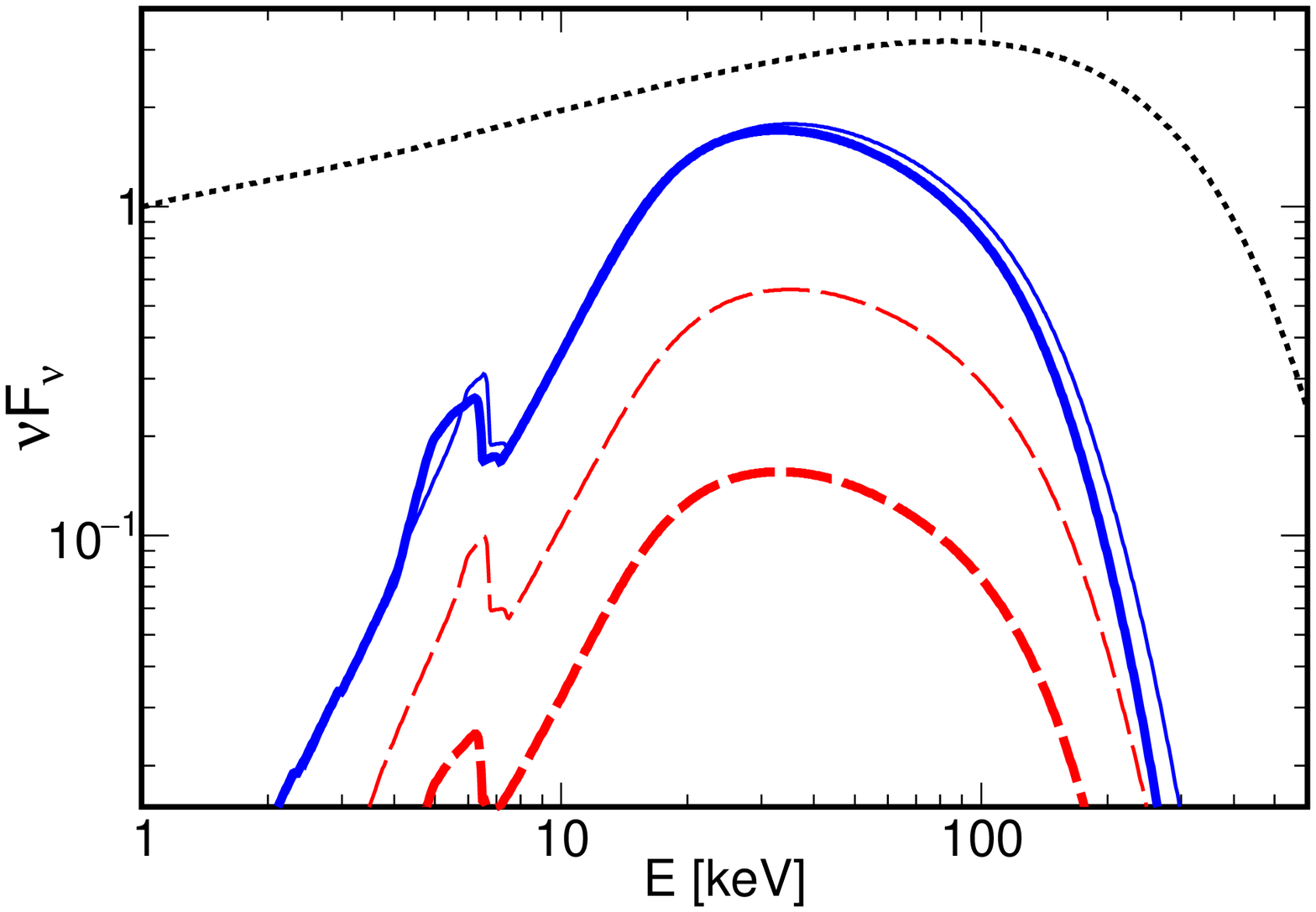}
\caption{An example illustration of the effect of the emission of the bottom lamp and of the emission of the top lamp circling around the BH (i.e., crossing twice the equatorial plane). The blue solid and red dashed curves show the observed reflection spectra computed with \texttt{reflkerr\_lp} for $\delta=0$ and $\delta=1$, respectively, for $h=2$, $a=0.998$, $r_{\rm in} = 4.9$, $r_{\rm out} = 1000$, $\Gamma=1.7$, $kT_{\rm e}=200$ keV, $\xi = 1$, $kT_{\rm bb}=50$ eV, $\theta_{\rm o} = 9\degr$ (thick curves) and $30\degr$ (thin curves). The black dotted curve shows the observed primary spectrum. We see the actual reflected spectra may be much weaker with respect to the observed primary emission than those in the approximate ($\delta = 0$) case. In the shown case, this is due mostly to the primary emission of the bottom lamp gravitationally focused toward the observer, which enhances the observed primary flux by as much as a factor of 10 in the case of $\theta_{\rm o} = 9\degr$. Since our plotted primary spectra are normalized to unity at 1 keV, this enhancement is seen here as a decrease of the amplitude of the reflection.
}
\label{bottom_lamp}
\end{center}
\end{figure}

\subsection{Reflection stength}

The \texttt{reflkerr\_lp} model takes into account the emission of the bottom lamp, neglected in \texttt{relxilllpCp}, which will strongly affect the reflection strength in models with $r_{\rm in} \ga 3$. Fig.\ \ref{bottom_lamp} illustrates the importance of this effect. In the shown examples, the primary emission is strongly enhanced when the (gravitationally focused) emission of the bottom lamp is included, while the reflected emission remains almost the same. However, since we plot the incident spectra normalized to unity at 1 keV, this is seen as a decrease of the reflection amplitude. 

We emphasize that properly assessed reflection amplitude gives an important constraint on the lamppost model, where both the fraction and the radial distribution of photons illuminating the disc surface are fully determined by the height of the primary X-ray source. Then, the radial emission profile is strictly related to assumed geometry, and treating the fraction of reflected photons as a free parameter leads to results which are not self-consistent. Still, our model allows the user to make it free.

\section{\texttt{hreflect}}
\label{hreflect}

\begin{figure}
\centerline{\includegraphics[width=7.5cm]{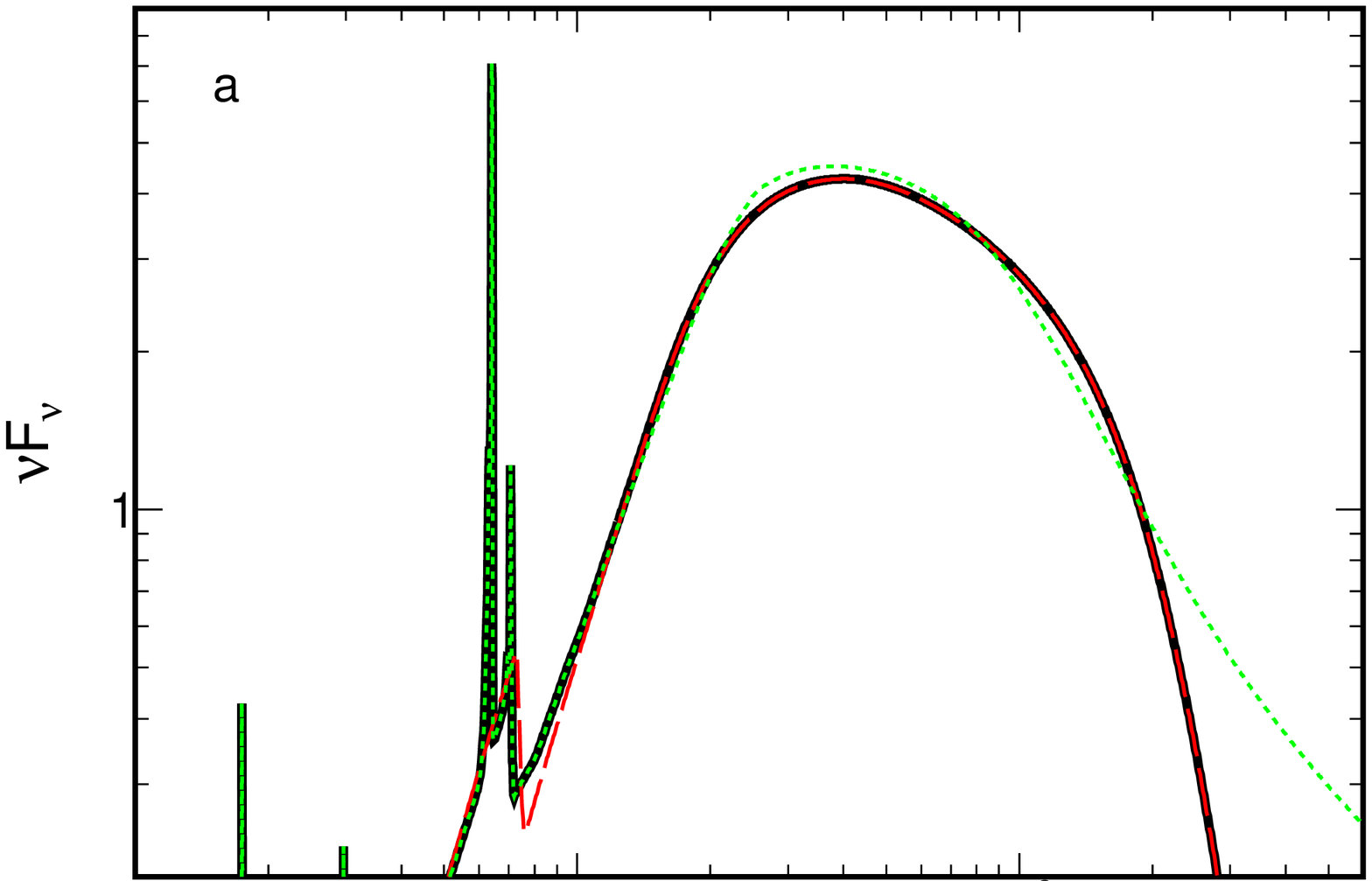}}
\centerline{\includegraphics[width=7.5cm]{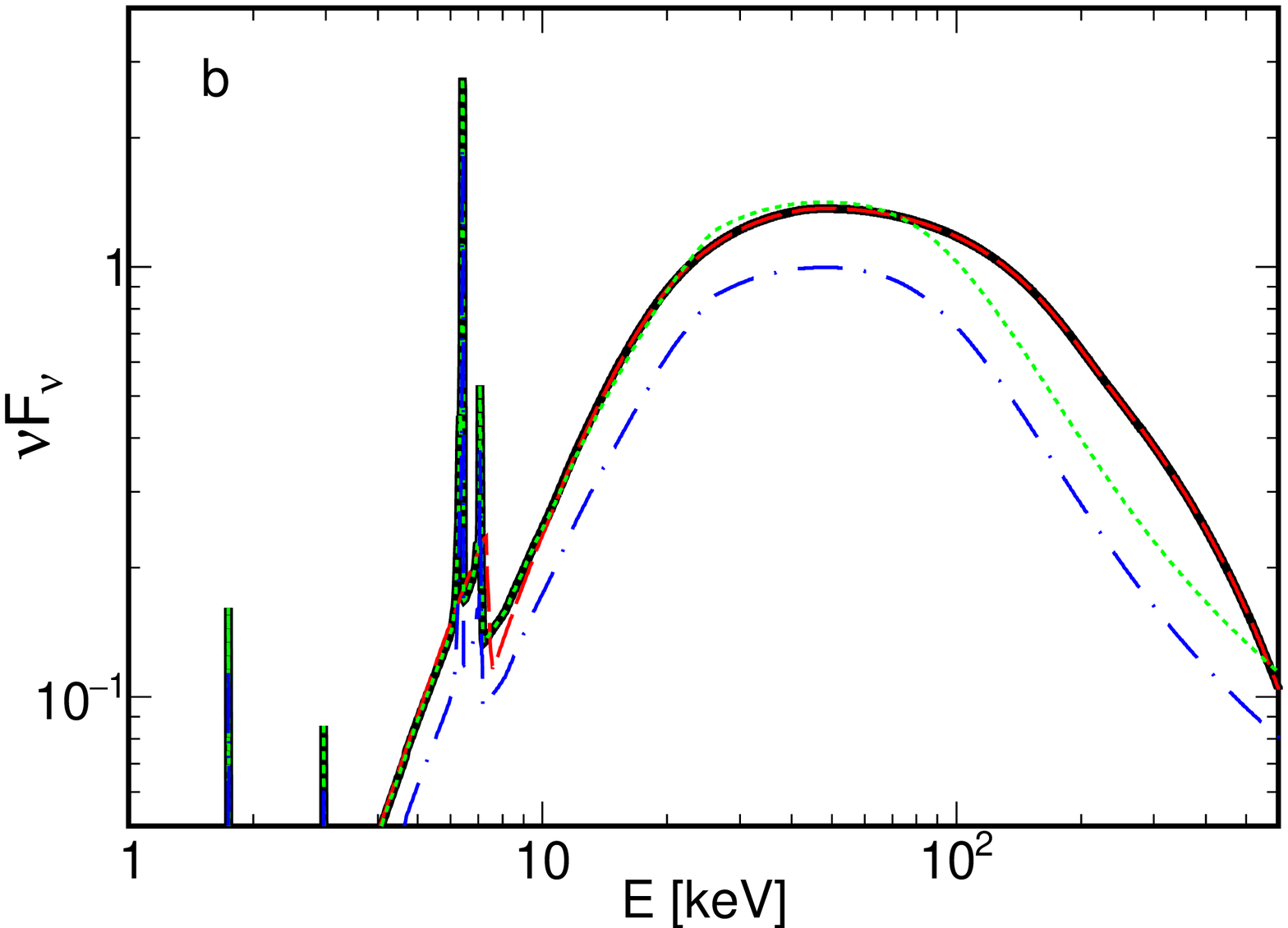}}
\caption{The solid black curves show the \texttt{hreflectExp} reflection for an e-folded power-law with $\Gamma=1.6$, $E_{\rm cut}=300$ keV, $\xi=1$ and (a) $\theta_{\rm obs} = 9^\circ$ and (b) $\theta_{\rm obs} = 80^\circ$.  The dotted green curves show the \texttt{xillver} spectra rescaled by $f_{\theta}$ (see text). The dot-dashed blue curve in (b)  shows the \texttt{xillver} spectrum with the original normalization; for $\theta_{\rm obs} = 9^\circ$ in (a) $f_{\theta}\simeq1$. 
The dashed red curves show the \texttt{pexriv} spectra with $\xi$ fitted to match the \texttt{xillver} spectrum in the 12--22 keV range (see text).
The primary spectra are normalized at 1 keV.
} 
\label{hreflect1}
\end{figure}

\begin{figure}
\centerline{\includegraphics[width=7.5cm]{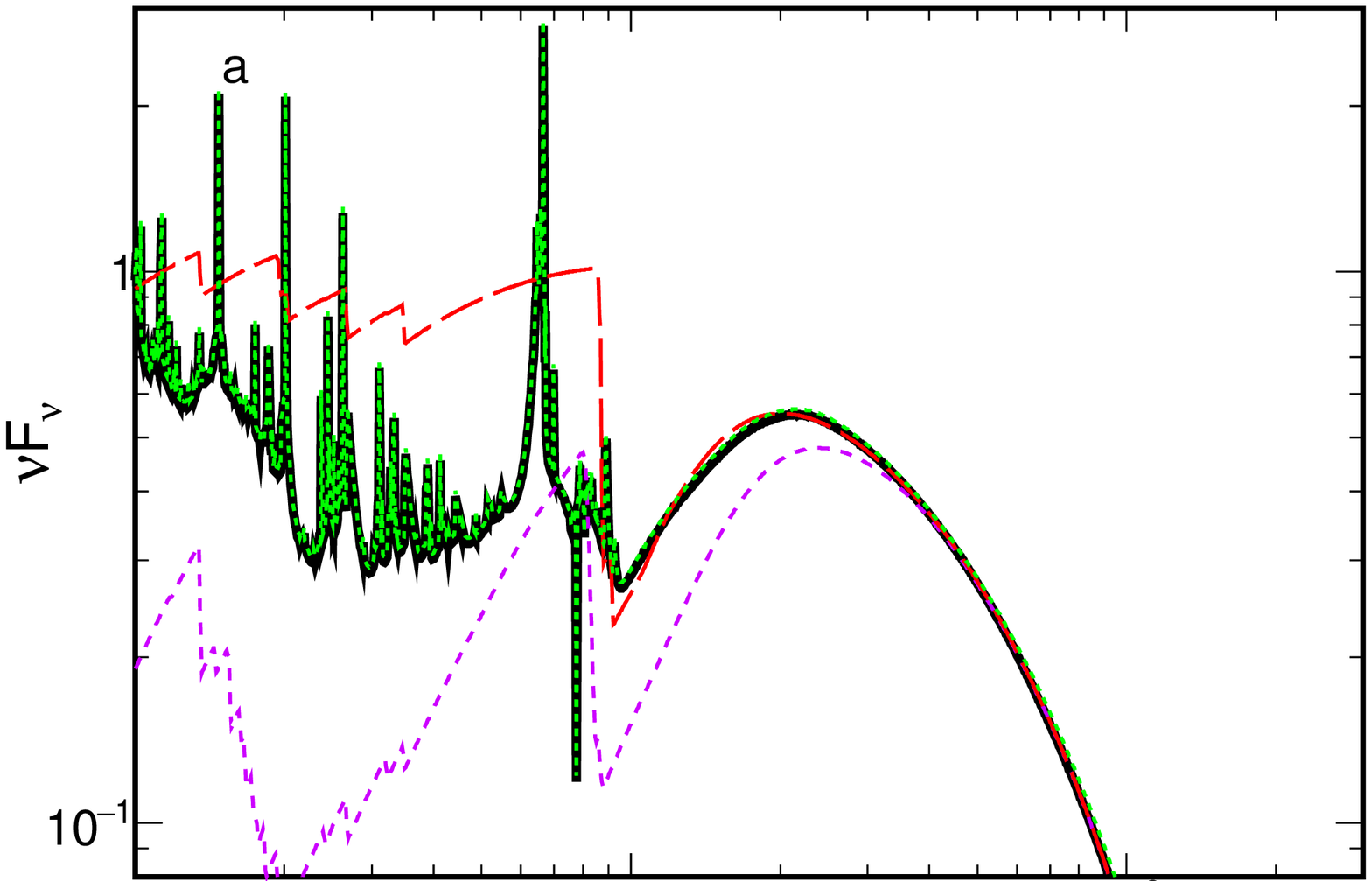}}
\centerline{\includegraphics[width=7.5cm]{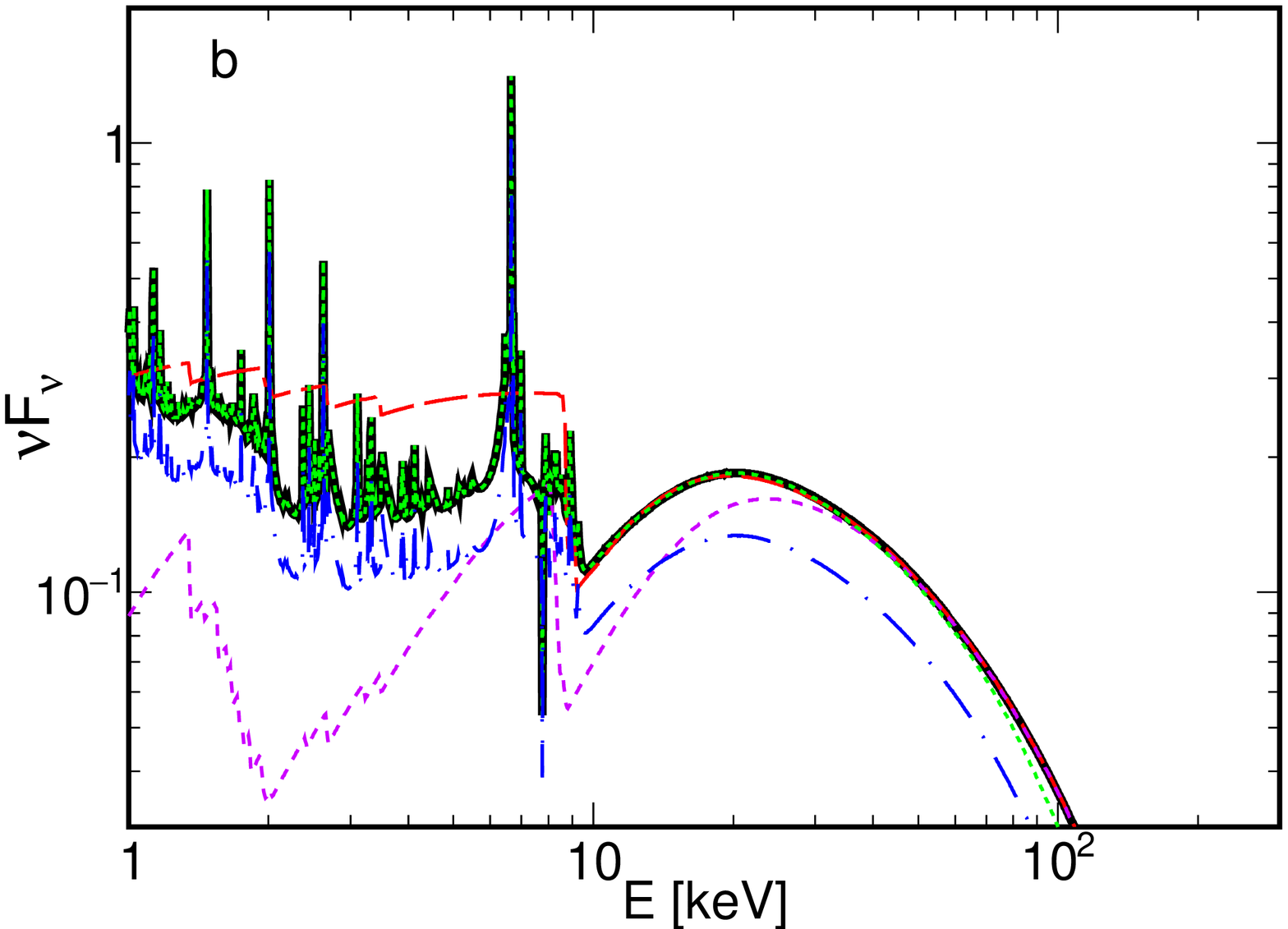}}
\caption{Similar to Fig.\ \ref{hreflect1} but for $\Gamma=2.1$, $E_{\rm cut}=90$ keV and $\xi=1000$. The short-dashed magenta curves show the \texttt{ireflect} spectra for $\xi=1000$ (i.e.\ $=\xi_{\rm h}$). 
The primary spectra are normalized at 1 keV.
} 
\label{hreflect2}
\end{figure}

\begin{figure}
\centerline{\includegraphics[width=7.5cm]{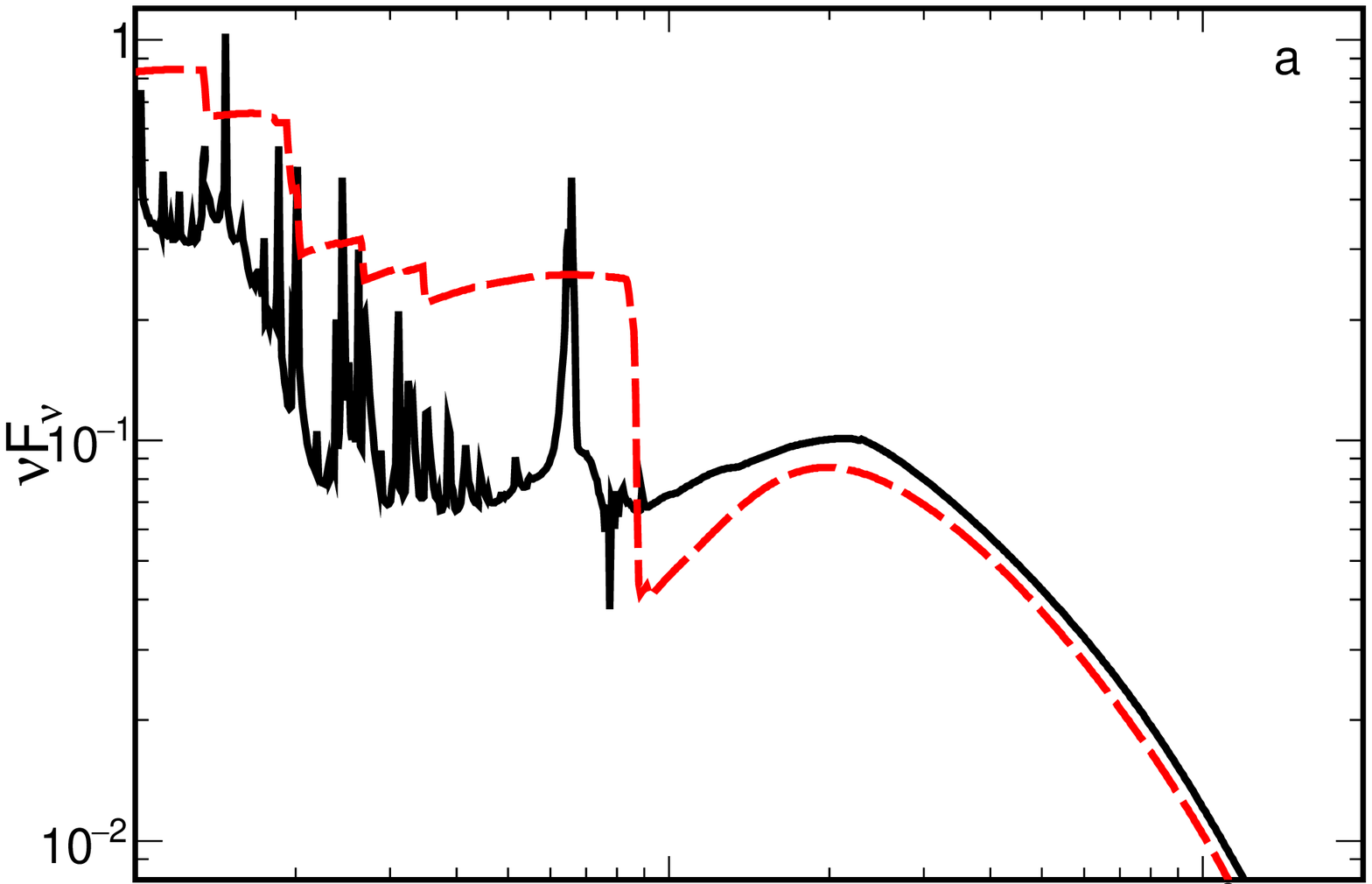}}
\centerline{\includegraphics[width=7.5cm]{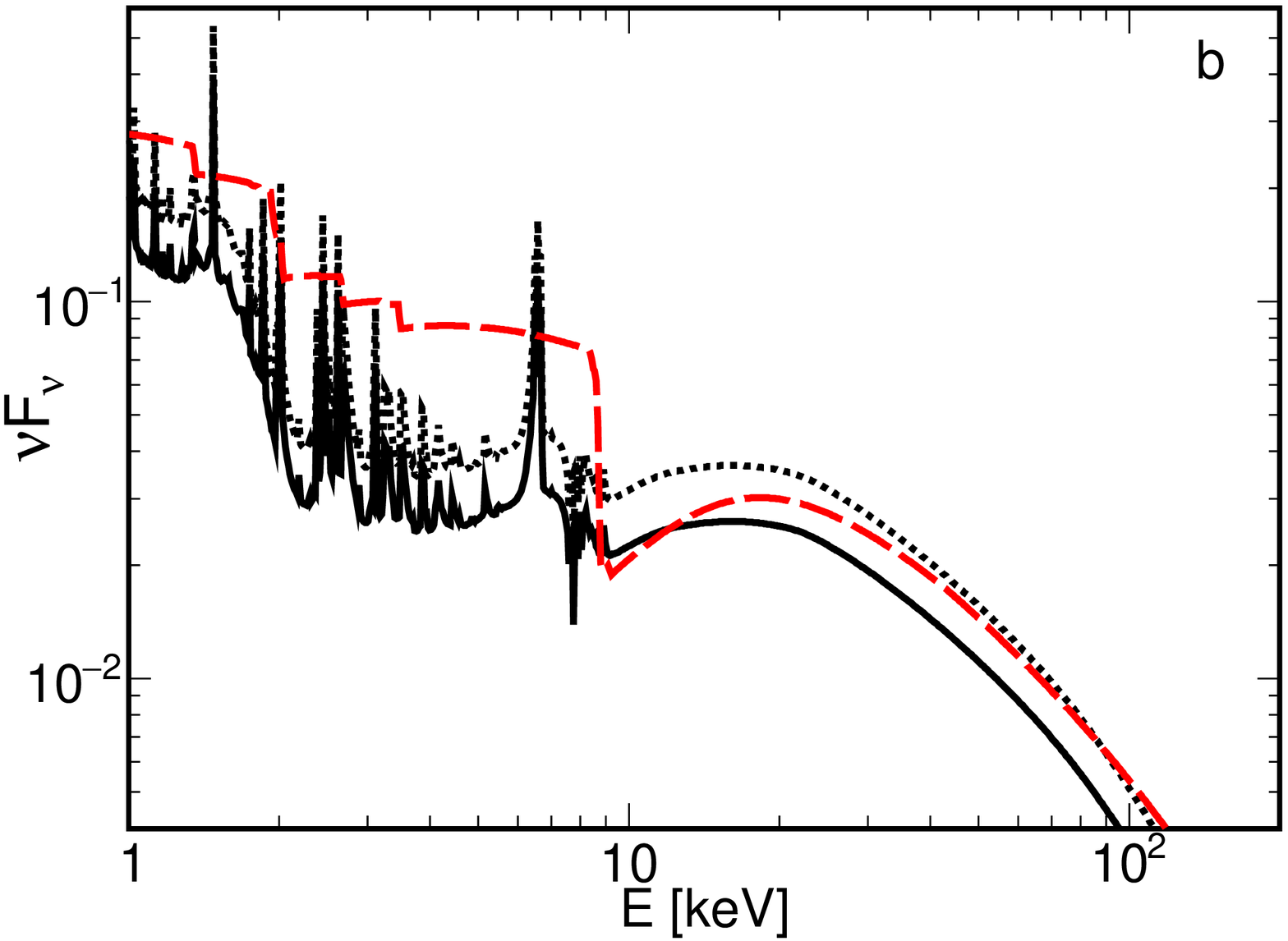}}
\caption{The black solid and red dashed curves show  the spectra of \texttt{pexriv} for $\xi = 10^5$ (i.e.\ the maximum value which can be used in this model)  and \texttt{xillver} for $\xi = 10^3$, respectively. The other parameters are  $\Gamma=2.7$, $E_{\rm cut}=300$ keV, (a) $\theta_{\rm obs} = 9\degr$ and (b) $\theta_{\rm obs} = 80\degr$. The dotted black curve in (b) shows the \texttt{xillver} spectrum rescaled by $f_{\theta}$. 
} 
\label{g27exp}
\end{figure}

\begin{figure}
\centerline{\includegraphics[width=7.5cm]{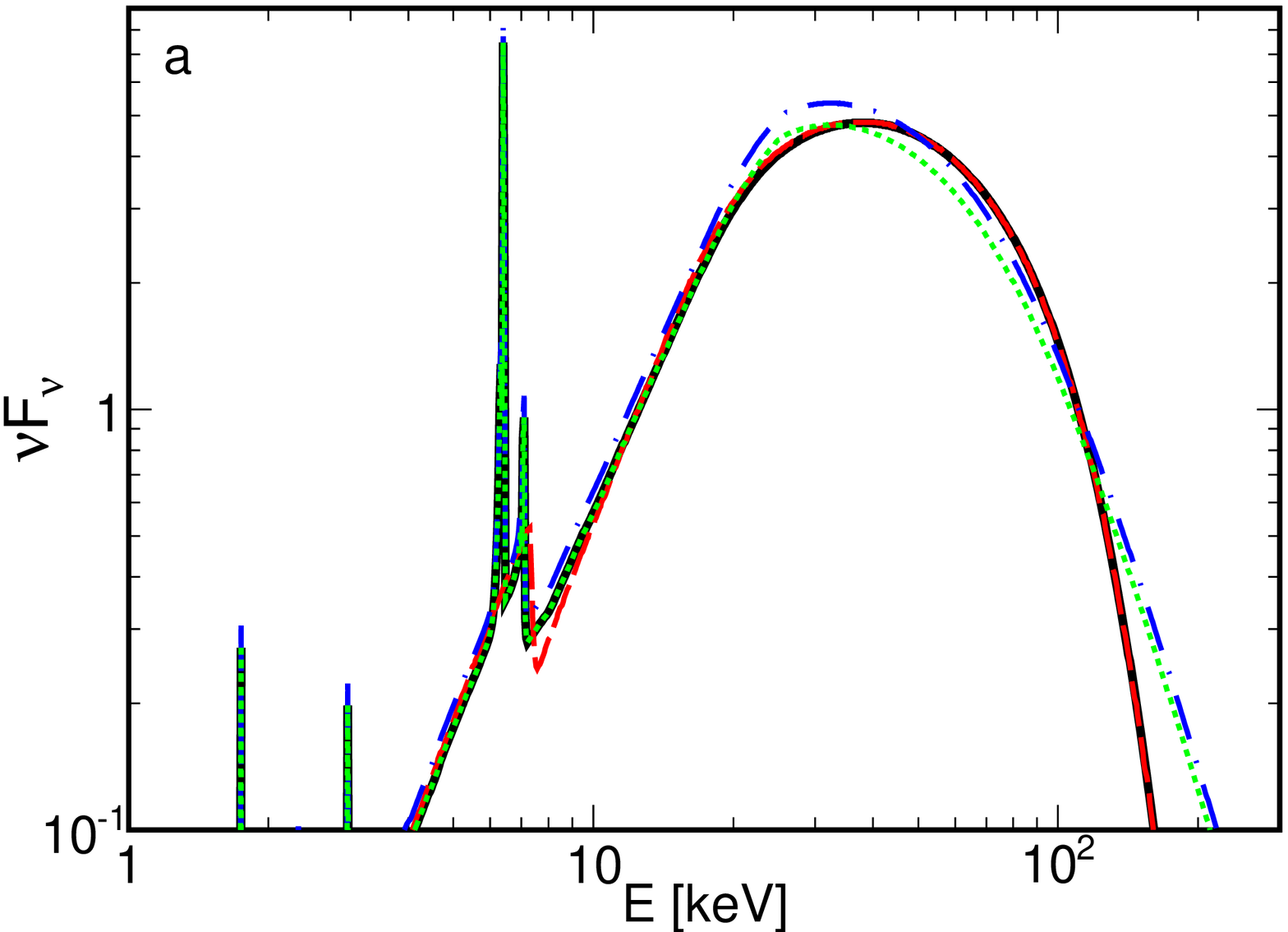}}
\centerline{\includegraphics[width=7.5cm]{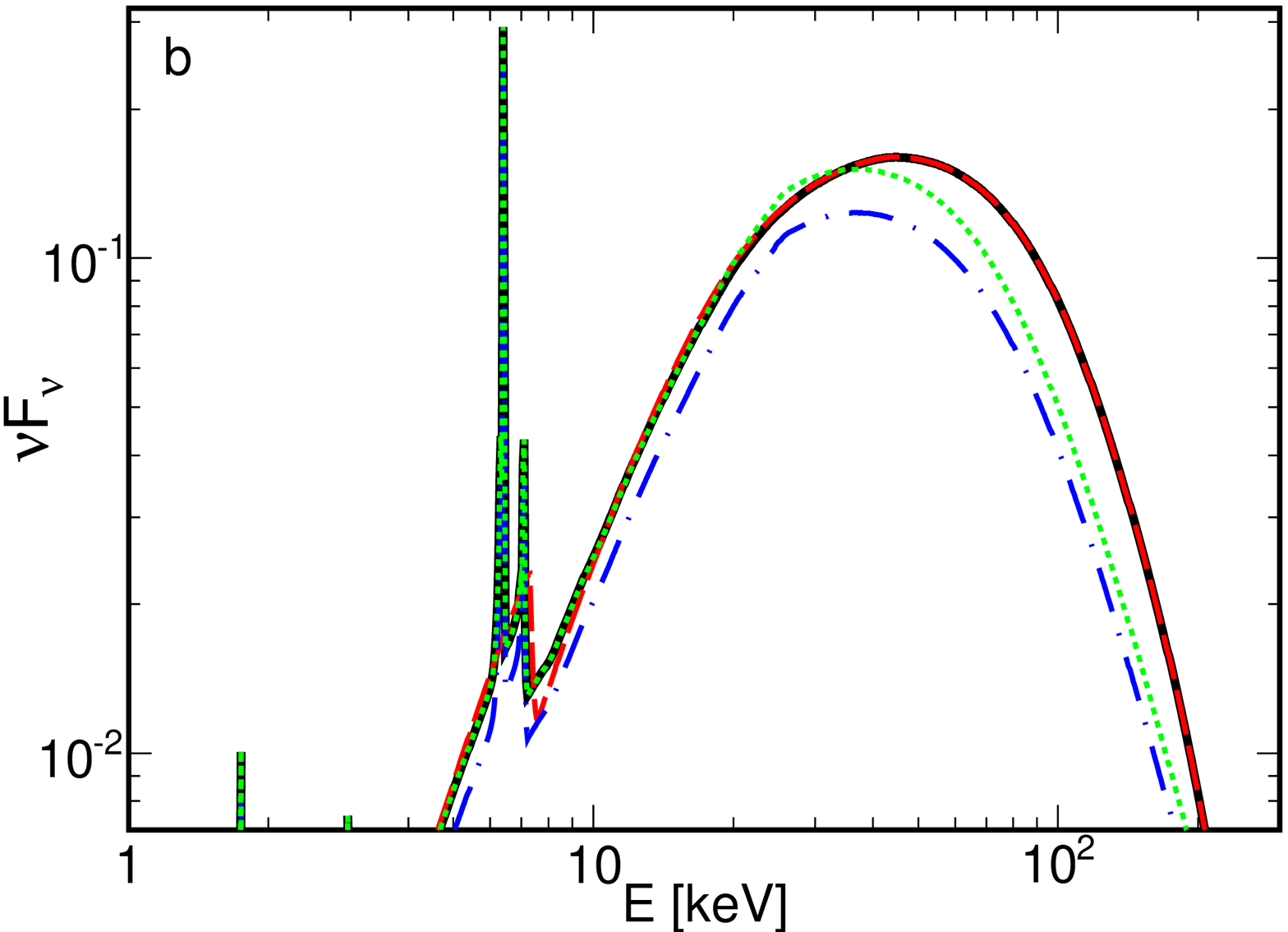}}
\caption{The solid black curves show the \texttt{hreflect} reflection for thermal Comptonization incident spectra with $\Gamma=1.6$, $kT_{\rm e}=30$ keV, $\xi=1$ and (a) $\theta_{\rm obs} = 9^\circ$ and (b) $\theta_{\rm obs} = 80^\circ$.  The dotted green curves show the \texttt{xillverCp} spectra for $kT_{\rm e}=34$ keV (fitted to \texttt{compps}, see Fig.\ \ref{compps_nthcomp}(a)) rescaled by $f$; the dot-dashed blue curves show \texttt{xillverCp} with the original normalization. The dashed red curves show the \texttt{ireflect} fitted to \texttt{xillverCp} in the 12--22 keV range.
The primary spectra are normalized at 1 keV.
} 
\label{hreflect3}
\end{figure}

\begin{figure}
\centerline{\includegraphics[width=7.5cm]{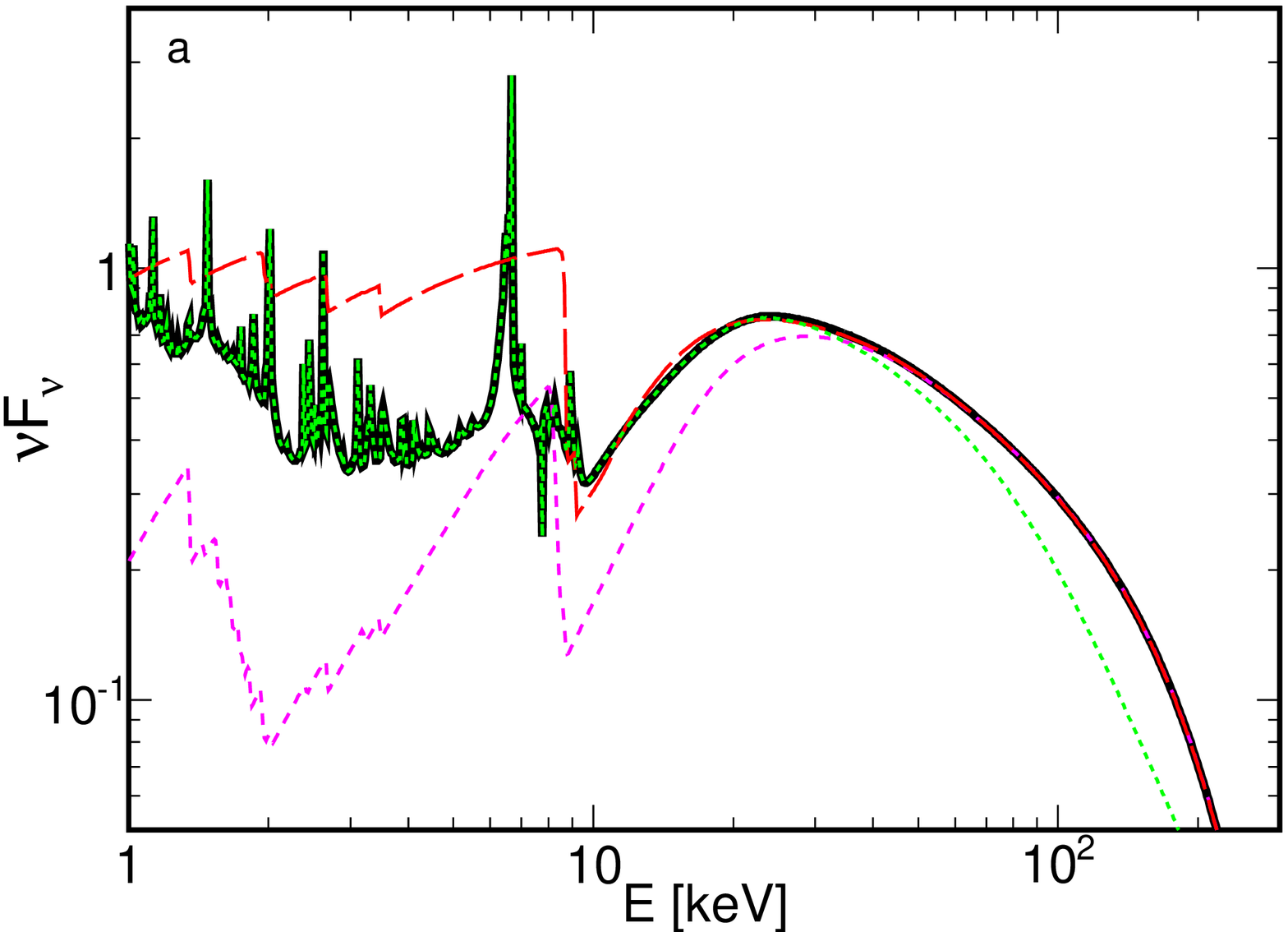}}
\centerline{\includegraphics[width=7.5cm]{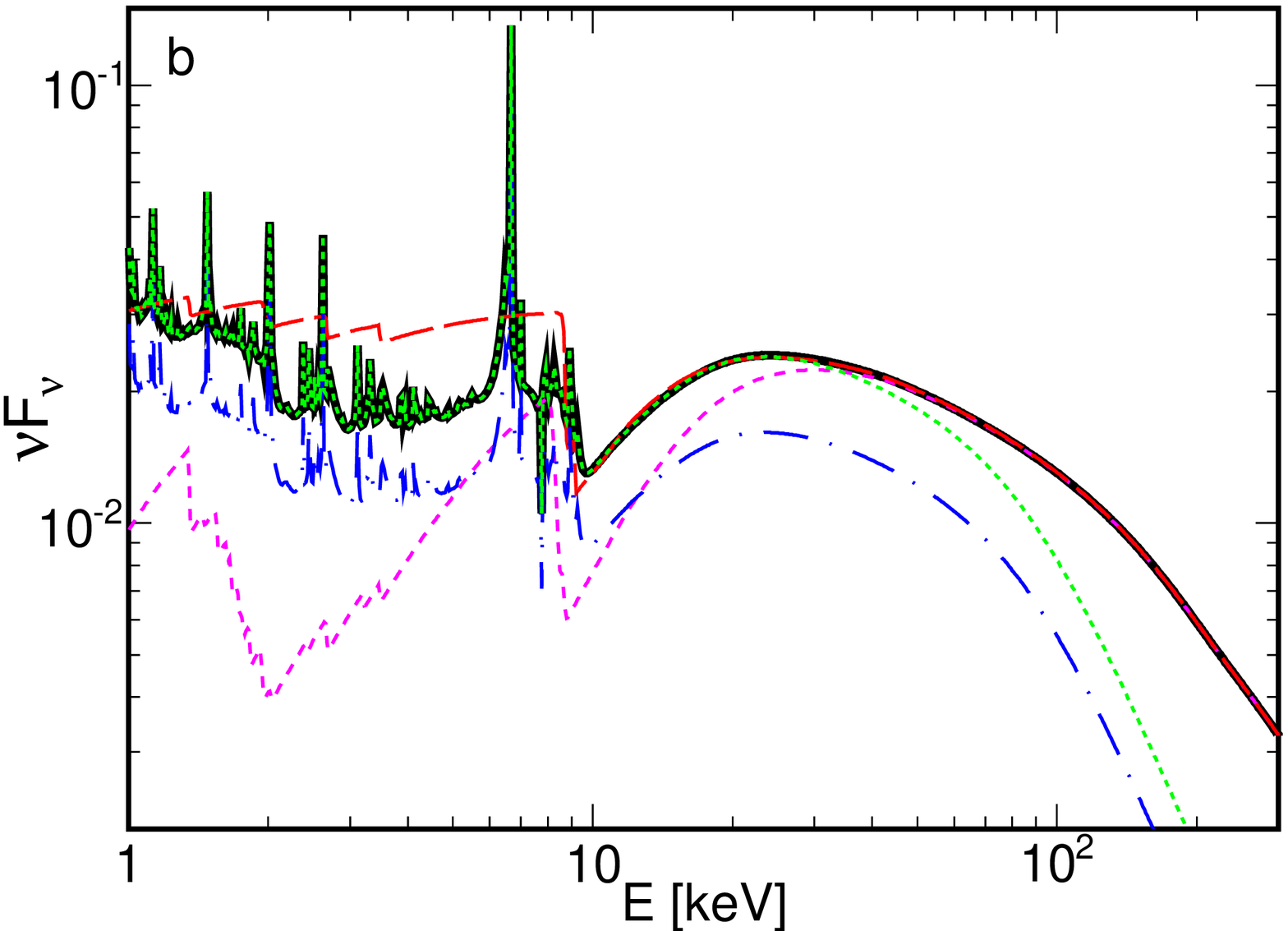}}
\caption{Similar to Fig.\ \ref{hreflect3} but for $\Gamma=2.1$, $kT_{\rm e}=200$ keV and $\xi=1000$. The \texttt{xillverCp} spectra are for $kT_{\rm e}=450$ keV. The short-dashed magenta curves show the \texttt{ireflect} spectra for $\xi=1000$  (i.e.\ $=\xi_{\rm h}$). 
The primary spectra are normalized at 1 keV.
} 
\label{hreflect4}
\end{figure}

\begin{figure}
\centerline{\includegraphics[height=5.5cm]{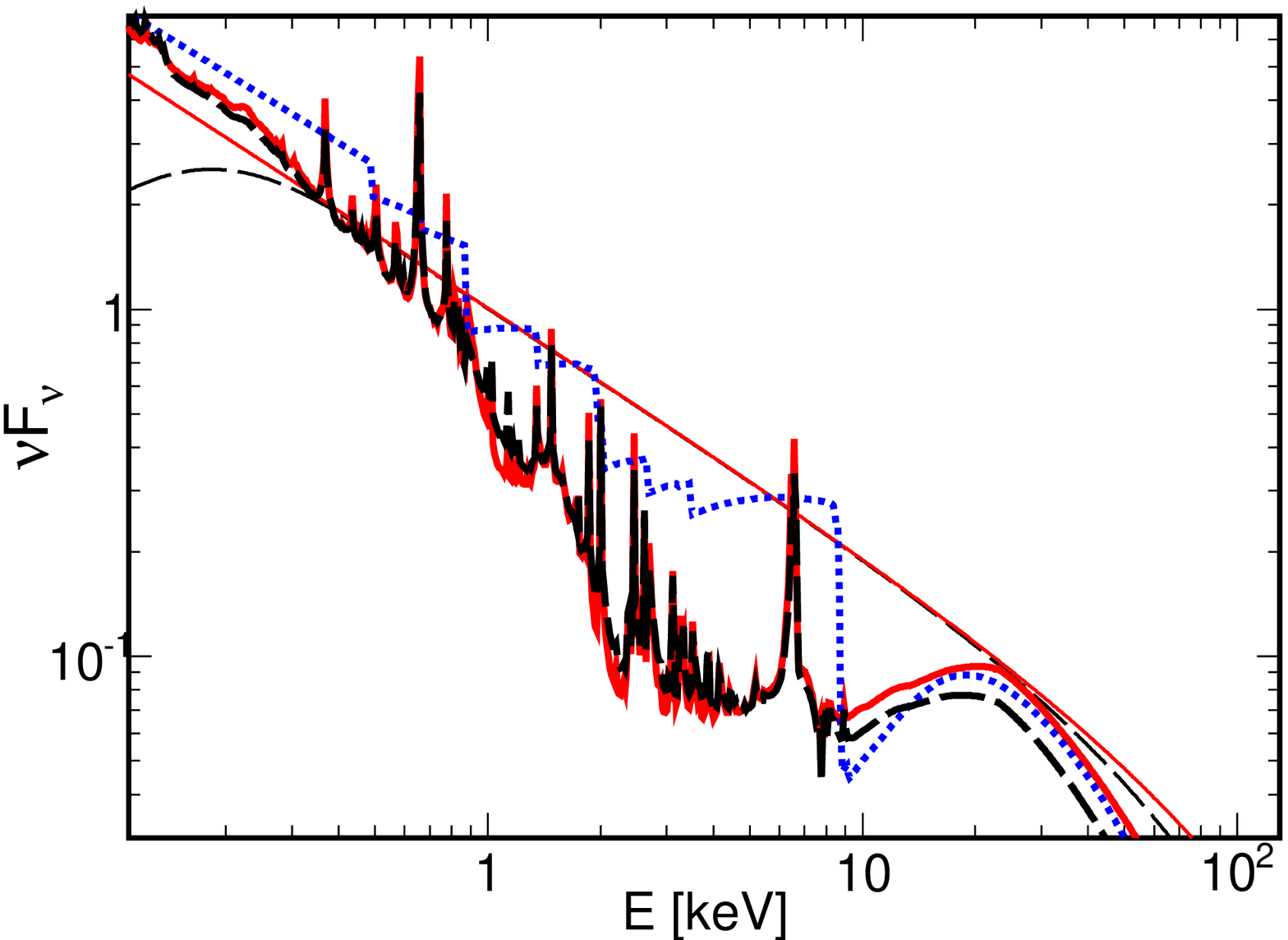}}
\caption{Comparison of reflection spectra for $\Gamma=2.7$, $\theta_{\rm obs} = 9\degr$ and high ionization parameters.
The blue dotted curve shows the \texttt{ireflect} spectrum for $\xi = 10^5$ and $kT=40$ keV. The thick solid red and dashed black curves show the \texttt{xillver} and \texttt{xillverCp}, respectively, reflection spectra for $\xi=1000$. The thinner black and red curves show the corresponding incident spectra (with  $kT_{\rm e}=110$ keV for \texttt{nthcomp} and $E_{\rm cut}=154$ keV for the e-folded power-law)
to illustrate their difference at low energies, where \texttt{xillver} assumes a sharp cutoff at 100 eV, while \texttt{xillverCp} has the low-energy cutoff at $\sim 300$ eV corresponding to $kT_{\rm bb}=50$ eV assumed for \texttt{nthcomp}. 
} 
\label{g27}
\end{figure}

\begin{figure}
\centerline{\includegraphics[height=7.5cm]{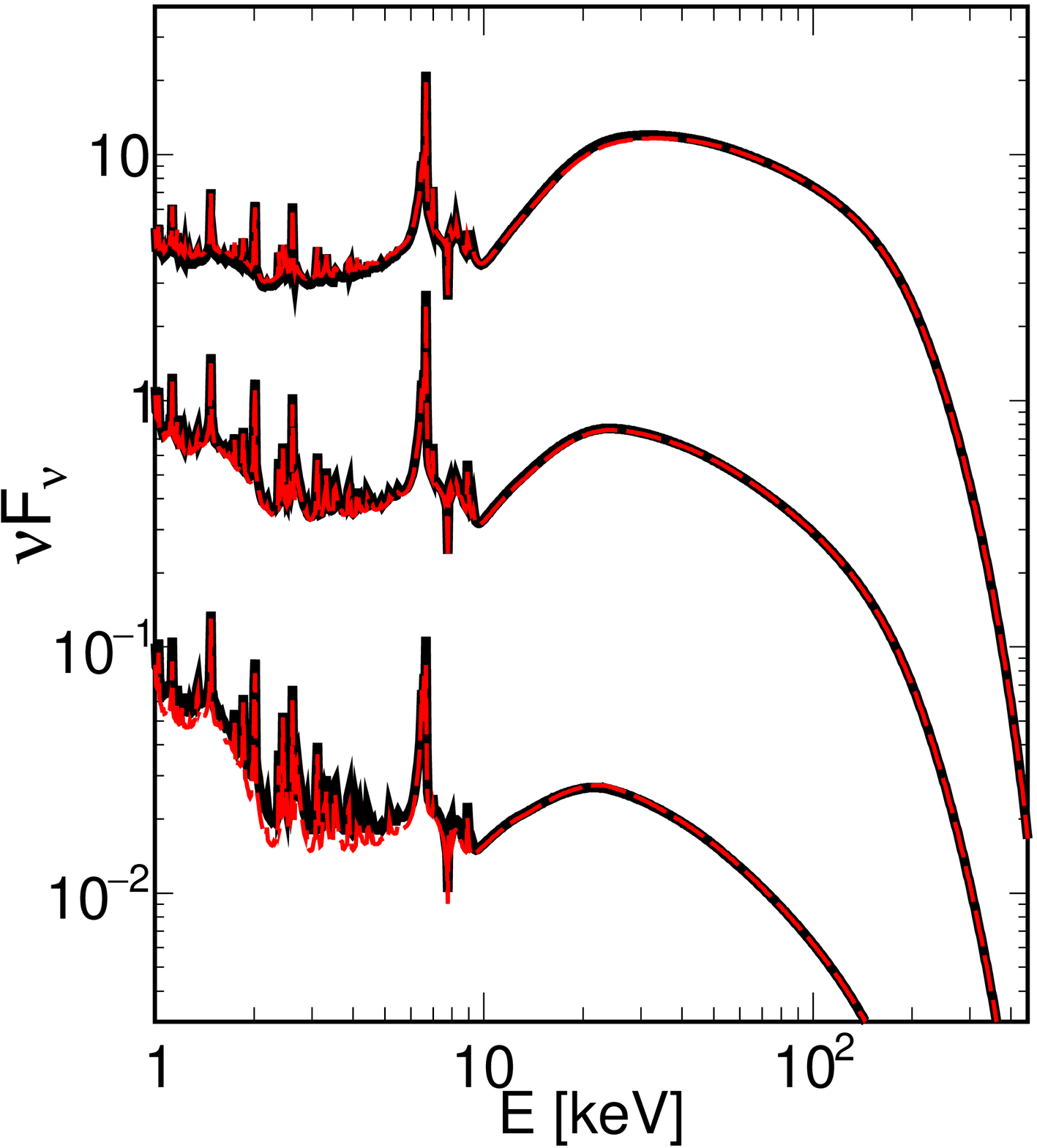}}
\caption{Comparison of the \texttt{hreflect} reflection spectra for $kT_{\rm e}=200$ keV, $\xi=1000$, $\theta_{\rm obs} = 9\degr$ and $\Gamma=1.8$ (scaled by 5), 2.1 and 2.4 (scaled by 0.2) from top to bottom, formed using either \texttt{xillverCp} with $kT_{\rm e}=450$ keV (black solid) or \texttt{xillver} with $E_{\rm cut}=444$, 461 and 485 keV, respectively (see Appendix \ref{hreflectEc5}; red dashed). The primary spectra are normalized at 1 keV.
} 
\label{exp_cp}
\end{figure}

\begin{figure}
\centerline{\includegraphics[width=6.5cm,scale=0.1]{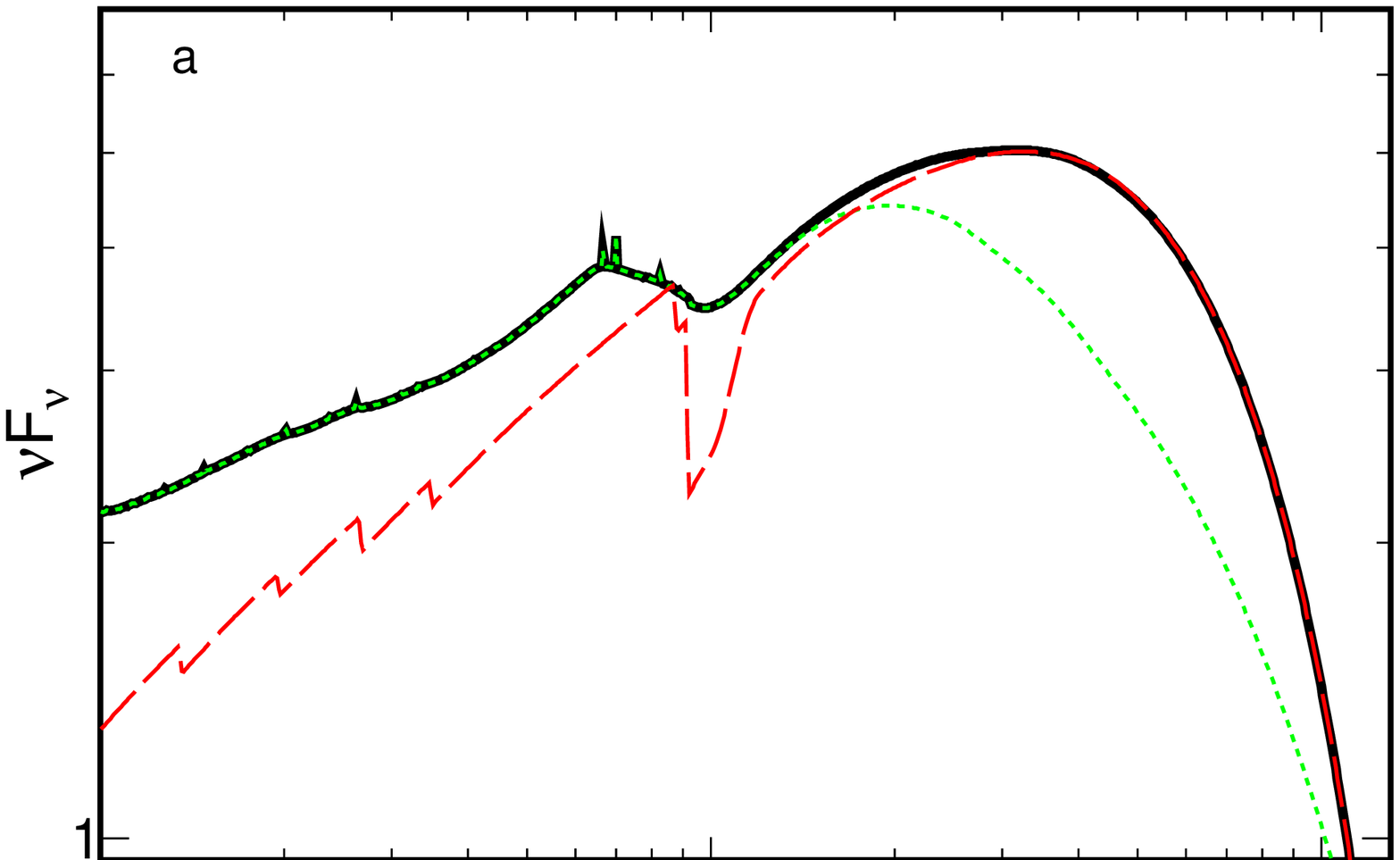}}
\centerline{\includegraphics[width=6.5cm,scale=0.1]{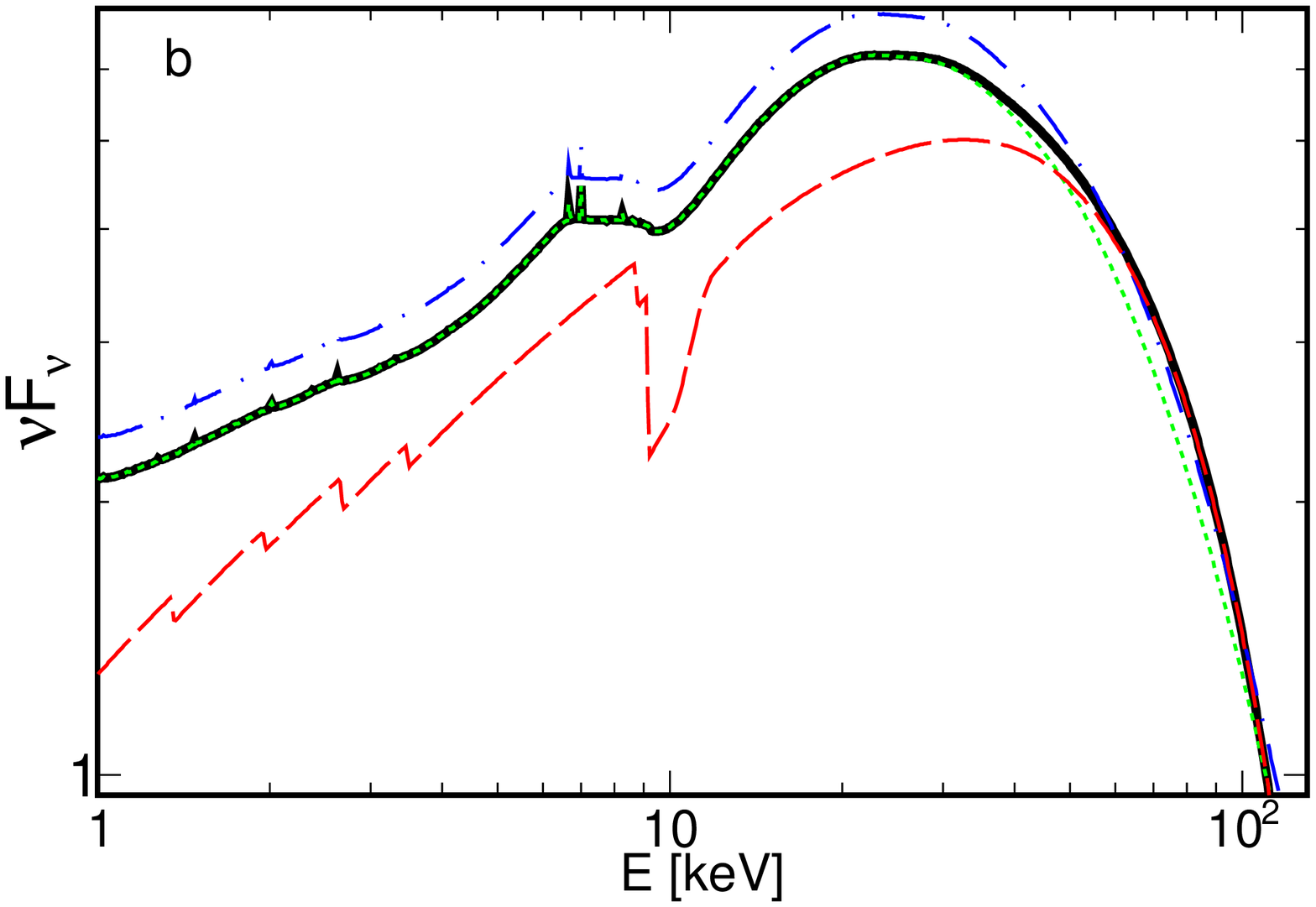}}
\caption{Comparison of the \texttt{hreflect} spectra (black solid curves) for $\xi=6000$, $\Gamma=1.6$, $kT_{\rm e}=30$ keV and $\theta_{\rm o} = 9\degr$, formed using (a) \texttt{xillver} and (b) \texttt{xillverCp}; the \texttt{xillver} components are shown by the dotted green curves. In both panels the red dashed curve shows the \texttt{ireflect} spectrum for $\xi = 10^5$, $\Gamma=1.6$, $kT_{\rm e}=30$ keV and $\theta_{\rm o} = 9\degr$. The blue dot-dashed curve in (b) shows the original (i.e.\ not scaled by $f_{\rm prim}$) \texttt{xillverCp} spectrum. The primary spectra are normalized at 1 keV.
} 
\label{xi6000}
\end{figure}

We describe here details of our hybrid reflection model. In particular, we apply rescaling of the \texttt{xillver} spectra to account for an angular distribution of reflection predicted by this model, which is different than that of \texttt{ireflect} (with the reflection strength in \texttt{ireflect} larger by a factor of $\simeq 1.5$ for edge-on directions). In some cases we apply also a minor rescaling related with differences of the incident spectra. We explain the criteria which we use to establish the rescaling factor, denoted below by $f$.    We follow an approach similar to that of \cite{dg06}.  We use \texttt{xillver}, which applies a more accurate absorption model, as a reference for the spectral shape below $\sim 20$ keV,  and \texttt{ireflect}, which correctly describes the angular distribution of the reflected radiation using the exact Klein-Nishina cross-section,  as a reference for the reflection amplitude.  The absorption model applied in \texttt{ireflect} assumes that the absorption coefficient above 12 keV is $\propto E^{-3}$, then, the depth of its absorption edge at these energies can be matched to that predicted by \texttt{xillver} by modifying any parameter (we choose $\xi$) affecting the absorption level at 12 keV.  As can be seen below, the shapes of \texttt{ireflect} spectra with such an adjusted $\xi$ typically agree very well with those of \texttt{xillver} in the $\sim 10$--30 keV range. Parameters of \texttt{hreflect} are denoted below with subscript 'h', in particular $\xi_{\rm h}$ denotes the ionization parameter in \texttt{hreflect}.

\subsection{E-folded power-law model}
\label{hreflectexp}

We first consider \texttt{hreflectExp}, which uses two reflection models, \texttt{xillver} and \texttt{pexriv}, assuming the same incident spectra parametrized by $\Gamma_{\rm h}$ and $E_{\rm cut,h}$. In both models we also use the same values of A$_{\rm Fe,h}$ and $\theta_{\rm obs,h}$ and we set $\xi=\xi_{\rm h}$ in \texttt{xillver}. Then, we  fit the \texttt{xillver} spectrum in the 12--22 keV range using \texttt{ireflect} with both $\xi$ and the normalization left as free parameters. The fitted $\xi$ of \texttt{ireflect}  is typically by a factor of several larger than $\xi_{\rm h}$. For low ionization parameters, the difference between the \texttt{pexriv} spectra for the fitted $\xi$ and for $\xi=\xi_{\rm h}$  is small and in Fig.\ \ref{hreflect1} we show only the former; in this regime (of low $\xi$) a very good agreement  between \texttt{xillver} and \texttt{pexriv} in the 12--22 keV energy range occurs for any A$_{\rm Fe}$ and $\Gamma$. For large $\xi$ the fitted \texttt{pexriv} spectrum differs significantly from  that for $\xi=\xi_{\rm h}$, see Fig.\ \ref{hreflect2}, and we note that in this regime (i.e.\ for large $\xi$) there are two cases where \texttt{xillver} and \texttt{pexriv} appear inconsistent and we could not find a robust method for merging them. Firstly, thermal Comptonization effects in the reflecting medium are very strong in all \texttt{xillver} models for $\Gamma \la 1.7$ and $\xi \ga 3000$, we discuss this case below. Secondly, for $\Gamma \ga 2.4$ and $\xi \ga 1000$ the spectra predicted by \texttt{xillver} are much flatter below $\sim 20$ keV than those of \texttt{pexriv}, see Fig.\ \ref{g27exp}. In this case, \texttt{xillver} produces also a stronger Compton hump than \texttt{pexriv}, which property does not occur at smaller $\Gamma$. 

The inverse of the fitted normalization of \texttt{pexriv} gives the scaling factor for \texttt{xillver} (i.e.\ $f$). We find that the only rescaling needed to adjust the reflection strength of these two models concerns the dependence on $\theta_{\rm obs}$ and it is given by $f_{\theta} = 1+0.1 \theta^{1.5}_{\rm obs}+2\sin^{10}(0.7 \theta_{\rm obs})$, where $\theta_{\rm obs}$ is measured  in radians. This scaling factor of  \texttt{xillver} does not depend on $\xi$, A$_{\rm Fe}$ or $\Gamma$.

\subsection{Thermal Comptonization: \texttt{ireflect} + \texttt{xillverCp}}
\label{hreflectCp}

Here we consider \texttt{ireflect} with the incident spectrum of \texttt{compps} and \texttt{xillverCp}, using in the latter model either $T_{\rm e}$ found by fitting \texttt{nthcomp} to \texttt{compps} or $kT_{\rm e}=450$ eV, if $T_{\rm e}$ exceeds the limit shown in Fig.\ \ref{map}. Similarly as above, we fit the \texttt{xillverCp} spectra in the 12--22 keV range using \texttt{ireflect} with $\xi$ and normalization treated as free parameters. Example results are shown in Figs \ref{hreflect3} and \ref{hreflect4}. Overall properties are similar to those found above for \texttt{pexriv} and \texttt{xillver}. In particular, we find that the difference of the angular distributions can be described by the $f_{\theta}(\theta_{\rm obs})$ function which has  the same form as defined above. However, here we find also an additional minor scaling by a factor $0.9 \la f_{\rm prim} \la 1.1$, which appears to be related with the difference of the incident spectra (note in Fig.\ \ref{compps_nthcomp}(a) the small difference between the fitted \texttt{nthcomp} and \texttt{compps} above $\sim 5$ keV). We find that these $f_{\rm prim}$ factors, found by fitting the reflection spectra, are approximately equal to the ratio of the \texttt{compps} to \texttt{nthcomp}  spectra at 30 keV and we apply such defined rescaling to form the \texttt{hreflect} spectra. Then, the total scaling factor is $f = f_{\theta} f_{\rm prim}$. The scaling by $f_{\rm prim}$ is illustrated by the small difference between the blue dot-dashed and green dotted spectra of \texttt{xillverCp} in Fig.\ \ref{hreflect3}(a).

\subsection{Thermal Comptonization: \texttt{ireflect} + \texttt{xillver}}
\label{hreflectEc5}

Here we consider \texttt{ireflect} convolved with \texttt{compps} and \texttt{xillver}. For the latter model, we use $E_{\rm cut}$, for which the e-folded power-law gives the same Compton temperature as \texttt{compps} (in both cases the lower limit of integration is 100 eV, corresponding to the low energy cutoff used for computing the \texttt{xillver} tables). Again, we find that the angular distribution of \texttt{xillver} and \texttt{ireflect} models  agree after rescaling the former by the $f_{\theta}$ function defined above. This version of the model is intended for high $T_{\rm e}$, in particular those giving departures from a power-law shape, therefore we do not use scaling by $f_{\rm prim}$, which would introduce artificial effects in this regime.

\subsection{Overview}

In all versions of the model we found that the angular distribution of \texttt{xillver} can be adjusted to that of \texttt{ireflect} using the same function $f_{\theta}$. As noted in Section \ref{comparison}, this rescaling insignificantly affects the reflection spectra of our relativistic models, because contribution from large emission angles in the disc frame is weak at small $r$. 

While the above method (i.e.\ involving fitting the $\xi$ parameter of \texttt{ireflect}) allowed us to establish the most accurate procedure for merging the reflection models, implementing it in the version of \texttt{hreflect} used for X-ray fitting is not feasible because the computation of \texttt{ireflect} spectra is the most time-consuming part of it and increasing the number of such computations (by a factor of several) would make it too slow. Therefore, in the X-ray fitting model we use \texttt{ireflect} with $\xi = \xi_{\rm h}$. The $E_{\rm merge}$ is then defined 
as the energy (larger than 12 keV) of the intersection of this \texttt{ireflect} (with $\xi_{\rm h}$) and \texttt{xillver}, being $E_{\rm merge} \simeq 15$ keV for $\xi=1$ and $E_{\rm merge} \simeq 30$ keV for $\xi=1000$. At $E > E_{\rm merge}$ the shape of \texttt{ireflect} is independent of $\xi$ and then using $\xi_{\rm h}$ we get the same \texttt{hreflect} spectrum (to within a few per cent) as that using the fitted $\xi$; we checked this for a large range (corresponding to the tabulation grid of \texttt{xillver}) of A$_{\rm Fe}$, $\xi$, $\Gamma$. In principle, \texttt{hreflect} using \texttt{xillverCp} and \texttt{xillver} should be more accurate at low ($\la 60$ keV) and high ($\ga 60$ keV) electron temperatures, respectively, as their incident spectra better approximate \texttt{compps} in such cases. However, for $T_{\rm e}$ below the limit shown in Fig.\ \ref{boundary}, we typically find that the difference between these two versions does not exceed about 10 per cent, see Fig.\ \ref{exp_cp}. Above this limit (i.e.\ at $T_{\rm e}$ of a few hundred keV, depending on $\Gamma$ and $T_{\rm bb}$) the version with \texttt{xillver} should be used, because arguments underlying the rescaling of \texttt{xillverCp} by $f_{\rm prim}$ are invalid in this regime.

Similarly as for \texttt{hreflectExp}, we note significant uncertainty regarding the reflection of thermal Comptonization in two ranges of parameters. For $\Gamma \ga 2.4$ and large $\xi$ ($ \ga 1000$),  both versions of \texttt{xillver} appear inconsistent with \texttt{ireflect}, see Fig.\ \ref{g27}. It is not clear which spectrum is closer to the true one in this range of parameters, as inaccuracies in the description of both the downscattering in  \texttt{xillver} and the absorption effects in \texttt{ireflect} may contribute to this disagreement. 

Fig.\ \ref{xi6000} shows the reflection spectra for a hard incident spectrum with low electron temperature, which parameters are relevant to observations of luminous hard states in some X-ray binaries. For $\Gamma \la 1.7$ and $\xi \ga 3000$, \texttt{xillver} predicts a very large temperature in the upper layers of the reflecting material, $kT \ga 10$ keV, in some cases even larger than the corresponding Compton temperature, see e.g.\ figures 4 and 8 in \cite{garcia13}. This, with the (assumed in \texttt{xillver}) Gaussian redistribution of energy including the temperature term for Compton scattering, see equations (6) and (7)  in \cite{garcia13}, most likely strongly overestimates thermal Comptonization effects in the reflected spectrum. As we see in Fig.\ \ref{xi6000}, the \texttt{hreflect} using \texttt{xillver}  discards a strong enhancement of the Compton hump by thermal Comptonization, which occurs in the model using \texttt{xillverCp}. Then, the former may present a better approximation of the (very uncertain) true reflection spectra for these parameters.

\label{lastpage}
\end{document}